\documentclass[11pt,oldtoc,a4paper]{cernart}  
\usepackage{times}
\usepackage{amsmath} 
\usepackage{amsfonts}
\usepackage{amssymb}
\usepackage{pifont}
\usepackage{graphicx}
\usepackage{cite}
\usepackage{multicol}
\usepackage{color} 
\PassOptionsToPackage{linktocpage}{hyperref}
\begin{document}
\setlength{\textheight}{24cm} 
\setcounter{tocdepth}{3}

\renewcommand{\t}{\ensuremath{\tau }}
\renewcommand{\i}{\ensuremath{\mathrm{i}}}
\renewcommand{\a}{\ensuremath{\alpha }}
\newcommand{\PL}{Phys.\ Lett.\ }
\newcommand{\EPJ}{Eur.\ Phys.\ J.\ }
\newcommand{\re}{\ensuremath{\mathrm{Re}}}
\newcommand{\im}{\ensuremath{\mathrm{Im}}}
\newcommand{\e}{\ensuremath{\mathrm{e}}}
\newcommand{\s}{\ensuremath{\mathrm{s}}}
\newcommand{\nn}{\ensuremath{\mathrm{n}}}
\newcommand{\p}{\ensuremath{\mathrm{p}}}
\newcommand{\gev}{\ensuremath{\mathrm{GeV}/c}}
\newcommand{\mev}{\ensuremath{\mathrm{MeV}/c}}
\newcommand{\geve}{\ensuremath{\mathrm{GeV}}}
\newcommand{\br}{\ensuremath{\mathrm{BR}}}
\newcommand{\brsf}{\ensuremath{\mathrm{BR}_f^{\mathrm{S}}}}
\newcommand{\brlf}{\ensuremath{\mathrm{BR}_f^{\mathrm{L}}}}
\newcommand{\cs}{\ensuremath{\mathrm{cos}}}
\newcommand{\sn}{\ensuremath{\mathrm{sin}}}
\newcommand{\sbraket}[2]{\langle #1 | #2 \rangle}
\newcommand{\ra}{\ensuremath{\rightarrow}}
\newcommand{\pvec}{\ensuremath{\vec{p}}}
\newcommand{\pt}{\ensuremath{p_{\mathrm T}}}
\newcommand{\dt}{\ensuremath{d_{\mathrm T}}}
\newcommand{\gspm}{\ensuremath{\mathrm{\Gamma_S^{\pipi}}}}
\newcommand{\gsppp}{\ensuremath{\mathrm{\Gamma_S^{\pipb}}}}
\newcommand{\glppp}{\ensuremath{\mathrm{\Gamma_L^{\pipb}}}}
\newcommand{\gspmn}{\ensuremath{\mathrm{\Gamma_S^{+-0}}}}
\newcommand{\glpmn}{\ensuremath{\mathrm{\Gamma_L^{+-0}}}}
\newcommand{\eps}{\ensuremath{\epsilon}}
\newcommand{\epn}{\ensuremath{\varepsilon}}
\newcommand{\epnb}{\ensuremath{\bar{\varepsilon}}}
\newcommand{\et}{\ensuremath{\mathrm{\epn_T}}}
\newcommand{\el}{\ensuremath{\mathrm{\epn\ - \den}}}
\newcommand{\es}{\ensuremath{\mathrm{\epn_S}}}
\newcommand{\rl}{\ensuremath{r_L}}
\newcommand{\rs}{\ensuremath{r_S}}
\newcommand{\tel}{\ensuremath{\mathrm{\tilde{\epn}_L}}}
\newcommand{\tes}{\ensuremath{\mathrm{\tilde{\epn}_S}}}
\newcommand{\epsp}{\ensuremath{\epn^{\prime}}}
\newcommand{\ree}{\ensuremath{\re (\epn )}}
\newcommand{\reel}{\ensuremath{\re (\el )}}
\newcommand{\rees}{\ensuremath{\re\ (\es )}}
\newcommand{\del}{\ensuremath{\delta}}
\newcommand{\delb}{\ensuremath{\bar{\delta}}}
\newcommand{\delcpt}{\ensuremath{\delta_{\mathrm CPT}}}
\newcommand{\den}{\ensuremath{\delta}}
\newcommand{\dell}{\ensuremath{\delta_{\ell}}}
\newcommand{\dper}{\ensuremath{\del _{\perp}}}
\newcommand{\dpar}{\ensuremath{\del _{\parallel}}}
\newcommand{\red}{\ensuremath{\re (\del )}}
\newcommand{\imd}{\ensuremath{\im (\del )}}
\newcommand{\itapm}{\ensuremath{\eta_{+-}}}
\newcommand{\mitapm}{\ensuremath{|\eta_{+-}|}}
\newcommand{\fpm}{\ensuremath{\phi_{+-}}}
\newcommand{\itaoo}{\ensuremath{\eta_{00}}}
\newcommand{\mitaoo}{\ensuremath{|\eta_{00}|}}
\newcommand{\foo}{\ensuremath{\phi_{00}}}
\newcommand{\ita}{\ensuremath{\eta}}
\newcommand{\mita}{\ensuremath{|\eta|}}
\newcommand{\fsw}{\ensuremath{\phi_\mathrm{SW}}}
\newcommand{\reitapmo}{\ensuremath{\re (\eta_{+-0})}}
\newcommand{\imitapmo}{\ensuremath{\im (\eta_{+-0})}}
\newcommand{\reitaooo}{\ensuremath{\re (\eta_{000})}}
\newcommand{\imitaooo}{\ensuremath{\im (\eta_{000})}}
\newcommand{\itapmo}{\ensuremath{\eta_{+-0}}}
\newcommand{\mitapmo}{\ensuremath{|\eta_{+-0}|}}
\newcommand{\fpmo}{\ensuremath{\phi_{+-0}}}
\newcommand{\itaooo}{\ensuremath{\eta_{000}}}
\newcommand{\mitaooo}{\ensuremath{|\eta_{000}|}}
\newcommand{\fooo}{\ensuremath{\phi_{000}}}
\newcommand{\itapp}{\ensuremath{\eta_{\pipa}}}
\newcommand{\mitapp}{\ensuremath{|\eta_{\pipa}|}}
\newcommand{\fpp}{\ensuremath{\phi_{\pipa}}}
\newcommand{\itappp}{\ensuremath{\eta_{\pipa\pi}}}
\newcommand{\mitappp}{\ensuremath{|\eta_{\pipa\pi}|}}
\newcommand{\fppp}{\ensuremath{\phi_{\pipa\pi}}}
\newcommand{\amz}{\ensuremath{{\cal A}}}
\newcommand{\amb}{\ensuremath{\overline{{\cal A}}}}
\newcommand{\afz}{\ensuremath{{\cal A}_f}}
\newcommand{\afb}{\ensuremath{\overline{{\cal A}}_f}}
\newcommand{\afzc}{\ensuremath{{\cal A}_f^*}}
\newcommand{\afbc}{\ensuremath{\overline{{\cal A}}_f^*}}
\newcommand{\aiz}{\ensuremath{{\cal A}_I}}
\newcommand{\aib}{\ensuremath{\overline{{\cal A}}_I}}
\newcommand{\ailz}{\ensuremath{{\cal A}_{I,l}^{3\pi}}}
\newcommand{\ailb}{\ensuremath{{\bar{\cal A}}_{I,l}^{3\pi}}}
\newcommand{\apz}{\ensuremath{{\cal A}_I^{\pipa}}}
\newcommand{\apb}{\ensuremath{\overline{{\cal A}_I^{\pipa}}}}
\newcommand{\anp}{\ensuremath{{\cal A_+}}}
\newcommand{\abp}{\ensuremath{\overline{{\cal A}}_+}}
\newcommand{\anm}{\ensuremath{{\cal A_-}}}
\newcommand{\abm}{\ensuremath{\overline{{\cal A}}_-}}
\newcommand{\ams}{\ensuremath{{\cal A}_{f{\mathrm S}}}}
\newcommand{\amsc}{\ensuremath{{\cal A}_{f{\mathrm S}}^*}}
\newcommand{\aml}{\ensuremath{{\cal A}_{f{\mathrm L}}}}
\newcommand{\Aa}{\ensuremath{\mathrm{A}_0}}
\newcommand{\Ab}{\ensuremath{\mathrm{A}_2}}
\newcommand{\Ba}{\ensuremath{\mathrm{B}_0}}
\newcommand{\Bb}{\ensuremath{\mathrm{B}_2}}
\newcommand{\aab}{\ensuremath{A_0^*\overline{A_0}}}
\newcommand{\reaa}{\ensuremath{\re (A_0)}}
\newcommand{\imaa}{\ensuremath{\im (A_0)}}
\newcommand{\reba}{\ensuremath{\re (B_0)}}
\newcommand{\reab}{\ensuremath{\re (A_1)}}
\newcommand{\rebb}{\ensuremath{\re (B_1)}}
\newcommand{\reac}{\ensuremath{\re (A_2)}}
\newcommand{\imac}{\ensuremath{\im (A_2)}}
\newcommand{\rebc}{\ensuremath{\re (B_2)}}
\newcommand{\rebi}{\ensuremath{\re (B_I)}}
\newcommand{\reai}{\ensuremath{\re (A_I)}}
\newcommand{\imbi}{\ensuremath{\im (B_I)}}
\newcommand{\yy}{\ensuremath{y}}
\newcommand{\xp}{\ensuremath{x_{+}}}
\newcommand{\xm}{\ensuremath{x_{-}}}
\newcommand{\xx}{\ensuremath{x}}
\newcommand{\xb}{\ensuremath{\overline{x}}}
\newcommand{\rex}{\ensuremath{\re (x)}}
\newcommand{\imx}{\ensuremath{\im (x)}}
\newcommand{\rexb}{\ensuremath{\re (\overline{x})}}
\newcommand{\imxb}{\ensuremath{\im (\overline{x})}}
\newcommand{\rey}{\ensuremath{\re (y)}}
\newcommand{\imxm}{\ensuremath{\im (x_{-})}}
\newcommand{\imxp}{\ensuremath{\im (x_{+})}}
\newcommand{\rexm}{\ensuremath{\re (x_{-})}}
\newcommand{\rexp}{\ensuremath{\re (x_{+})}}
\newcommand{\aklil}{\ensuremath{{\cal A}_{\mathrm L}^{l,I}}}
\newcommand{\aklis}{\ensuremath{{\cal A}_{\mathrm S}^{l,I}}} 
\newcommand{\akli}{\ensuremath{a_{l,I}}}             
\newcommand{\askli}{\ensuremath{a^{*}_{l,I}}}        
\newcommand{\deltai}{\ensuremath{\delta_{I}}}        
\newcommand{\edelti}{\ensuremath{\e^{\i\delta_{I}}}} 
\newcommand{\asxy}{\ensuremath{{\cal A}^{+-0}_\mathrm{S}(X,Y)}}   
\newcommand{\asxym}{\ensuremath{{\cal A}^{+-0}_\mathrm{S}(-X,Y)}} 
\newcommand{\alxy}{\ensuremath{{\cal A}^{+-0}_\mathrm{L}(X,Y)}}   
\newcommand{\alxym}{\ensuremath{{\cal A}^{+-0}_\mathrm{L}(-X,Y)}} 
\newcommand{\asxyo}{\ensuremath{{\cal A}^{000}_\mathrm{S}(X,Y)}}  
\newcommand{\alxyo}{\ensuremath{{\cal A}^{000}_\mathrm{L}(X,Y)}}  
\newcommand{\asxya}{\ensuremath{{\cal A}^{3\pi}_\mathrm{S}(X,Y)}} 
\newcommand{\alxya}{\ensuremath{{\cal A}^{3\pi}_\mathrm{L}(X,Y)}} 
\newcommand{\al}{\ensuremath{{\cal A}^{+-0}(X,Y)}}                
\newcommand{\all}{\ensuremath{{\cal A}}}
\newcommand{\alcpc}{\ensuremath{{\cal A}_\mathrm{L}^{3\pi(\mathrm{CP=-1})}
                                (X,Y)}} 
\newcommand{\alcpv}{\ensuremath{{\cal A}_\mathrm{L}^{3\pi(\mathrm{CP=+1})}
                                (X,Y)}} 
\newcommand{\alcpcs}{\ensuremath{{\cal A}_\mathrm{L}^{*\ 3\pi(\mathrm{CP=-1})}
                                (X,Y)}} 
\newcommand{\ascpc}{\ensuremath{{\cal A}_\mathrm{S}^{3\pi (\mathrm{CP=+1})}
                                (X,Y)}} 
\newcommand{\ascpcm}{\ensuremath{{\cal A}_\mathrm{S}^{3\pi (\mathrm{CP=+1})}
                                (-X,Y)}} 
\newcommand{\ascpv}{\ensuremath{{\cal A}_\mathrm{S}^{3\pi(\mathrm{CP=-1})}
                                (X,Y)}} 
\newcommand{\apm}{\ensuremath{A_{+-}(\t )}}     
\newcommand{\apmz}{\ensuremath{A_{+-0}(\t )}}   
\newcommand{\aepmz}{\ensuremath{A_{+-0}^\mathrm{exp}(\t )}}  
\newcommand{\apmzxp}{\ensuremath{A_{+-0}(X>0, \t )}}   
\newcommand{\aepmzxp}{\ensuremath{A_{+-0}^\mathrm{exp}(X>0, \t )}}          
\newcommand{\apmzxm}{\ensuremath{A_{+-0}(X<0, \t )}} 
\newcommand{\aepmzxm}{\ensuremath{A_{+-0}^\mathrm{exp}(X<0, \t )}}       
\newcommand{\relam}{\ensuremath{\re (\lz )}}
\newcommand{\imlam}{\ensuremath{\im (\lz )}}
\newcommand{\xin}{\mbox{$\xi_\mathrm{N}$}}             
\newcommand{\xinxp}{\mbox{$\xi_\mathrm{N}^{X>0}$}}     
\newcommand{\xinxm}{\mbox{$\xi_\mathrm{N}^{X<0}$}}     
\newcommand{\xixy}{\mbox{$\xi_{XY}$}}                  
\newcommand{\zeb}{\ensuremath{\zeta_\mathrm{B}}}       
\newcommand{\n}{\ensuremath{n}}
\newcommand{\m}{\ensuremath{m}}
\newcommand{\x}{\ensuremath{X}}
\newcommand{\y}{\ensuremath{Y}}
\newcommand{\PPb}{\ensuremath{\mathrm{\overline{p}p}}}
\newcommand{\Pp}{\ensuremath{\mathrm{p}}}
\newcommand{\Pb}{\ensuremath{\mathrm{\overline{p}}}}
\newcommand{\pb}{\ensuremath{\mathrm{\overline{p}}}}
\newcommand{\pp}{\ensuremath{\mathrm{p}}}
\newcommand{\ppb}{\ensuremath{\mathrm{\overline{p}p}}}
\newcommand{\pipa}{\ensuremath{\pi\pi }}
\newcommand{\pipb}{\ensuremath{\pi\pi\pi }}
\newcommand{\pipi}{\ensuremath{\pip\pim }}
\newcommand{\pip}{\ensuremath{\mathrm{\pi^+}}}
\newcommand{\pim}{\ensuremath{\mathrm{\pi^-}}}
\newcommand{\pipm}{\ensuremath{\mathrm{\pi^{\pm}}}}
\newcommand{\pimp}{\ensuremath{\mathrm{\pi^{\mp}}}}
\newcommand{\pin}{\ensuremath{\mathrm{\pi^{0}}}}
\newcommand{\pinn}{\ensuremath{\mathrm{\pi^0\pi^0}}}
\newcommand{\pinnn}{\ensuremath{\mathrm{\pi^0\pi^0\pi^0}}}
\newcommand{\pipmn}{\ensuremath{\mathrm{\pi^+\pi^-\pi^0}}}
\newcommand{\semi}{\ensuremath{\ell \pi \nu }}
\newcommand{\pen}{\ensuremath{\e \pi \nu }}
\newcommand{\penp}{\ensuremath{\elp \pim \net }}
\newcommand{\penm}{\ensuremath{\elm \pip \netb }}
\newcommand{\pmn}{\ensuremath{\mu \pi \nu }}
\newcommand{\elp}{\ensuremath{\mathrm{e^+}}}
\newcommand{\elm}{\ensuremath{\mathrm{e^-}}}
\newcommand{\elpm}{\ensuremath{\mathrm{e^{\pm}}}}
\newcommand{\elmp}{\ensuremath{\mathrm{e^{\mp}}}}
\newcommand{\net}{\ensuremath{\mathrm{\nu}}}
\newcommand{\netb}{\ensuremath{\mathrm{\overline{\nu}}}}
\newcommand{\dsb}{\ensuremath{\mathbf{d\overline{s}}}}
\newcommand{\dbs}{\ensuremath{\mathbf{\overline{d}s}}}
\newcommand{\sbb}{\ensuremath{\mathbf{\overline{s}}}}
\newcommand{\sss}{\ensuremath{\mathbf{s}}}
\newcommand{\kk}{\ensuremath{\mathrm{K}}}
\newcommand{\kpi}{\ensuremath{\mathrm{K}\pi}}
\newcommand{\kpipm}{\ensuremath{\mathrm{K}^{\pm}\pi ^{\mp}}}
\newcommand{\kpimp}{\ensuremath{\mathrm{K}^{\mp}\pi ^{\pm}}}
\newcommand{\kp}{\ensuremath{\mathrm{K^{+}}}}
\newcommand{\km}{\ensuremath{\mathrm{K^{-}}}}
\newcommand{\kpm}{\ensuremath{\mathrm{K^{\pm }}}}
\newcommand{\kmp}{\ensuremath{\mathrm{K^{\mp }}}}
\newcommand{\kn}{\ensuremath{\mathrm{K^0}}}
\newcommand{\knb}{\ensuremath{\mathrm{\overline{K}}{}^0}}
\newcommand{\ks}{\ensuremath{\mathrm{K_S}}}
\newcommand{\kl}{\ensuremath{\mathrm{K_L}}}
\newcommand{\kls}{\ensuremath{\mathrm{K_{L,S}}}}
\newcommand{\knin}{\ensuremath{\mathrm{K}{}^{0\text{in}}}}
\newcommand{\kbin}{\ensuremath{\mathrm{\overline{K}}{}^{0\text{in}}}}
\newcommand{\knut}{\ensuremath{\mathrm{K}{}^{0\text{out}}}}
\newcommand{\kbut}{\ensuremath{\mathrm{\overline{K}}{}^{0\text{out}}}}
\newcommand{\ksin}{\ensuremath{\mathrm{K_S}{}^{\text{in}}}}
\newcommand{\klin}{\ensuremath{\mathrm{K_L}{}^{\text{in}}}}
\newcommand{\ksut}{\ensuremath{\mathrm{K_S}{}^{\text{out}}}}
\newcommand{\klut}{\ensuremath{\mathrm{K_L}{}^{\text{out}}}}
\newcommand{\klsin}{\ensuremath{\mathrm{K_{L,S}}{}^{\text{in}}}}
\newcommand{\klsut}{\ensuremath{\mathrm{K_{L,S}}{}^{\text{out}}}}
\newcommand{\gls}{\ensuremath{\mathrm{\Gamma_{L,S}}}}
\newcommand{\gl}{\ensuremath{\mathrm{\Gamma_{L}}}}
\newcommand{\gs}{\ensuremath{\mathrm{\Gamma_{S}}}}
\newcommand{\gm}{\ensuremath{\overline{\Gamma}}} 
\newcommand{\gz}{\ensuremath{\gamma}}
\newcommand{\tls}{\ensuremath{\mathrm{\tau_{L,S}}}}
\newcommand{\ts}{\ensuremath{\mathrm{\tau_S}}}
\newcommand{\tl}{\ensuremath{\mathrm{\tau_L}}}
\newcommand{\mkz}{\ensuremath{m_{\kn }}}
\newcommand{\mkb}{\ensuremath{m_{\knb }}}
\newcommand{\mls}{\ensuremath{m_{\mathrm{L,S}}}}
\newcommand{\ml}{\ensuremath{m_{\mathrm{L}}}}
\newcommand{\ms}{\ensuremath{m_{\mathrm{S}}}}
\newcommand{\dm}{\ensuremath{\Delta m}}
\newcommand{\dg}{\ensuremath{\mathrm{\Delta\Gamma}}}
\newcommand{\lz}{\ensuremath{\mathrm{\lambda}}}
\newcommand{\lzs}{\ensuremath{\mathrm{\lambda_S}}}
\newcommand{\lzl}{\ensuremath{\mathrm{\lambda_L}}}
\newcommand{\lzls}{\ensuremath{\mathrm{\lambda_{L,S}}}}
\newcommand{\dlz}{\ensuremath{\mathrm{\Delta\lambda}}}
\newcommand{\Rb}{\ensuremath{\overline{R}}}
\newcommand{\Rz}{\ensuremath{R}}
\newcommand{\delm}{\ensuremath{\mathrm{M}_{\kn\kn }-\mathrm{M}_{\knb\knb }}}
\newcommand{\delg}{\ensuremath{\Gamma_{\kn\kn }-\Gamma_{\knb\knb }}}
\newcommand{\Lz}{\ensuremath{\Lambda }}
\newcommand{\Vl}{\ensuremath{V_L }}
\newcommand{\Vr}{\ensuremath{V_R }}
\newcommand{\PLaa}{\ensuremath{\mathrm{\Lambda_{\kn\kn }}}}
\newcommand{\PLab}{\ensuremath{\mathrm{\Lambda_{\kn\knb }}}}
\newcommand{\PLba}{\ensuremath{\mathrm{\Lambda_{\knb\kn }}}}
\newcommand{\PLbb}{\ensuremath{\mathrm{\Lambda_{\knb\knb }}}}
\newcommand{\PLij}{\ensuremath{\mathrm{\Lambda_{\alpha\alpha^{\prime}}}}}
\newcommand{\PLik}{\ensuremath{\mathrm{\Lambda_{\alpha\alpha^{\prime}}}}}
\newcommand{\Mz}{\ensuremath{\mathrm{M}}}
\newcommand{\Mzaa}{\ensuremath{\mathrm{M_{\kn\kn }}}}
\newcommand{\Mzab}{\ensuremath{\mathrm{M_{\kn\knb }}}}
\newcommand{\Mzba}{\ensuremath{\mathrm{M_{\knb\kn }}}}
\newcommand{\Mzbas}{\ensuremath{\mathrm{M^*_{\knb\kn }}}}
\newcommand{\Mzbb}{\ensuremath{\mathrm{M_{\knb\knb }}}}
\newcommand{\Mzij}{\ensuremath{\mathrm{M_{\alpha\alpha^{\prime}}}}}
\newcommand{\fm}{\ensuremath{\mathrm{\phi_{M}}}}

\newcommand{\Gz}{\ensuremath{\mathrm{\Gamma}}}
\newcommand{\Gzaa}{\ensuremath{\mathrm{\Gamma_{\kn\kn }}}}
\newcommand{\Gzab}{\ensuremath{\mathrm{\Gamma_{\kn\knb }}}}
\newcommand{\Gzba}{\ensuremath{\mathrm{\Gamma_{\knb\kn }}}}
\newcommand{\Gzbas}{\ensuremath{\mathrm{\Gamma^*_{\knb\kn }}}}
\newcommand{\Gzbb}{\ensuremath{\mathrm{\Gamma_{\knb\knb }}}}
\newcommand{\Gzij}{\ensuremath{\mathrm{\Gamma_{\alpha\alpha^{\prime}}}}}
\newcommand{\fg}{\ensuremath{\mathrm{\phi_{\Gamma}}}}

\newcommand{\Xz}{\ensuremath{\mathrm{X}}}
\newcommand{\Iz}{\ensuremath{\mathrm{I}}}
\newcommand{\CPx}{\ensuremath{{\pmb {\mathcal CP}}}}
\newcommand{\CPz}{\ensuremath{\mathrm{\cal CP}}}
\newcommand{\Cz}{\ensuremath{\mathrm{\cal C}}}
\newcommand{\Pz}{\ensuremath{\mathrm{\cal P}}}
\newcommand{\Tz}{\ensuremath{\mathrm{\cal T}}}
\newcommand{\CPTz}{\ensuremath{\mathrm{\cal CPT}}}

\newcommand{\Hst}{\ensuremath{\mathrm{{\cal H}_{st}}}}
\newcommand{\Hem}{\ensuremath{\mathrm{{\cal H}_{em}}}}
\newcommand{\Hwk}{\ensuremath{\mathrm{{\cal H}_{wk}}}}
\newcommand{\Hz}{\ensuremath{\mathrm{{\cal H}}}}
\newcommand{\Vz}{\ensuremath{\mathrm{V}}}

\newcommand{\mean}[1]{\langle #1 \rangle}
\newcommand{\AT}{\ensuremath{A_\mathrm{T}}}
\newcommand{\ATl}{\ensuremath{A_\mathrm{T}^{\ell }}}
\newcommand{\ATexp}{\ensuremath{A_\mathrm{T}^{\text{exp}}}}
\newcommand{\avATexp}{\ensuremath{\langle A_\mathrm{T}^{\text{exp}}\rangle }}
\newcommand{\ACPT}{\ensuremath{A_\mathrm{CPT}}}
\newcommand{\ACPTl}{\ensuremath{A_\mathrm{CPT}^{\ell }}}
\newcommand{\ACPTexp}{\ensuremath{A_\mathrm{CPT}^{\text{exp}}}}
\newcommand{\Ad}{\ensuremath{A_{\del }}}
\newcommand{\Adl}{\ensuremath{A_{\del}^{\ell}}}
\newcommand{\Adexp}{\ensuremath{A_{\del }^{\text{exp}}}}
\newcommand{\avAdexp}{\ensuremath{\langle A_{\del }^{\text{exp}}\rangle }}
\newcommand{\avedm}{\ensuremath{\langle \dm \rangle }}
\newcommand{\avets}{\ensuremath{\langle \ts \rangle }}
\newcommand{\Adm}{\ensuremath{A_{\dm }}}
\newcommand{\Admexp}{\ensuremath{A_{\dm }^{\text{exp}}}}
\newcommand{\Admc}{\ensuremath{A_{\dm }}}
\newcommand{\Adml}{\ensuremath{A_{\dm }^{\ell }}}
\newcommand{\Admcexp}{\ensuremath{A_{\dm }^{\text{exp}}}}
\newcommand{\Aexp}{\ensuremath{A^{\text{exp}}}}
\newcommand{\Afexp}{\ensuremath{A_f^{\text{exp}}}}
\newcommand{\Af}{\ensuremath{A_f}}
\newcommand{\ACP}{\ensuremath{A_\mathrm{CP}}}
\newcommand{\ACPf}{\ensuremath{A_\mathrm{CP}^f}}
\newcommand{\Aast}{\ensuremath{A_{+-}^*}}
\newcommand{\Apm}{\ensuremath{A_{+-}}}
\newcommand{\Apmexp}{\ensuremath{A_{+-}^{\text{exp}}}}
\newcommand{\Ann}{\ensuremath{A_{00}}}
\newcommand{\Annexp}{\ensuremath{A_{00}^{\text{exp}}}}
\newcommand{\Apmn}{\ensuremath{A_{+-0}}}
\newcommand{\Apmnexp}{\ensuremath{A_{+-0}^{\text{exp}}}}
\newcommand{\Annn}{\ensuremath{A_{000}}}
\newcommand{\Annnexp}{\ensuremath{A_{000}^{\text{exp}}}}
\newcommand{\modulus}[1]{\left| #1 \right|}
\newcommand{\rrf}{\ensuremath{R_f}}
\newcommand{\rrfb}{\ensuremath{\overline{R}_f}}
\newcommand{\rrpm}{\ensuremath{R_{\pipi}}}
\newcommand{\rrpmb}{\ensuremath{\overline{R}_{\pipi}}}
\newcommand{\rr}{\ensuremath{R}}
\newcommand{\rrb}{\ensuremath{\overline{R}}}
\newcommand{\rrp}{\ensuremath{R_+}}
\newcommand{\rrm}{\ensuremath{R_-}}
\newcommand{\rrpb}{\ensuremath{\overline{R}_+}}
\newcommand{\rrmb}{\ensuremath{\overline{R}_-}}
\newcommand{\nrfb}{\ensuremath{\overline{N}_f}}
\newcommand{\nrf}{\ensuremath{N_f}}
\newcommand{\nrpb}{\ensuremath{\overline{N}_+}}
\newcommand{\nrp}{\ensuremath{N_+}}
\newcommand{\nrmb}{\ensuremath{\overline{N}_-}}
\newcommand{\nrm}{\ensuremath{N_-}}
\newcommand{\nrpm}{\ensuremath{N_{\pm}}}
\newcommand{\nrpmb}{\ensuremath{\overline{N}_{\pm}}}
\newcommand{\brr}{\ensuremath{B}}
\newcommand{\brrb}{\ensuremath{\overline{B}}}
\newcommand{\brpb}{\ensuremath{\overline{B}_+}}
\newcommand{\brp}{\ensuremath{B_+}}
\newcommand{\brmb}{\ensuremath{\overline{B}_-}}
\newcommand{\brm}{\ensuremath{B_-}}
\newcommand{\brpm}{\ensuremath{B_{\pm}}}
\newcommand{\brpmb}{\ensuremath{\overline{B}_{\pm}}}
\newcommand{\nwb}{\ensuremath{\overline{N}_{w}}}
\newcommand{\nw}{\ensuremath{N_{w}}}
\newcommand{\nwpb}{\ensuremath{\overline{N}_{w+}}}
\newcommand{\nwp}{\ensuremath{N_{w+}}}
\newcommand{\nwmb}{\ensuremath{\overline{N}_{w-}}}
\newcommand{\nwm}{\ensuremath{N_{w-}}}
\newcommand{\nwpmb}{\ensuremath{\overline{N}_{w\pm }}}
\newcommand{\nwpm}{\ensuremath{N_{w\pm }}}
\newcommand{\nwrpb}{\ensuremath{\overline{N}_{+w_{r}}}}
\newcommand{\nwrp}{\ensuremath{N_{+w_{r}}}}
\newcommand{\nwrmb}{\ensuremath{\overline{N}_{-w_{r}}}}
\newcommand{\nwrm}{\ensuremath{N_{-w_{r}}}}
\newcommand{\nwrpmb}{\ensuremath{\overline{N}_{\pm w_{r}}}}
\newcommand{\nwrpm}{\ensuremath{N_{\pm w_{r}}}}
\newcommand{\nwxpb}{\ensuremath{\overline{N}_{+w_{\xi}}}}
\newcommand{\nwxp}{\ensuremath{N_{+w_{\xi}}}}
\newcommand{\nwxmb}{\ensuremath{\overline{N}_{-w_{\xi}}}}
\newcommand{\nwxm}{\ensuremath{N_{-w_{\xi}}}}
\newcommand{\nwxpmb}{\ensuremath{\overline{N}_{\pm w_{\xi}}}}
\newcommand{\nwxpm}{\ensuremath{N_{\pm w_{\xi}}}}
\newcommand{\wra}{\ensuremath{w_r}}
\newcommand{\wrb}{\ensuremath{\overline{w}_r}}
\newcommand{\wt}{\ensuremath{w}}
\newcommand{\wtb}{\ensuremath{\overline{w}}}
\newcommand{\api}{\ensuremath{(1+4\reel )\xi}}
\renewcommand{\exp}{\ensuremath{\mathrm{exp}}}
\newcommand{\expm}{\ensuremath{\mathrm{E_-}}}
\newcommand{\expp}{\ensuremath{\mathrm{E_+}}}
\newcommand{\exppm}{\ensuremath{\mathrm{E_{\pm}}}}
\newcommand{\eff}{\ensuremath{\epsilon}}
\newcommand{\DS}{\ensuremath{\Delta S}}
\newcommand{\DQ}{\ensuremath{\Delta Q}}
\newcommand{\Dt}{\ensuremath{\Delta t}}
\newcommand{\kef}{\ensuremath{\mathrm{K}_{\e 3}^0}}
\newcommand{\dd}{\ensuremath{\mathrm{d}}}
\newcommand{\bce}{\begin{center}}
\newcommand{\ece}{\end{center}}
\newcommand{\bfig}{\begin{figure}}
\newcommand{\efig}{\end{figure}}
\newcommand{\scbs}{\ensuremath{\mathrm{S1\overline{C}S2}}}
\newcommand{\pro }{\ensuremath{{\cal P}}}
\newcommand{\Ppl}{\ensuremath{{\cal P}_+}}
\newcommand{\Pbp}{\ensuremath{\overline{\cal P}_+}}
\newcommand{\Pm}{\ensuremath{{\cal P}_-}}
\newcommand{\Pbm}{\ensuremath{\overline{\cal P}_-}}
\newcommand{\Ppm}{\ensuremath{{\cal P}_{\pm }}}
\newcommand{\Pbpm}{\ensuremath{\overline{\cal P}_{\pm }}}

\newcommand{\Np}{\ensuremath{N_+}}
\newcommand{\Nbp}{\ensuremath{\overline{N}_+}}
\newcommand{\Nm}{\ensuremath{N_-}}
\newcommand{\Nbm}{\ensuremath{\overline{N}_-}}
\newcommand{\Npm}{\ensuremath{N_{\pm}}} 
\newcommand{\Nbpm}{\ensuremath{\overline{N}_{\pm}}}
\newcommand{\Nb}{\ensuremath{{\overline{N}}}}

\pagenumbering{roman}
\topmargin 3cm
\thispagestyle{empty}
\title{ 
THE FUNDAMENTAL  SYMMETRIES IN THE
       NEUTRAL KAON SYSTEM \\ {\it -- a pedagogical choice --}}
\begin{center}
\vspace{6mm}
Maria Fidecaro\\
CERN, CH--1211 Gen\`eve 23, Switzerland \\[4mm]
and\\[4mm]
Hans--J\"{u}rg Gerber\\
Institute for Particle Physics, ETH, CH--8093 Z\"urich, Switzerland\\[10mm]
\end{center}
\begin{abstract}
During the recent years experiments with neutral kaons have yielded remarkably 
sensitive results which are pertinent to such fundamental phenomena as \CPTz\ 
invariance (protecting causality), time-reversal invariance violation, 
coherence 
of wave functions, and entanglement of kaons in pair states. We describe the 
phenomenological developments and the theoretical conclusions drawn from the 
experimental material. An outlook to future experimentation is indicated.
\end{abstract}
\begin{center}
\medskip
March 1, 2006
\end{center} 
\vspace{4mm}
\tableofcontents
\maketitle

\cleardoublepage
\topmargin 20mm
\pagenumbering{arabic}

\section{Introduction}

\it Symmetries\rm , here, designate certain properties of theoretically formulated laws of physics. For the experimental investigation of these laws, symmetries are of great utility. They establish simple and reliable relations between the abstract law and specific, in practice, observable quantities. In this article, we shall encounter symmetry properties of the experimental data, that directly reflect symmetry properties of the physical law. Examples are given in Table \ref{tab:01}.

\it Fundamental symmetries\rm\ shall be those ones which concern space and time in our world, or in the unavoidable 
antimatter world. Since many laws govern evolutions in space and time, fundamental symmetries have a wide range of 
applications and are valid under very general provisions, most often conveniently independent on unknown details.
\\
In the search for the formal properties of a law, experimentally observed symmetries are most valuable guide lines.

For a causal description of Nature, where the future is connected to the past, it is necessary, that the law of time evolution is endowed with an invariance, with respect to such a symmetry operation, that contains, at least, a reversal of the arrow of time.
Res Jost \cite{jost}, pioneer
of the\ \CPTz\ theorem \cite{schw, bellcpt, lued, paulicpt, jostcpt}, writes '... : eine kausale Beschreibung
ist nur m\"oglich, wenn man den Naturgesetzen eine Symmetrie zugesteht, welche den Zeitpfeil
umkehrt'.
\\
The only suitable symmetry for this purpose, not found violated (yet), is a combination: the symmetry with respect to the \CPTz\ transformation, where
time and space coordinates are reflected at their origins (\Tz\ and \Pz), and particles become antiparticles (\Cz). (We remember that each and every one alone of the symmetries with respect to \ \Cz,\ \Pz\ or \Tz\ are not respected by weak interactions).
\\\\
The property of \ \CPTz\ symmetry entails the existence of antimatter, and the equality of the masses and the decay widths of a particle and of its antiparticle.
\\\\
We shall describe the cleverness, the favorable natural circumstances, and the experimental ingenuity, that allowed one to achieve the extraordinary insight, that the masses and the decay widths of \ \kn\ and \knb\ are really equal, within about one in $10^{18}$ (!).
\\

\CPTz\ symmetry is not implied by quantum mechanics (QM). If wanted, it has to be implemented as an extra property of the Hamiltonian in the Schr\"odinger equation. This allows one to also formulate hypothetical \ \CPTz\ symmetry violating processes within QM, and to devise appropriate experimental searches.
\\
We shall describe experimental results that could have contradicted \CPTz\ invariance, but
did not.
\\

\it Two-state systems\rm , treated by quantum mechanics, introduce an additional hierarchy among the symmetry violations:
the detection of a violation of the \ \CPTz\ or (and) \ of the \Tz\ symmetry implies a \ \CPz\ symmetry violation.
This is exemplified by the third column of Table \ref{tab:01}.
Furthermore, they exhibit the difficulty, that a \ \Tz\ symmetry violation can, in these systems, only be defined,
if a decay process exists, which is, by its nature, inherently \ \Tz\ \it asymmetric\rm.
This has brought on the question, whether the observed effect of \ \Tz\ symmetry violation in the \ \mbox{\kn\ \knb}\ system is a genuine property of the weak interaction, or whether it is just an artifact of the decaying two-state system.
\\
We shall describe the way out of this maze. It consists of formulating \ \Tz\ symmetry not on the level of the two-state system, but in the complete environment, which comprises all the states involved, including the decay products, and where the Hamiltonian is (thus) hermitian; and then to show that a \ \Tz\ symmetric Hamiltonian would not be able to reproduce the experimental data.
\\

There is only one place in particle physics, where a \ \Tz\ asymmetry has been 
detected \cite{schub,pen2}:
\\
the \ \knb\ develops faster towards a \kn\ than does a \ \kn\ towards a \knb .
\\

The interpretation of non-vanishing '\Tz - odd' quantities as signals for \ \Tz\ asymmetry, is plagued by \ \Tz\ symmetric effects, such as final state interactions, which also are able to produce non-vanishing \Tz - odd results. We discuss an experiment which has achieved to yield a sizable \Tz - odd correlation under a \ \Tz\ \it symmetric\rm\ law, even in the absence of final state interactions. (We know of no other case in particle physics, where this happens).
\\

Particles with two states occur at various, independent places in physics, where they exhibit widely different phenomena. However, when a quantum mechanical description based on two-component wave functions is appropriate, then these phenomena display formally identical patterns, with common properties, such as the number of free parameters, or as the fact that there is only one basic observable at choice, whose expectation value can be determined in a single measurement with an unambiguous result.

The insight into the nature of these common properties is most elegantly achieved with the 'Pauli matrix toolbox', which we freely use in this article. (The Appendix gives a description at an introductory level).
\\

Could we test quantum mechanics ? Could we find out whether \ \CPTz\ symmetry could appear violated, not because of a \ \CPTz\ asymmetry of the Hamiltonian, but because the way, we apply the common rules of QM, is inadequate ?
\\
We shall describe how it has been possible to conclude, that a hypothetical, hidden QM violating energy would have to be tinier than about $10^{-21}$ \geve , in order to have escaped the existing experiments with neutral kaons\cite{Ad95}. (The extension of the Pauli matrix toolbox to density matrices will be introduced).
\\

One of the most sensational predictions of QM is \it teleportation\rm, which appears between two distant particles in an entangled state. As experimentation at \ $\phi$ factories with entangled neutral kaons is growing, we shall give the basic formal toolbox for two-particle systems, where each particle is a decaying two-state system, and where QM rules are not necessarily respected.
\\

The main body of this article explains, how the unique properties of the neutral kaons have been used to investigate the properties of the fundamental symmetries. To do justice to those students, who want to worry about the weak points in the thoughts, we present the most important doubts about side effects, which could have endangered the validity of the conclusions.
\\

A neutral kaon oscillates forth and back between itself and its antiparticle, because the
physical quantity, \it strangeness\rm , which distinguishes antikaons from kaons, is not conserved
by the interaction which governs the time evolution \cite{gell,pais}. A neutral kaon is by itself a two-state system.
\\
The oscillation here can be viewed in analogy to the spin precession of a nucleon in a magnetic field, but where the r\^ole of the field is carried out by the inherent weak interaction of the kaon. The spin manipulation in NMR experiments by radio wave pulses corresponds to the application of \ \it coherent regeneration\rm , a phenomenon which occurs, when neutral kaons traverse matter.
\\

The experimentation with neutral kaons provides the researcher with an impressing scene. The interference pattern of matter waves becomes visible in macroscopic dimensions !
\\\\
Think of two highly stable damped oscillators with a relative frequency difference of \ $\approx 7 \times 10^{-15}$.
The corresponding beat frequency of \ $ \approx 5.3\times 10^9 \ \mathrm{s}^{-1}$, and the scale of the
resulting spatial interference pattern of \ $\approx 0.36$ m\ are ideally adapted to the technical capabilities of detectors in high-energy physics.

The extreme sensitivity of the neutral kaon system comes from the tiny mass difference \dm\ of the exponentially decaying states. Some fraction of the size of $\dm \approx 3.5 \times 10^{-15}\geve$, depending on the
measurement's precision, determines the level for effects, which may be present inside of the
equation of motion, to which the measurements are sensitive: $10^{-18}$ to $10^{-21}\geve$.
\\

The history of the unveilings of the characteristics of the neutral kaons is a sparkling
succession of brilliant ideas and achievements, theoretical and  experimental
\cite{cronin, femn, kbir, jrlg, wo, cago, bloi, klei}.\\
We have reasons to assume that neutral kaons will enable one to make more basic discoveries.
Some of the kaons' properties (e. g. entanglement) have up to now only scarecely been exploited.
\\\\

\section{The neutral-kaon system}\label{sec:motiv}
\subsection{Time evolution}\label{sub:evol}

The time evolution of a neutral kaon and of its decay products may be 
represented by the state vector
\begin{equation}\label{eq:2.01}
\ket{\psi} = \psi_{\kn }(t)\ket{\kn } + \psi_{\knb }(t)\ket{\knb }
           +  \sum_m c_m (t)\ket{m}
\end{equation}
which satisfies the Schr\"odinger equation
\begin{equation}
\label{eq:2.02}
\i \frac{\dd \ket{\psi }}{\dd t} = \Hz \ket{\psi }.
\end{equation}
In the Hamiltonian, $\Hz = \Hz _0 + \Hwk $, $\Hz _0$ governs the strong and 
electromagnetic interactions. It is invariant with respect to the 
transformations \Cz , \Pz , \Tz , and it conserves the strangeness $S$. 
The states $\ket{\kn }$ and $\ket{\knb }$ are common stationary eigenstates 
of $\Hz _0$ and $S$, with the mass $m_0$ and with opposite strangeness: 
$\Hz _0\ket{\kn } = m_0 \ket{\kn }$, $\Hz _0\ket{\knb } = m_0 \ket{\knb }$,
$S\ket{\kn } = \ket {\kn } $, $S\ket{\knb } = -\ket{\knb} $. The states 
$\ket{m}$ are continuum eigenstates of $\Hz _0$ and represent the decay 
products. They are absent at the instant of production $(t=0)$ of the neutral 
kaon. The initial condition is thus
\begin{equation}\label{eq:2.03}
\ket{\psi_0} = \psi_{\kn }(0)\ket{\kn } + \psi_{\knb }(0)\ket{\knb } \ .
\end{equation}
\Hwk\  governs the weak interactions. Since these do not conserve strangeness, 
a neutral kaon will, in general,  change its strangeness as time evolves.

Equation (\ref{eq:2.02}) may be solved for the unknown functions $\psi _{\kn}
(t)$ and $\psi_{\knb}(t)$ , by using a perturbation
approximation~\cite{wigner} which yields~\cite{trei, lee2}
\begin{equation}\label{eq:2.04}
\psi = \e^{-\i\Lz t} \psi_0 \ .
\end{equation}
$\psi$ is the column vector  with components $\psi_{\kn}(t)$ and 
$\psi_{\knb}(t)$, $\psi_0$ equals $\psi$ at $t=0$, and \Lz\ is the 
time-independent $2 \times 2$ matrix $\left(\PLik \right)$, whose components
refer to the two-dimensional basis $\ket{\kn },\ket{\knb}$ and may be written
as \ \PLik\ = \ $\bra{\alpha}\Lambda \ket{\alpha^{\prime}}$ with
$\alpha , \alpha^{\prime} = \kn , \knb$.
\\\\
Since the kaons decay, we have 
\begin{equation}\label{eq:2.05}
0 > \frac{\dd |\psi|^2}{\dd t}  = -\i\psi^{\dagger}\left(\Lz-\Lz^{\dagger}\right)\psi \,.
\end{equation}
\Lz \ is thus not hermitian, $\e^{-\i\Lz t}$ is not unitary, in general.
\\\\
This motivates the definition of $\Mz$ and $\Gz$ as
\begin{subequations}\label{eq:2.06}
\begin{eqnarray}
\Lz &=& \Mz - \frac{\i}{2} \Gz \;,\label{eq:2.06a}\\ 
\Mz &=& \Mz ^{\dagger},\quad \Gz = \Gz ^{\dagger}\;.\label{eq:2.06b}
\end{eqnarray}
\end{subequations}
We find
\begin{equation}\label{eq:2.07}
0 > \frac{\dd |\psi|^2}{\dd t}  = - \psi^{\dagger}\Gz\psi \;.
\end{equation}
\\
This expresses that \Gz\ has to be a positive matrix.
\\\\
The perturbation approximation also establishes the relation
from \Hwk\ to \Lz \ (including second order in \Hwk ) by
\begin{subequations}\label{eq:2.08}
\begin{align}
\Mz _{\alpha\alpha^{\prime}} & =   
   m_0 \del_{\alpha\alpha^{\prime}}
 + \sbraket{\alpha|\Hwk }{\alpha^{\prime}} +
 \Pz  \sum_{\beta} \left( \frac{\sbraket{\alpha|\Hwk }{\beta}
                                \sbraket{\beta|\Hwk }{\alpha^{\prime}}}
                               {m_0 - E_{\beta}}\right)\;, \label{eq:2.08a} \\
\Gz _{\alpha\alpha^{\prime}} & = 2\pi\sum_{\beta}
        \sbraket{\alpha|\Hwk }{\beta}\sbraket{\beta|\Hwk }{\alpha^{\prime}}
                              \del (m_0 - E_{\beta})\;,      \label{eq:2.08b}\\
(\alpha , \alpha^{\prime} &= \kn , \knb )\,.              \notag 
\end{align}
\end{subequations}
\\
Equations (\ref{eq:2.08a}, \ref{eq:2.08b}) enable one now to state directly 
the symmetry properties of \Hwk\  in terms of experimentally observable 
relations among the elements of \Lz , see Table \ref{tab:01} and Ref. 
\cite{leebook} . We remark
that \CPTz\ invariance imposes no restrictions on the off-diagonal elements,
and that \Tz\ invariance imposes no restrictions on the diagonal elements
of \Lz\ . \ \CPz\ invariance is violated, whenever one, at least, of these invariances is
violated.\\

\begin{table}
\begin{center}
\caption{The symmetry properties of \Hwk\ induce symmetries in observable quantities.}
The last column indicates asymmetries of quantities which have been measured\\
by the CPLEAR experiment at CERN. More explanations are given in the text.\\
\medskip
{\small
\begin{tabular}{llll}
\hline\hline \\[-0.2cm]
If \Hwk\ has the property   & called        &   then    & or \\[0.2cm]
\hline \\[0.2cm]
$\Tz ^{-1} \ \Hwk\ \ \Tz = \Hwk $ &\Tz\ invariance &
                                  $|\PLab | = |\PLba |$ & $\AT   = 0$ \\[0.2cm]
$(\CPTz )^{-1} \ \Hwk\ \ (\CPTz ) = \Hwk $ & \CPTz\ invariance & 
                                  $ \ \ \PLaa   = \ \PLbb  $ & $\ACPT = 0$ \\[0.2cm]
$(\CPz  )^{-1} \ \Hwk\ \ (\CPz  ) = \Hwk $ & \CPz\  invariance &
                                  $ \ \ \PLaa   = \ \PLbb   $  \ and &      \\[0.2cm]
                           & &    $|\PLab | = |\PLba |$ & $\ACP  = 0$ \\[0.2cm]
\hline\hline \\[-0.2cm] 
\end{tabular}}
\label{tab:01}
\end{center}
\end{table}

The definitions of $\ket{\kn }$ and $\ket{\knb }$ leave a real phase
$\vartheta $ undetermined: \\\\ Since $ S ^{-1} \ \Hz _0 \ S = \ \Hz _0 $, the states

\begin{subequations}\label{eq:2.09}
\begin{eqnarray}
\ket{\kn '} & = & \e^{\i\vartheta S}  \ket{\kn } \ = \ \e^{\i\vartheta} \ \ \ket{\kn } \ , \label{eq:2.09a}\\
\ket{{\knb} ^{'}}  & = & \e^{\i\vartheta S} \ket{\knb } \ = \ \e^{-\i\vartheta} \ket{\knb } \ . \label{eq:2.09b}
\end{eqnarray}
\end{subequations}
fulfil the definitions of $\ket{\kn}$ and $\ket{\knb}$ as well, and 
constitute thus an equivalent basis which is related to the original basis by
a unitary transformation. As the observables are always invariant with respect
to unitary base transformations, the parameter $\vartheta$ cannot be measured, and
remains undetermined. This has the effect that expressions which depend on
$\vartheta$ are not suited to represent experimental results, unless $\vartheta$ has
beforehand been fixed to a definite value by convention. Although such a
convention may simplify sometimes the arithmetic, it risks to obscure the
insight as to whether a certain result is genuine or whether it is just an
artifact of the convention.\\As an example we consider the elements of $\Lz $,
$\PLik $ = \ $\bra{\alpha}\Lambda \ket{\alpha^{\prime}}$ which refer to the basis
$\alpha , \alpha^{\prime} = \kn , \knb $. With respect to the basis
$\e^{\i \vartheta}\kn $, $\e^{-\i \vartheta}\knb $ we obtain the same diagonal elements,
whereas the off-diagonal elements change into
\begin{subequations}\label{eq:2.10}
\begin{eqnarray}
\PLab  & \longmapsto  &  \Lambda'_{\kn\knb }  \ = \ \e^{-2\i\vartheta} \ \ \PLab \ , \label{eq:2.10a}\\
\PLba  & \longmapsto  &  \Lambda'_{\knb\kn }  \ = \ \e^{2\i\vartheta} \ \ \ \ \PLba \  , \label{eq:2.10b}
\end{eqnarray}
\end{subequations}
and are thus convention dependent. However, their product, their absolute values,
the trace tr\{\Lz\}, its determinant, and its eigenvalues (not so its eigenvectors),
but also the partition into $M$ and $\Gz$, are convention independent ~\cite{lavoura}. (We will
introduce a phase convention later in view of comparing experimental
results).

The definition of the operations \CPz\ and \CPTz\ allows one to define two additional phase
angles. We select them such that \ $\cal{O}\kn =$ \knb\ and \ $\cal{O}$\knb = \kn, where
$\cal{O}$ stands for \CPz\ or \CPTz. See e. g. \cite{tya}.\\

In order to describe the time evolution of a neutral kaon, the matrix exponential
$\e^{-\i\Lz t}$ has to be calculated.\\
If the exponent matrix has $two$ dimensions, a generalized Euler formula gives a
straightforward answer. \\ Be $\Lz$ represented as a superposition of Pauli matrices
\begin{equation}\label{eq:2.11}
\Lz = \Lz^\mu \sigma^\mu
\end{equation}
with $\sigma^0$ = \ unit matrix, $\sigma^k$ = \ Pauli matrices, $\Lz^\mu$ complex.
(Summation over multiple indices: Greek 0 to 3, Roman 1 to 3). \\\\Then
\begin{equation}\label{eq:2.12}
\e^{-\i\Lz t} = \e^{-\i\Lz^\mu \sigma^\mu t} = \e^{-\i\Lz^0 t}( \sigma^0\cos(\Omega t)
- \i\Lz^m \sigma^m t \sin(\Omega t)/(\Omega t)),
\end{equation}\\
where \ $\Omega ^2 = \ \Lz^m \Lz^m $. Noting that $\Lz^\mu=\frac{1}{2}$tr$\{\sigma^\mu\Lz\}$,
we see that Eq. (\ref{eq:2.12}) expresses $\e^{-\i\Lz t}$ entirely in terms of the elements of $\Lz$. Since the
(complex) eigenvalues $\lzl$, $\lzs$ of $\Lz$ turn out to be observable (and are thus doubtless
phase transformation invariant) we introduce them into (\ref{eq:2.12}).\\
They fulfil $\lzl\lzs = \mathrm{det }(\Lz)=\Lz^0\Lz^0 -\Lz^m\Lz^m$, and $\lzl+\lzs=\ $tr$\{\Lz\}=2\Lz^0$,
and thus, with $\dlz \equiv \lzl - \lzs$,
\begin{equation}\label{eq:2.13}
\Omega=\dlz/2 \;.
\end{equation}
We note here (with relief) that the calculation of the general time evolution, expressed
in Eq. (\ref{eq:2.04}), does not need the knowledge of the $eigenstates$ of $\Lz$, whose
physical interpretation needs special attention ~\cite{enz,gau1}.\\
The following corollary will be of interest:\\ The off-diagonal elements of a $2\times 2$ exponent matrix
factorize the off-diagonal elements of its matrix exponential, with equal factors:
\begin{equation}\label{eq:2.14}
(\e^{-\i \Lz t})_{j \neq k} = 
(-i \Lz t)_{j \neq k} \ \e^{-\frac{1}{2}\i (\PLaa + \PLbb) t} \ \sin (\Omega t)/(\Omega t) \ \ .
\end{equation}

Eq. (\ref{eq:2.11}) also establishes an analogy between the spin states of particles of spin 1/2 and the eigenstates of neutral kaons. We discuss this analogy in the Appendix, where Table \ref{ana} \ gives a summary.
\\

Independent of the dimension $n$ of the exponent matrix,
diagonalization allows one to calculate the matrix exponential. Find the two vectors
$\ket{\kls}$ which $\Lz$ transforms into a multiple of themselves
\begin{equation}\label{eq:2.15}
\Lz \ket{\kls} = \lzls \ket{\kls}\;.
\end{equation}
The eigenvalues $\lzls$ need to be
\begin{equation}\label{eq:2.16}
\lzls = \frac{1}{2}\ \mathrm{tr}\{\Lz\}\pm \sqrt{(\mathrm{tr}\{\Lz\})^2/4 - \mathrm{det}(\Lz)} \ \;.
\end{equation}\\ We may express the solutions of (\ref{eq:2.15}) in the basis
$\ket{\kn},\ket{\knb}$ as
\begin{eqnarray}
\ket{\ks}&=&V^{11}\ket{\kn}+V^{21} \ket{\knb} \ \hat{=} \
\left ( \begin{array}{c} V^{11} \\ V^{21} \end{array} \right )\label{eq:2.17}\\
\ket{\kl}&=&V^{12}\ket{\kn}+V^{22} \ket{\knb} \ \hat{=} \
\left ( \begin{array}{c} V^{12} \\ V^{22} \end{array} \right )\label{eq:2.18}
\end{eqnarray}
and form the matrix $V=(V^{ij})$ whose columns are the components of the eigenvectors,
and also $W=V^{-1}$. The matrix $\Lz$ can now be represented as
\begin{equation}\label{eq:2.19}
\Lz = VDW
\end{equation}
where $D$ is diagonal,
\[ D= \left ( \begin{array}{cc} \lzs & 0 \\ 0 & \lzl \end{array} \right ). \] \\
Since we need to extract $V$ and $W$ from the exponent to obtain
\begin{equation}\label{eq:2.20}
\e^{-\i \Lz t}=\e^{-\i VDW t}=V\e^{-\i D t}W,
\end{equation}
it is important that \ $W=V^{-1}$ (and $not \ W=V^{\dagger} \ne V^{-1}$). Since $WV=1$
or \ $W^{ij}V^{jk}=\delta^{ik}$, the rows of $W$ are orthogonal to the columns of
$V$. A convenient solution is
\begin{equation}\label{eq:2.21}
W=\left(W^{ij}\right)=\frac{1}{\modulus{V}}\left ( \begin{array}{cc}
V^{22} & -V^{12} \\ -V^{21} & V^{11} \end{array} \right ).
\end{equation}\\
Inserting $V$, $D$, $W$ into (\ref{eq:2.20}) allows one to express $\e^{-\i \Lz t}$
in terms of the eigenelements $\ket{\kls}$, and $\lzls$. (Eq. (\ref{eq:2.19})
also shows how to construct a matrix with prescribed (non-parallel) eigenvectors and
independently prescribed eigenvalues).\\
If we define the vectors
\begin{eqnarray}
\bra{\widetilde{\ks}}=W^{11}\bra{\kn}+W^{12}\bra{\knb}\label{eq:2.22}\\
\bra{\widetilde{\kl}}=W^{21}\bra{\kn}+W^{22}\bra{\knb}\label{eq:2.23}
\end{eqnarray}
then we have 
\begin{equation}\label{eq:2.24}
\sbraket{\widetilde{\mathrm{K}_\kappa}}{\mathrm{K}_{\kappa'}}=\delta_{\kappa \kappa'}\;,\ \ \ \kappa,\kappa'=\mathrm{L,S}
\end{equation}
in contrast to
\begin{equation}\label{eq:2.25}
\sbraket{{\mathrm{K}_\kappa}}{\mathrm{K}_{\kappa'\ne \kappa}}\ne 0.
\end{equation}\\
The difficulty to interprete these vectors as states is discussed in \cite{enz}.
Eq. (\ref{eq:2.25}) shows that there is no clear state $\kl$, because the vector
$\ket{\kl}$ always has a component of $\ket{\ks}$, with the probability $\modulus{\sbraket{\ks}{\kl}}^2$.\\\\
We now solve (\ref{eq:2.15}): $\Lz^{ij}V^{j\kappa}=\lambda_{\kappa}V^{i\kappa}$, (no sum $\kappa$)
for $V^{j\kappa} \hat{=} \ket{\mathrm{K}_\kappa}$, $\kappa = (1,2) \ \hat{=}$ (S,L), and regain
(\ref{eq:2.12}) in a different form:\\
\begin{equation}\label{eq:2.26}
\e^{- \i\Lz t} = \left(
\begin{array}{cc}
f_+ + 2 \den f_-  & -2\PLab f_-/\dlz \\
- 2 \PLba f_-/\dlz & f_+ - 2 \den f_-
\end{array}\right)
= \left(
\begin{array}{cc}
f_+ + 2 \den f_-  & -2 (\sigma - \epn) f_- \\
- 2 (\sigma + \epn) f_- & f_+ - 2 \den f_-
\end{array}\right)
\end{equation}\\
with
\begin{equation}\label{eq:2.27}
f_{\pm}(t) = \dfrac{\e^{-\i \lzs t }\pm \e^{-\i\lzl t }}{2}\;,
\end{equation}
\begin{subequations}\label{eq:2.28}
\begin{eqnarray}
\den   & \equiv & (\PLbb - \PLaa ) /(2 \dlz ) \,, \label{eq:2.28a}\\
\epn   & \equiv & (\PLba  -\PLab ) /(2 \dlz ) \,, \label{eq:2.28b}\\
\sigma & \equiv & (\PLba + \PLab ) /(2 \dlz ) \,. \label{eq:2.28c}
\end{eqnarray}
\end{subequations}
We have set
\begin{equation}\label{eq:2.29}
\lzls = \mls - \frac{\i}{2}\gls
\end{equation}
and
\begin{equation}\label{eq:2.30}
\dlz = \lzl -\lzs = \ml-\ms + \frac{\i}{2}(\gs-\gl)\equiv \dm + \frac{\i}{2}\dg =
\modulus{\dlz}\ \e^{\i(\frac{\pi}{2}-\fsw)} 
\end{equation}
with
\begin{eqnarray}\label{eq:2.31}
\dm \equiv \ml -\ms \;,\ \ \ 
\dg \equiv \gs -\gl\;,
\end{eqnarray}
and with \fsw\ defined by
\begin{equation}\label{phisw}
\tan(\fsw)=(2\dm/\dg) \ .
\end{equation}
%
We shall also use
\begin{equation}
\gm \equiv (\gs +\gl)/2 \ .
\end{equation}
The parameters in (\ref{eq:2.28}) satisfy the identity
\begin{equation}\label{ouridentity}
\sigma^2 - \epn^2 + \den^2  \equiv 1/4\ 
\end{equation}
which entails
\begin{equation}\label{modourid}
\zeta \equiv\ {\modulus{\sigma}}^2 + \modulus{\epn}^2 + \modulus{\den}^2 -1/4 \ \ge 0 \ .
\end{equation}
From Table \ref{tab:01} we can deduce that $\zeta $ signifies the violations of \Tz\ and \CPTz\ .\\\\
The positivity of the matrix \Gz\ requires the determinant $|\Gz|$ to be positive
\begin{equation}\label{ourpositivity}
0 < |\Gz| =
{\modulus{\dlz}}^2 (\frac{\gs\gl}{{\modulus{\dlz}}^2}-2\zeta)\ ,
\end{equation}
which needs
\begin{equation}\label{limmods}
\zeta\ <\
\frac{\gs\gl}{{2\modulus{\dlz}}^2} \approx \frac{\gl}{\gs} \ .
\end{equation}
The last approximation is valid for neutral kaons where, experimentally,
$2{\modulus{\dlz}}^2 \approx (\gs)^2$.\\
We see from
(\ref{limmods}), that the ratio \ $\gl/\gs\ (\approx 1.7\times 10^{-3})$ provides a general limit for
the violations of \Tz\ and \CPTz\ invariance.
\\\\
The eigenstates can now be expressed by the elements of $\Lz$
\begin{eqnarray}
\ket{\ks}=N_S \left(\PLab \ket{\kn} \ +\ (\lzs-\PLaa)\ket{\knb}\right)\;,\label{eq:2.34}\\ 
\ket{\kl}=N_L \left((\PLbb-\lzl)\ket{\kn} \ -\ \PLba \ket{\knb}\right)\;,\label{eq:2.35}
\end{eqnarray}
with suitable normalization factors $N_S$, $N_L$.\\
They develop in time according to 
\begin{equation*}
\ket{\kls} \rightarrow \e^{-\i\lzls t}\ket{\kls}\ .
\end{equation*}
\gls\ thus signify the decay widths of the eigenstates with mean lifes $\tls=1/\gls $, and \mls\
are the rest masses. \\
These quantities are directly measurable. The results show that \ $\tl\ \gg \ts$ and $\ml>\ms$\ .
We therefore have
\begin{eqnarray*}
0 \le \fsw\ \le \pi/2\ .\\
& &
\end{eqnarray*}
In the limit \ $\epn \rightarrow 0$, $\den \rightarrow 0$, the eigenstates are
\begin{eqnarray}
\ket{\ks}\rightarrow \ket{\mathrm{K}_1}=\frac{1}{\sqrt{2}} \left(\ket{\kn} \ +\ \ket{\knb}\right)\;,\notag\\ 
\ket{\kl}\rightarrow \ket{\mathrm{K}_2}=\frac{1}{\sqrt{2}} \left(\ket{\kn} \ -\ \ket{\knb}\right)\;.\notag
\end{eqnarray}

The differences \Mzaa\ $-$\ \Mzbb\ and \Gzaa\ $-$\ \Gzbb\ , which may be interpreted as \ (\CPTz\
violating) mass and decay width differences between the \kn\ and the \knb , 
are related to \den\ and to \dlz\ as follows: \\\\
Define the reals \ \dpar\ and \ \dper\ by
\begin{equation}\label{dpardsen}
\dpar + \i \dper\ = \den\ \e^{-\i \fsw}
\end{equation}
then
\begin{eqnarray}
\Delta\Mzaa\ &\equiv\ &\Mzaa\ -\ \Mzbb\ =  2\modulus{\dlz}\dper\label{eq:2.32} \\
\Delta\Gzaa\ &\equiv\ &\Gzaa\ \ -\  \Gzbb \phantom{x}  = 4\modulus{\dlz}\dpar\ .\label{eq:2.33}
\end{eqnarray}
We wish to remark that, given the constants $\modulus{\dlz}$\ and \fsw , the information contained in the mass and
decay width differences (\ref{eq:2.32},\ref{eq:2.33}) is identical to the one in \den .

\subsection{Symmetry}\label{symmetry}

The measurement of particularly chosen transition rate asymmetries concerning the neutral kaon's time
evolution exploit properties of \Hwk\ in an astonishingly direct way.

To explain the principle of the choice of the observables we make the temporary assumption that the
detected decay products unambigously mark a relevant property of the kaon at the moment of its
decay: The decay into two pions indicates a \CPz\ eigenstate with a positive eigenvalue, a semileptonic
decay, (\ra\ e$\pi\nu$ \ or \ \ra\ $\mu\pi\nu$) , indicates the kaon's strangeness to be equal to the
charge of the lepton (in units of positron charge).

We will later show that previously unknown symmetry properties of the decay mechanism
($\Delta S = \Delta Q$ rule, $\CPTz$ violation 'in decay') or practical experimental
conditions (efficiencies, interactions with the detector material, regeneration of $\ks$
by matter) do not change the conclusions of this section.

\subsubsection{\Tz\ violation}\label{tviol}

Compare the probability for an antikaon to develop into a kaon, $| \bra{\kn }\e^{-\i\Lz t}\ket{\knb } |^2$ , with the one for
a kaon to develop into an antikaon, $| \bra{\knb }\e^{-\i\Lz t}\ket{\kn } |^2$, within the 
same time interval $t$. Intuition wants the probabilities for these mutually reverse processes to
be the same, if time reversal invariance holds.

We now show that the experimentally observed difference ~\cite{pen2} formally contradicts 
\Tz\ invariance in $\Hwk$.

Following ~\cite{leebook}, time reversal invariance, defined by $\Tz ^{-1} \ \Hwk\ \ \Tz = \Hwk $,
requires 
\begin{equation}\label{eq:2.36}
\Gzbas/\Gzba = \Mzbas/\Mzba
\end{equation}
which is equivalent to $\modulus{\PLab}^2 =\modulus{\PLba}^2$. This is measurable !\\
The normalized
difference of these quantities\\
\begin{equation}\label{eq:2.37}
\AT \equiv \frac{|\PLab |^2 - |\PLba |^2 }{|\PLab |^2 + |\PLba |^2}
\end{equation}\\
is a theoretical measure for time reversal violation ~\cite{kabirb}, and we find, using
(\ref{eq:2.14}), the identity
\begin{equation}\label{eq:2.38}
\AT  \equiv     \frac{| \bra{\kn }\e^{-\i\Lz t}\ket{\knb } |^2 -  
                      | \bra{\knb }\e^{-\i\Lz t}\ket{\kn } |^2} 
                     {| \bra{\kn }\e^{-\i\Lz t}\ket{\knb } |^2 + 
                      | \bra{\knb }\e^{-\i\Lz t}\ket{\kn } |^2} \ ,
\end{equation}
which expresses the different transition probabilities for the mutually reverse processes \
$\knb \Longleftrightarrow  \kn$ , as a formal consequence of the property of \Hwk \ not to commute
with \Tz .
\\\\
A visualization of the terms in the numerator of Eq. (\ref{eq:2.38}) is outlined in the Appendix.
\\\\
The value of \AT\ is predicted as follows
\begin{equation}\label{eq:2.39}
\AT = \frac{-2\re (\epn\sigma^*)}{\modulus{\epn}^2+\modulus{\sigma}^2} \approx 4\ree
\ \ \ \ \ \ (\mathrm{for} \modulus{\epn}\ll \modulus{\sigma}\ \mathrm{and}\ \
\sigma \approx -\frac{1}{2}) \;.
\end{equation}
\\
We add some general remarks:\\
The directness of the relation between \AT \ and \Hwk \ rests partly on the fact that the
neutral kaons are described in a two dimensional space, (\kn, \knb), in which the corollary
(\ref{eq:2.14}) is valid. This is also the origin for the time independence of \AT ~\cite{gerberEPJ}.\\
\AT \ is a \Tz-odd quantity insofar as it changes its sign under the interchange of the initial and final states \
$\knb \Longleftrightarrow  \kn$.\\
Eqs. (\ref{eq:2.36}) and (\ref{eq:2.37}) describe time reversal invariance in an explicitly
phase transformation invariant form. In Eq. (\ref{eq:2.39}), both, the numerator and the
denominator, have this invariance. The approximations concerning
$\modulus{\epn}\ $ and $ \modulus{\sigma}\ $ correspond to the phase convention to be introduced later. (We will choose a phase angle $\vartheta $, neglecting
$\modulus{\epn}^2 \ll 1$ and $\modulus{\den}^2 \ll 1$, such that
 $\sigma  =  -1/2$). 
\\
Eq. (\ref{eq:2.36}) shows that the
present two dimensional system can manifest time reversal violation only, if \Gz \ is not the null
matrix, i. e. if there is decay. However, since the absolute value $\modulus{\Gzba}$ does not
enter Eq. (\ref{eq:2.36}), the definition of time reversal invariance would stay intact if the
decay rates $\gs$ and $\gl$ would (hypothetically) become the same, contradicting ~\cite{wolfe1,wolfe}.\\
\AT \ has been measured ~\cite{pen2} not to vanish, $\AT \ne 0$. Since only the relative phase
of \Gzba \ and \Mzba , $arg(\Gzba)-arg(\Mzba)$, and not the absolute values, determines time
reversal violation, Eq. (\ref{eq:2.36}) does not give any prescription as to what extent the
violation should be attributed to M or to \Gz.

\subsubsection{\CPTz\ invariance}\label{cpt}

The \CPTz invariance of \Hwk \ requires the equality of the probabilities, for a kaon and for an antikaon,
to develop into themselves.
\begin{equation}\label{eq:2.40}
\ACPT  \equiv \frac{| \bra{\knb} \e^{-\i \Lz t} \ket{\knb }|^2 -
                    | \bra{\kn } \e^{-\i \Lz t} \ket{\kn  }|^2} 
                   {| \bra{\knb} \e^{-\i \Lz t} \ket{\knb }|^2 +
                    | \bra{\kn } \e^{-\i \Lz t} \ket{\kn  }|^2}\
\end{equation}
is thus a measure for a possible \CPTz violation.
We note (from Ref.~\cite{leebook}), indicated in Table \ref{tab:01}, that \CPTz invariance entails $\PLaa =\PLbb $,
or \ $\den = 0$. Using (\ref{eq:2.12}), we obtain, with $\modulus{\den}\ll 1$,
\begin{equation}\label{eq:2.41}
\ACPT = \frac{4\red \sinh (\frac{1}{2}\dg t )+ 4\imd \sin (\dm t) } 
{\cosh(\frac{1}{2} \dg t ) + \cos (\dm t ) } \;,
\end{equation}
and confirm that $\ACPT \neq 0$ at any time, i. e. \den $\neq$ 0, would contradict the property of
\Hwk \ to commute with \CPTz \;.

\subsection{Decays}\label{decays}

We assume that the creation, the evolution, and the decay of a neutral kaon can be considered as
a succession of three distinct and independent processes. Each step has its own amplitude with its
particular properties. It determines the initial conditions for the succeeding one. (For a refined
treatment which considers the kaon as a virtual particle, see ~\cite{sachsd, lipu},\ with a comment
in \cite{enz}).

The amplitude for a kaon characterized by $\ket{\psi_0}$ at $t=0$, decaying at time $t$,
into a state $\ket{\psi_f}$ is given by
\begin{eqnarray}
{\cal A}^f = \bra{\psi_f}\Hwk \ket{\psi} = \bra{\psi_f}\Hwk \ \e^{-\i \Lz t}\ket{\psi_0} & = &
\bra{\psi_f}\Hwk \ket{s'}\bra{s'}\e^{-\i \Lz t}\ket{s} \sbraket{s}{\psi_0}\nonumber\\
& = & {\cal A}^f_{s'}\ (\e^{-\i \Lz t})_{s's}\ \psi_s(0).\label{eq:2.42}
\end{eqnarray}
The sum over $s'$, $s$ \ includes all existent, unobservable, (interfering) paths.\\
It is
\begin{equation}\label{eq:2.43}
{\cal A}^f_{s'} = \bra{\psi_f}\Hwk \ket{s'}
\end{equation}
the amplitude for the instantaneous decay of the state with
strangeness $s'$ into the state $\ket{\psi_f}$, and $\psi_s(0) = \sbraket{s}{\psi_0}$,
($s=\kn$, $\knb$) are the components of $\psi_0$. The $(\e^{-\i \Lz t})_{s's}$ are taken from
(\ref{eq:2.12}) or (\ref{eq:2.26}).\\
The probability density for the whole process becomes
\begin{equation}\label{eq:2.44}
\modulus{{\cal A}^f}^2 =
D^f_{s't'} \ (\e^{-\i \Lz t})^*_{s's}(\e^{-\i \Lz t})_{t't} \ \psi_s^*(0)\psi_t(0),
\end{equation}
or, for an initial $\kn (s=1)$ or \ $\knb (s=-1)$, it becomes
\begin{equation}\label{eq:2.45}
R^f_s = D^f_{s't'} \ \e_{s't',s}\;.
\end{equation}
Here, the contributions from the kaon's time development ($\e_{s't',s}$) and those from the
decay process ($D^f_{s't'}$) are neatly separated.\\
We have set
\begin{eqnarray}
D^f_{s't'} & = & {\cal A}^{*f}_{s'}{\cal A}^f_{t'}\label{eq:2.46}\\
\e_{s't',s} & = & (\e^{-\i \Lz t})^*_{s's}(\e^{-\i \Lz t})_{t's}\;.\label{eq:2.47}
\end{eqnarray}

\subsubsection{Semileptonic Decays}\label{semilep}

For the instant decay of a kaon to a final state $(\ell\pi\nu)$ we
define the four amplitudes
\begin{equation}\label{eq:2.48}
{\cal A}^q_{s'} = \bra{\ell\pi\nu}\Hwk \ket{s'}, \ \ \ q, s' = \pm 1 \ ,
\end{equation}
with
$q$: lepton charge (in units of positron charge), \ $s'$: strangeness of the decaying kaon.\\
We assume lepton universality. The amplitudes in (\ref{eq:2.48})  thus must not depend
on whether $\ell$ is an electron or a muon.

Known physical laws impose constraints on these amplitudes:\\
The $\DS = \DQ$ rule allows only decays where the strangeness of the kaon equals the
lepton charge, ${\cal A}^q_q$\;, and $\CPTz$ invariance requires (with lepton spins ignored)
${\cal A}^{-q}_{-s'} = {\cal A}^{*q}_{s'}$ \cite{lee2}. The violation of these laws will be
parameterized by the quantities \xx \ , \xb \ , and \rey , \ posing \\\\
${\cal A}^{1}_{-1}$ = \ \xx \ ${\cal A}^{1}_{1}$, \ \  ${\cal A}^{-1}_{1}$ = \
{\xb}* ${\cal A}^{-1}_{-1}$, \ \
$(\modulus{{\cal A}^{1}_{1}}^2 - \modulus{{\cal A}^{-1}_{-1}}^2 ) \ = \ -2\rey \
(\modulus{{\cal A}^{1}_{1}}^2 + \modulus{{\cal A}^{-1}_{-1}}^2 )$, \\\\
or by \ \ $\xp = (x+\xb)/2$ \ and \ $\xm=(x-\xb)/2$ \ .\\\\
$\xp $ describes the violation of the $\DS=\DQ$ rule in \CPTz 
-invariant amplitudes, $\xm $ does so in \CPTz -violating amplitudes.\\

The four probability densities for neutral kaons of strangeness $s = \pm 1$, born at $t=0$, to decay at time $t$
into $\ell\pi\nu$ with the lepton charge $q = \pm 1$ are, according to (\ref{eq:2.45}),
\begin{equation}\label{eq:2.49}
R^q_s = D^q_{s't'} \ \e_{s't',s}\;,\ \ \ \mathrm{with}
\ \ \ D^q_{s't'}= {\cal A}^{*q}_{s'}{\cal A}^q_{t'} \ .
\end{equation}
The decay rates, proportional to $R^q_s$, are given in \cite{pr}.\\\\
We discuss here the asymmetries \AT \ and \ACPT \ with possible, additional symmetry breakings
in the decay taken into account. The completed expressions are denoted by \ $\AT (t)$
and $\ACPT (t)$.
Using Eqs. (\ref{eq:2.49}) and (\ref{eq:2.12}), we obtain\\
\begin{eqnarray}
\AT   (t ) & = & \frac{R^{1}_{-1} (t ) - R^{-1}_{1} (t )}{R^{1}_{-1} (t )  + R^{-1}_{1} (t )}
            \notag \\
            & = & \AT - 2\re (y+x_{-})
             +  2 \frac{\rexm (\e^{-\frac{1}{2}\dg t } -
                         \cos(\dm t )) + \imxp \sin(\dm t )}
                  {\cosh(\frac{1}{2} \dg t ) - \cos(\dm t )}\label{eq:2.50}\\
 &\ra & \AT -2\re (y+x_{-})\; \text{for} \; t \gg \ts , \label{eq:2.51}\\
 &  & \notag \\
\ACPT (t ) & = &\frac{R^{-1}_{-1} (t ) - R^{1}_{1} (t )}{R^{-1}_{-1} t ) + R^{1}_{1} (t )} 
                    \notag\\
            & = &\frac{4\re (\delta+x_- /2) \sinh (\frac{1}{2}\dg \t )+4\im (\delta+x_+/2) \sin (\dm \t) }
                         {\cosh(\frac{1}{2} \dg \t ) + \cos (\dm \t )}
                  + 2\rey\ \label{eq:2.52}\\
 &\ra & 4\red + 2\re (y+x_{-}) \;\text{for}\; t \gg \ts .\label{eq:2.53}
\end{eqnarray}
We compare (\ref{eq:2.52}) with (\ref{eq:2.41}), and we recognize that, besides the
additional term \rey\ , the new expression has the same functional behaviour, just
with the old variable $\delta$\ replaced by 
\begin{equation}\label{eq:2.54}
\den\ \ra \den\ + \frac{1}{2}(\rexm + \i\ \imxp).
\end{equation}
This shows that the analysis of a measurement, based on Eq. (\ref{eq:2.52}) alone, 
can not distinguish a possible \CPTz\ violation in the time development from possible violations due to \rexm\ or \imxp\ in the decay.\\\\
We note that the combination
\begin{equation}\label{eq:2.55}
\AT(t \gg \ts) + \ACPT(t \gg \ts)\ =\ \AT + 4\red \approx 4\ree + 4\red
\end{equation}
yields a particular result on the kaon's time evolution that is free from symmetry
violations in the decay~!
\\\\
Eqs. (\ref{eq:2.50}) to (\ref{eq:2.54}) are valid for $\modulus{\xx} \ll 1 \ ,
\modulus{\xb}\ll 1, \modulus{\rey} \ll 1$, and $\modulus{\den} \ll 1$. For the last
term in (\ref{eq:2.55}) we also assume $\modulus{\epn} \ll 1$ and $\sigma \approx -1/2.$\\

Additional information on ${\re (y+x_{-})}$\ is gained by measuring the charge asymmetry
\dell\ in the semi-leptonic decays.
\\
For \ $t \gg \ts$\ we obtain, up to first order in
$\modulus{\epn},\modulus{\den},\modulus{y}, \mathrm{and} \modulus{x_-}\ ,$
(and for $\sigma = -\frac{1}{2}$)
\begin{equation}\label{eq:2.56}
\dell\ = \frac{R^1_s - R^{-1}_s}{R^1_s + R^{-1}_s}=
2\re (\epn-\del)-2\re (y+x_{-})
\end{equation}
independent of \ $s = 1$ or $-1$\ .\\\\
An equivalent asymmetry $A_\mathrm{S}$ has been derived from the same rates integrated at short decay
time \cite{buch,misc}
\[A_\mathrm{S} = 2\re(\epn + \del) -2\re(y+x_{-})\ .\]

\subsubsection{Decays to two pions - Decay rate asymmetries}\label{twopi}

The amplitude ${\cal A}^{\pi\pi}$\ for the decay of the kaon $\ket{\psi_0}$\ into two pions
$\ket{\pi\pi}$ is, following (\ref{eq:2.42}),
\begin{equation}\label{2.57}
{\cal A}^{\pi\pi}\ = 
\bra{\pi\pi}\Hwk \ket{s'}(\e^{-\i \Lz t})_{s's}\ \psi_s(0)\ .
\end{equation}
We express $\ket{s'}$\ by the eigenstates $\ket{\mathrm{K}_\kappa}$,\ \
$\kappa =$ L or S,  using (\ref{eq:2.17}) to (\ref{eq:2.20}),
(\ref{eq:2.22}) and (\ref{eq:2.23}),\ and find $W^{\kappa s}\ =
\sbraket{\widetilde{\mathrm{K}_\kappa}}{s}$\ and\\
\begin{equation}\label{2.58}
\ket{s'}(\e^{-\i \Lz t})_{s's}\ =\ \ket{\mathrm{K}_\kappa}
\e^{-\i \lambda_\kappa t}\ \sbraket{\widetilde{\mathrm{K}_\kappa}}{s},
\end{equation}
and regain~\cite{enz}
\begin{equation}\label{2.59}
{\cal A}^{\pi\pi}\ = {\cal A}^{\pi\pi}_{\kappa}\ \e^{-\i \lambda_\kappa t}\
\sbraket{\widetilde{\mathrm{K}_\kappa}}{s}\ \psi_s(0)
\end{equation}
where
\begin{equation}\label{2.60}
{\cal A}^{\pi\pi}_{\kappa}=\bra{\pi\pi}\Hwk \ket{\mathrm{K}_\kappa},\ \ \ 
\kappa = \mathrm{L\ or\ S}\ .
\end{equation}\\
The decay rates are \ $\propto {\modulus{{\cal A}^{\pi\pi}}}^2\ .$\ For a
$\kn$ at\ \ $t=0$, we obtain\\
\begin{equation}\label{eq:2.61}
R^{\pi\pi}_{1}\ = \ {\modulus{{\cal A}^{\pi\pi}_{\kappa}\ 
\e^{-\i \lambda_\kappa t}\ W^{\kappa 1}  }}^2\ =
{\modulus{{\cal A}^{\pi\pi}_{\mathrm{S}}\ \e^{-\i \lzs t}\ W^{11}+
{\cal A}^{\pi\pi}_{\mathrm{L}}\ \e^{-\i \lzl t}\ W^{21}  }}^2\ .
\end{equation}
To calculate this expression it is convenient to use the following approximate
eigenvectors, derived from Eqs. (\ref{eq:2.34},\ \ref{eq:2.35}), with
$\sigma= -\frac{1}{2}$, and valid to
first order in \epn\ and \den\ ,
\begin{eqnarray}
\ket{\ks} &=& N^S \left((1+\epn\ +\den) \ket{\kn} \ +\ (1-(\epn\ +\den))\ket{\knb}\right)\;,\label{eq:2.62}\\ 
\ket{\kl} &=& N^L \left((1+\epn\ -\den)\ket{\kn} \ -\ (1-(\epn\ -\den)) \ket{\knb}\right)\;,\label{eq:2.63}
\end{eqnarray}\\
where $N^S = N^L \approx\ 1/\sqrt{2}$\ .\\
From Eqs. (\ref{eq:2.21}), (\ref{eq:2.62}), (\ref{eq:2.63})  we derive
\begin{equation}\label{eq:2.64}
W=\left(W^{ij}\right)\approx\ \frac{1}{\sqrt{2}}\left ( \begin{array}{cc}
(1-\epn\ +\den) & \ \ \ (1+\epn\ -\den) \\ (1-\epn\ -\den) & -(1+\epn\ +\den) \end{array} \right ),
\end{equation}\\
and obtain, from (\ref{eq:2.61}), the rates of the decays $\kn \ra  \pi\pi$ and
$\knb \ra  \pi\pi$,\\
\begin{eqnarray}\label{eq:2.65}
R^{\pi\pi}_{\pm 1}\ (t )            &=& \; \frac{[1\mp 2\re (\epn -\den )]}{2}\ \mathrm{\Gamma_S^{\pi\pi}}\ \notag\\
            & &  \times \bigl [\e^{-\gs t } + {\modulus{\eta_f} }^{2}\ \e^{-\gl t } 
            \pm\ 2{\modulus{\eta_f} }\ \e^{-\frac{1}{2}(\gs +\gl )t }\cos (\dm t - \phi_f )
            \bigr ]\,, 
\end{eqnarray}
where $\mathrm{\Gamma_S^{\pi\pi}}$\ is the partial decay width of $\ks \ra  \pi\pi $, and
where $\eta_f$\ equals
\begin{equation}\label{eq:2.66}
\eta_f\ = \frac{{\cal A}^{\pi\pi}_{\mathrm{L}}}
             {{\cal A}^{\pi\pi}_{\mathrm{S}}} = \modulus{\eta_f}\e^{\i\phi_f}\ .
\end{equation}\\
In (\ref{eq:2.65}), terms of the order $\modulus{\epn}^2$, $\modulus{\den}^2$, and
$\modulus{\eta_f\ \den}$ are neglected.\\
The term $2\re (\epn - \del )$ may be of special use in some measurements, for example in
the CPLEAR experiment, see Section \ref{sec:exp_1}.\\

A difference between the rates of the decays of \kn\ and of \knb\ to the \CPz\ 
eigenstates \pipa\ is an indication of \CPz\ violation.\\\\
 For its study, the following rate asymmetries have been formed:
\begin{eqnarray}
\ACPf   (t ) & = & \frac{R^{f}_{-1}\ (t ) - R^{f}_1\ (t )}{R^{f}_{-1}\ (t ) + R^{f}_1\ (t )}
                                                                  \notag \\  
            & = &2\re (\epn - \del ) - 2\frac{
             |\ita _f|\e^{\frac{1}{2}(\gs -\gl)t }\cos(\dm t - \phi_f)}
                  {1+ |\ita _f| ^2 \e^{(\gs -\gl) t }}\,,\label{eq:2.67}
\end{eqnarray}
where $f=\pipi$ (and $\ita _f=\itapm $) or $f=\pinn $ (and $\ita _f=\itaoo $). \\

\subsubsection{Decays to two pions - Isospin analysis}\label{twopiISO}

The final states $f$ may alternatively be represented by states with a definite total 
isospin $I$. 
The following three physical laws are expressed in terms of $I$, and can then be applied to the
neutral kaon decay \cite{kall}:
(i) the Bose symmetry of the two-pion states,
(ii) the empirical \ $\Delta I = \frac{1}{2}$ rule, and
(iii) the final state interaction theorem.
\\\\
The Bose symmetry of the two-pion states forbids the pion pair to have $I=1$.
The $\Delta I = \frac{1}{2}$ rule in turn identifies the dominant transition \ \kn\ (or \knb) $\ra \ket{I=0}
$.
The final state interaction theorem, together with the assumption of
\CPTz\ invariance, relates the amplitudes \ $\bra{I}\Hwk\ket{\kn}$ and
$\bra{I}\Hwk\ket{\knb}$. It then naturally suggests a parametrization of \CPTz\ violation in the decay
process.
\\
The relations
\begin{eqnarray}
\bra{\pipi} &=& \sqrt{2/3}\ \bra{I=0}+\sqrt{1/3}\ \bra{I=2}\label{brapipi}\\
\bra{\pinn} &=&\sqrt{1/3}\ \bra{I=0}-\sqrt{2/3}\ \bra{I=2} \label{brapinn}
\end{eqnarray}
transfer the implications of the laws mentioned to the observable final pion states.\\

We can now calculate the expressions of $\eta_f$ in (\ref{eq:2.66}) for \pipi\ and for \pinn , in terms
of the decay amplitudes to the states with \ $I=0,2$.\\
If we denote $\bra{I}\Hwk\ket{\kn}=A_I\ \e^{\i\del_I}$, then the final state interaction theorem
asserts, that, if \CPTz\ invariance holds, the corresponding amplitude for the antikaon decay is
$\bra{I}\Hwk\ket{\knb}=A^*_I\ \e^{\i\del_I}$. \ $\del_I$ is the phase angle for \pipa\ elastic
scattering of the two pions at their center of momentum energy.\\
Following \cite{barmin} we violate \CPTz\ invariance by intruding the parameters
$B_I\ ,\ I=0,2\ ,$
\begin{eqnarray}
\bra{I}\Hwk\ket{\kn}  &=& (A_I+B_I)\ \e^{\i\del_I}\label{IHK} \\
\bra{I}\Hwk\ket{\knb} &=& (A^*_I-B^*_I)\ \e^{\i\del_I}.\label{IHKb}
\end{eqnarray}
With the relations (\ref{brapipi}) to (\ref{IHKb}), and using (\ref{eq:2.62}), (\ref{eq:2.63}),
we find the amplitudes $\bra{\pipa}\Hwk\ket{\kls}$, and the observables
\itapm\ and \itaoo\ \cite{buch}.
We give them as follows
\begin{eqnarray}
\itapm\ &\approx& \left[\epn\ + \i\ \frac{\imaa }{\reaa }\right] -\ \den' + \epn' \label{epm}   \\
\itaoo\ &\approx& \itapm\ -\ 3\ \epn' \label{enn}
\end{eqnarray}
with
\begin{equation}
\den'\ = \ \left[\den\ - \frac{\reba}{\reaa}\ \right] \label{dprim}, \ \ \ \ \  \ \ \ \ \ \ \
\end{equation}
and with
\begin{eqnarray}\label{eprm}
\epsp\ &\equiv& \ \ \ \e^{i\phi_{\epsp}} \ \times \modulus{\epsp}\ \notag \\  
&=& 
\e^{i{{\phi}^{\mathrm{CPT}}_{\epsp}
}} \times \
\frac{1}{\sqrt{2}}
\frac{\reac }{\reaa }
         \left(\left[\frac{\imac }{\reac } - \frac{\imaa }{\reaa } \right]
           -\i \left[\frac{\rebc }{\reac } - \frac{\reba }{\reaa } \right] \right),
\end{eqnarray}
where
\begin{eqnarray*}
{\phi}^{\mathrm{CPT}}_{\epsp} \ = (\frac{\pi}{2}+\del _2 -\del _0) \ .\\
& &
\end{eqnarray*}
In (\ref{epm}) and (\ref{enn}), terms of second order in the \CPz\ and \CPTz\ parameters, and
of first order in \epsp\ 
multiplied by \reac/\ \reaa, are neglected \cite{kall}.\\ 

\subsubsection{Decays to two pions - With focus on \CPTz\ invariance}\label{twopiCPT}

Equation (\ref{eq:2.08b}) relates the decay amplitudes $A_0$ and $B_0$ to the
elements of \Gz.\\
With approximating the decay rates by the dominating partial rates
into the \pipa\ states with $I=0$, we have
\begin{eqnarray}
\frac{\imaa }{\reaa }&\approx& -\frac{\im\ \Gzab}{\dg}\label{imadreao} \\
\frac{\reba}{\reaa}&\approx& \frac{\Gzaa\ -\ \Gzbb}{2\dg}\ , \label{rebdreao}
\end{eqnarray}
and we recognize that 
\begin{equation}\label{dprimphase}
\den'\ =\ \modulus{\den'}\ \e^{\i(\fsw\ \pm\ \frac{\pi}{2})}
=\ \imd\ (-\tan{(\fsw) + \i})\ ,
\end{equation}
and
\begin{equation}
\frac{\reba}{\reaa}= \den\ -\den'= \red\ +\ \imd\ \times \tan{(\fsw)}\label{rbdra}\ ,
\end{equation}
and that the terms 
$ \left[\epn\ + \i\ \frac{\imaa }{\reaa }\right] \ $ and \ $\den'$ in (\ref{epm})
are out of phase by $\pi/2$. \ See \cite{buch}, as well as \cite{pr,dmdg}, with
\cite{misc} and \cite{NA}, for a justification of the neglect of the other decay modes .\
(In Eq. (\ref{dprimphase}) we had made use
of $\im \den' = \im \den$) \ .\\\\
Equation (\ref{rbdra}) relates the \CPTz\ violating amplitude, of the dominating decay process into
\pipa , with the \CPTz\ violating parameter in the time evolution. Originally, these processes have been treated as
independent.\\

$\epsp$ measures \CPz\ violation in the decay process. From (\ref{eprm})\ we see that it is independent of the parameters of the time evolution.
It is a sum of two terms. One of them is made exclusively of the decay amplitudes
$A_0$ and $A_2$\ , and is thus \CPTz\ invariant. The other one contains the amplitudes $B_0$ and
$B_2$\ , and is thus \CPTz\ violating. They are out of phase by $\frac{\pi}{2}$\ .\\

The value of the phase angle of the \CPTz\ respecting part,
${\phi}^{\mathrm{CPT}}_{\epsp} =(\frac{\pi}{2}+\del _2 -\del _0)$,  happens to be \cite{cola} roughly equal to \fpm\
\begin{equation}\label{fpmfsw}
{\phi}^{\mathrm{CPT}}_{\epsp} - \fpm \approx\ \mathrm{few\ degrees}\ . 
\end{equation}
From the sine theorem, applied to the triangle of Eq. (\ref{enn}),
\begin{equation}\label{sinth}
(\fpm-\foo)=\frac{3 \modulus{\epsp}}{\modulus{\itaoo}}(\phi_{\epsp}-\fpm)\ ,
\end{equation}
we conclude that \CPTz\ invariance in the decay process to two pions requires
\begin{equation}\label{fpmmoo}
\modulus{\fpm - \ \foo} \ll \mathrm{few\ degrees}\ .
\end{equation}
(We have used \ $\modulus{\phi_{\epn'}-\fpm} \ll 1$ \ and $\modulus{\epsp}/\modulus{\itaoo} \ll 1$).\\

On the other hand, the measured difference \ $\fpm -\foo\ $\ limits the \CPTz\ violating parameters
in $\epsp$ as follows.\\\\
From (\ref{eprm}) and (\ref{sinth}) we obtain
\begin{eqnarray}
\frac{\reac }{\reaa }\left[\frac{\rebc }{\reac } - \frac{\reba }{\reaa }\right]
&=&
- \mathrm{Im}(\sqrt{2}\modulus{\epsp}\e^{\i(\phi_{\epsp}- {\phi}^{\mathrm{CPT}}_{\epsp})})\notag
\\
&\approx&
\frac{\sqrt{2}}{3}\ \mitaoo(\foo -\fpm)\ ,\label{reb}
\end{eqnarray}
and finally, with the use
of the estimate \ $\frac{\reac }{\reaa }\approx \modulus{\frac{A_2}{A_0}}$
\ (see \cite{pr, dedi}), we arrive at
\begin{equation}
\frac{\rebc}{\reac}=\frac{\reba}{\reaa}\ +\frac{\sqrt{2}}{3}\
\modulus{\frac{A_0}{A_2}} \mitaoo(\foo -\fpm)\ .\label{B2A2}\\
\end{equation}
This equation, and (\ref{rbdra}), relate the \CPTz\ violating expressions in \epsp\ with
the measured quantities.\\

For any two complex numbers \itapm\ and \itaoo\ with similar phase angles,
$\modulus{\fpm - \foo} \ll 1$, we have to first order
\begin{equation}\label{epmmoo}
\itapm \mp \itaoo\ = \ \{\ \mitapm \mp\mitaoo\ \pm
\ \i \mitaoo(\fpm -\foo)\ \}\ \e^{\i \fpm} \ .
\end{equation}

If we allow for the approximation \ $\e^{\i(\fpm-\phi_\epn)}=1$, we obtain
from (\ref{enn}) and (\ref{epmmoo})
\begin{equation}\label{reepe}
\re{(\epsp/\epn)}=\frac{1}{3}\left(1-\frac{\mitaoo}{\mitapm}\right)
\frac{\mitapm}{\modulus{\epn}}\ .
\end{equation}
This quantity has been determined by a measurement of 
$\modulus{\itaoo/\ \itapm}^2$. See Section \ref{sec:exp_1}.\\\\
We apply (\ref{epmmoo}) to
\begin{equation}\label{et}
\ita\ \equiv \frac{2}{3}\itapm +\frac{1}{3}\itaoo
\end{equation}
and obtain \ (for $\modulus{\fpm - \foo} \ll 1$)
\begin{equation}\label{etx}
\ita = \ \{\ (\frac{2}{3}\mitapm\ +\frac{1}{3}\mitaoo)\ -
\ \frac{\i}{3} \mitaoo(\fpm -\foo)\ \}\ \e^{\i \fpm}=
\modulus{\ita}\e^{\i\phi_{\ita}}
\end{equation}
with
\begin{equation}\label{pheta}
\phi_{\ita}=\frac{2}{3}\fpm +\frac{1}{3}\foo\ .
\end{equation}
For the measured values see Section \ref{meas}.\\\\
We eliminate now \epsp\ from (\ref{epm}) and (\ref{enn}):
\begin{equation}\label{path}
\epn\ + \i\ \frac{\imaa }{\reaa } - \den +\frac{\reba}{\reaa}\ =\ \ita \ ,
\end{equation}
and simplify this equation by setting the arbitrary phase angle $\vartheta$\ in
(\ref{eq:2.10}) to have
\begin{equation}
\Gzab\ =  \mathrm{real}\ , \label{phco}
\end{equation}
making \imaa\ negligible
\begin{equation}
\imaa\ \approx\ \ \ \ 0\ .\ \ \label{aconv}
\end{equation}
This allows one, given \ree\ $> 0$, to fix the phase angle of \epn\ to \fsw 
\begin{equation}
\epn\ =  \modulus{\epn}\ \e^{\i \fsw} \label{efsw}
\end{equation}
and to set $\sigma = -1/2$ 
(having neglected $\modulus{\epn}^2\ll 1$, $\modulus{\den}^2 \ll 1$).\\\\
\epn\ has now obtained the property to vanish, if \Tz\ invariance holds.\\\\
The term $\frac{\reba}{\reaa}\ $ in (\ref{path}) remains, as seen from (\ref{rebdreao}),
uninfluenced by the phase adjustment. We then obtain
\begin{equation}\label{pathconv}
\epn\ - \den +\frac{\reba}{\reaa}\ =\ \ita \ .
\end{equation}
Under \CPTz\ invariance, this relation would be
\begin{equation}\label{pathcpt}
\epn\ = \ita\ = \frac{2}{3}\itapm +\frac{1}{3}\itaoo\ ,
\end{equation}
with
\begin{equation}\label{fecpt}
\phi_{\epn}\ \equiv\ \fsw\ =\phi_{\ita}\ \approx\ \fpm\ .
\end{equation}
Applying (\ref{epmmoo}) to \ $\epn - \ita\ $\ in (\ref{pathconv}) yields, with (\ref{dpardsen}),
\begin{eqnarray}
\dpar + \i \dper\ &=& \den\ \e^{-\i \fsw}\notag \\
&=&  \modulus{\epn}-\mita\ +
\ \i \modulus{\ita}(\fsw-\frac{2}{3}\fpm -\frac{1}{3}\foo)+
\ \frac{\reba}{\reaa} \ \e^{-\i \fsw}\ ,
\end{eqnarray}
and with (\ref{eq:2.32}, \ref{rbdra})
\begin{eqnarray}
\Mzaa - \Mzbb
&=& 2\modulus{\dlz}\dper \notag \\
&=& 2\modulus{\dlz}
\{\modulus{\ita}
(\fsw- \phi_{\ita})
\ -\frac{\reba}{\reaa}\ \sin{(\fsw)}\}\notag \\ 
&=& 2\modulus{\dlz}
\{\ \modulus{\ita}
(\fsw- \phi_{\ita})
- (\ \red + \imd \tan(\fsw)\ )\sin{(\fsw)} \}. \label{mmmb} \ \ \ \ \ \ \ \
\end{eqnarray}
All terms on the rhs are deduced from measurements. \\\\
\CPTz\ invariance requires \ $\Mzaa - \Mzbb = \den = 0$,\ and thus
$\ (\fsw- \phi_{\ita}) = 0$~.\\
The comparison of the values of \ $\fsw $ and of $\ \phi_{\ita}\equiv \frac{2}{3}\fpm +\frac{1}{3}\foo$ ,
done in Section \ref{meas},
will confirm \CPTz~invariance.\\

Finally, combining (\ref{pathconv}) with the semileptonic charge asymmetry (\ref{eq:2.56}),
we obtain
\begin{equation}\label{etdl}
\re{\ (\frac{2}{3}\itapm\ +\frac{1}{3}\itaoo)}\ -\frac{\dell}{2}
=\frac{\reba}{\reaa}+\re{(y+x_-)} \ .
\end{equation}
The terms on the rhs are \CPTz\ violating. 

\subsubsection{Unitarity}\label{uni}

The relation between the process of decay of the neutral kaon and the non-hermitian
part \Gz\ of \Lz , expressed in the Eqs. (\ref{eq:2.07}) and (\ref{eq:2.08b}) offers
the study of certain symmetry violations of \Hwk\ .\\

The terms in the sum for \Gzij\ with \ $\alpha\ \neq\ \alpha'$,\ \ \ 
$\sum_{\beta}\sbraket{\alpha|\Hwk }{\beta}\sbraket{\beta|\Hwk }{\alpha^{\prime}}$,\ \
express simultaneous transitions from different states $\ket{\alpha'}\ \neq\ \ket{\alpha}$\
to one single final state $\ket{\beta}$. If the quantum numbers \ $\alpha'\ \neq\ \alpha$\
represent conserved quantities, then the transitions to the single final state
$\ket{\beta}\ $ would violate the conservation law in question.\\

Based on the fact that the occurence of decay products requires a corresponding decrease
of the probability of existence of the kaon, the following relation \cite{bell} holds 
\begin{equation} \label{bell1}
\ree - \i\ \imd = \frac{1}{2\i\dm + \gs  + \gl} \times \sum  \langle f | \Hwk | \kl 
\rangle \langle f | \Hwk | \ks \rangle ^*,
\end{equation}
where the sum runs over all the final decay states $f$. \\ 

This equation has several remarkable aspects:\\
(i) It is of great generality. Having admitted the time evolution to be of the general form
(\ref{eq:2.04}), its validity is not restricted to perturbation theory or to
\CPTz\ invariance.\\
(ii) The left-hand side (lhs) refers uniquely to the symmetry violations in the time evolution of the
kaon, before decay, while the right-hand side (rhs) consists of the measurements, which include the
complete processes.\\
(iii) The rhs is dominated by the decays to \pipi\ and \pinn\ . The other processes enter with
the reduction factor \gs /\ \gl \ $\approx 580$, and, given their abundances, can often be
neglected \cite{pr,misc,NA}. What remains of the sum, is approximately \ \gs\ \itapp\ , with \ \itapp\
defined by
\begin{equation}\label{etpp}
\itapp\ =\mitapp\ \e^{\i \phi_{\pipa}}
=\itapm\ \mathrm{BR^S _{\pip\pim}}+\itaoo\ \mathrm{BR^S _{\pin\pin}}
\approx\ \frac{2}{3}\ \itapm\ + \frac{1}{3}\ \itaoo
\equiv\ \eta\ ,
\end{equation}
where the BR denote the appropriate branching ratios.\\
The measurements show
\ $\itapp \approx \itapm \approx \itaoo \approx \eta $.\\
(iv) The factor $1/(2\i \dm + \gs + \gl)$ may be approximated by \ 
$\cos(\fsw)\ \e^{-\i \fsw}/\gs$\ ,
and thus
\begin{equation}\label{bellapp}
\ree - \i \imd \approx\ \mitapp \cos(\fsw)\ \e^{-\i (\fsw -\phi_{\pipa})}.
\end{equation}
Besides the results from the semileptonic decays, it is thus the phase in the decay to
\pipa\ which reveals the extent, to which the \CPz\ violation
in the time development, is a \Tz\ violation, and/\ or a \CPTz\ violation.\\ Since the
measurements yield $\phi_{\pipa}\approx \fsw$, the \CPz\ violation is a \Tz\ violation
with \CPTz\ invariance\cite{schub}.\\
We will later consider the hypothetical outcome \ $\phi_{\pipa}\approx \fsw + \pi/2$, which would signal
a \ \CPz\ violation with \ \Tz\ invariance and \ \CPTz\ violation.\\
From Eq. (\ref{bellapp}) we note (since \ $\modulus{\fsw\ - \fpp} \ll 1$ and \ $\fpp \approx \phi_\eta$)
\begin{eqnarray}
\ree\ &\approx\ & \mitapp\ \cos(\fsw)\label{reappr} \\
\imd\ &\approx\ & \mitapp\ \cos(\fsw)\ (\fsw\ - \phi_\eta)\ \approx\ \ree\ (\fsw\ - \phi_\eta)\label{imappr}\ .
\end{eqnarray}
(v) It is straight forward to formally recognize that the measured value of
$\phi_{\pipa}\approx \fsw$ is in contradiction with $\Tz ^{-1} \ \Hwk\ \ \Tz = \Hwk\ $. However,
the experiment which measures $\phi_{\pipa}$ does not seem to involve any comparison of a process,
running forward, with an identical one, but running backward.\\
(vi) An analog mystery concerns \CPTz\ invariance, as the measurement of $\phi_{\pipa}$ also does
not obviously compare \CPTz\ conjugated processes.\\
(vii) Since the result of (\ref{bell1}) is independent on possible symmetry violations in the decay,
while the charge asymmetry \dell\ (\ref{eq:2.56}) contains such violations in the form of
$\re{(y+x_-)}$, we may combine Eqs. (\ref{eq:2.56}) and (\ref{bell1}) in view to evaluate this term.
\\\\
Details of the application of (\ref{bell1}) are found in \cite{pr,blois,phen2,NA}.

\subsection{\Tz\ violation and \CPTz\ invariance measured without assumptions on the decay processes}\label{ohneetwas}

Some following chapters explain the measurements of \AT ($t$)\ and of \ACPT ($t$),
performed in the CPLEAR experiment at CERN.
These quantities are designed as comparisons of processes with initial and final states
interchanged or, with particles replaced by antiparticles, and, as already shown above,
they are intimately related to the symmetry properties of \Hwk\ .
However, they include contributions from possible violations of symmetries in the decays, such as
of \CPTz\ invariance or of the $\Delta S = \Delta Q$ rule.\\
We will evaluate the sizes of \rey, $\re{(y+\xm)}$, and \imxp\ , which constrain such violations
to a negligible level.

As a preview, we recognize that the functions \AT ($t$)\ and \ACPT ($t$)\ consist of a
part which is constant in time, and of a part which varies with time. The varying parts are
rapidly decaying, and they become practically unimportant after $t\ \widetilde{>}\ 5\tau_S $.
The two parts depend differently on the unknowns.\\ The constant parts already, of \AT ($t$) and of
\ACPT ($t$), together with \dell\ , constitute three equations which show the feasibility
to evaluate \AT\ , \red\ , and $\re (y+x_{-})$, and thus to determine an \AT\ , which is independent
from assumptions on \CPTz\ symmetry or from the $\Delta S = \Delta Q$ rule in the semileptonic
decays.\\ This \AT\ depends uniquely on the time-reversal violation in the $evolution$ of the
kaon, and it is thus the direct measure for $\Tz ^{-1} \ \Hwk\ \ \Tz \neq\ \Hwk $ searched for.\\
The \red\ in turn is a limitation of a hypothetical violation of
$(\CPTz )^{-1} \ \Hwk\ \ (\CPTz ) = \Hwk $ , also uniquely concerning the time evolution.

\subsection{Time reversal invariance in the decay to \ $\pi\pi\e^+\e^-$ \ ?}\label{ppee}

The decay of neutral kaons into $\pip\pim\gamma\ $ has been studied in view of gaining
information on symmetry violations which could not be obtained from the decay into \pip\pim\ ,
especially on \CPz\ violation of a different origin than the kaon's time evolution 
\cite{dopo,sewo,coka,linv}.

The existence of a sizable linear $\gamma$ polarization and the possibility of its detection by internal
pair conversion \cite{dali,krwa}, as well as the presence of a \Tz\ noninvariant term have been
pointed out in \cite{dopo}.

Experiments have detected the corresponding \Tz- odd intensity modulation with respect to the
angle between the planes of (\pip\pim) and ($\e^+\e^-$) in the decay \kl \ra \pip\pim$\e^+\e^-$
\cite{ktev,wahl}. As expected, the decay \ks \ra \pip\pim$\e^+\e^-$ shows isotropy \cite{wahl}.
The data confirm a model \cite{svl}, where, as usual, the \CPz\ violation is also \Tz\ violating,
and localized entirely in the time evolution of the kaon.

We discuss this result here, because its interpretation as a genuine example of a time-reversal
noninvariance \cite{svl}, or as a first direct observation of a time asymmetry in neutral kaon
decay \cite{fnew} has triggered critical comments
\cite{gau1,elma,wolfe,gerberEPJ},
with the concurring conclusion that, in the absence of final state interactions, the KTEV experiment
at FNAL would find the same asymmetry when we assume there is no \Tz\ violation \cite{wolfe}.

The enigma is explained in Ref. \cite{elma}, whose authors remind us that a \Tz\ - odd term
does not involve switching `in` and `out` states, and so is not a direct probe of \Tz\ violation.

As a complement, we wish to show that the model of \cite{svl} is an example, that a
\Tz\ odd effect may well persist within \Tz\ invariance, even in the absence of final state
interactions.

The $\gamma$ radiation of \kls\ \ra\ $\pi\pi\gamma$ has basically only two contributions, allowed
by gauge invariance (up to third order in momenta) \cite{linv}, which we refer to as E and as M.
They have opposite space parity, and their space parity is opposite to their \CPz\ parity. Since
\CPz($\pi\pi$) $= +1$, we have \CPz($\pi\pi\gamma$) $= -\Pz(\gamma$).\\
In detail
\begin{equation}\label{cpppg}
\CPz(\pi\pi\gamma) = - \Pz(\gamma)= \ \
\left\{ \begin{array}{rl} 
                         +1 \ &\mathrm{E \phantom{a} radiation}\\
                         -1 \ &\mathrm{M \phantom{l} radiation}\ .
\end{array} \right.
\end{equation}
We thus see that the decays from the \CPz\ eigenstates 
$\mathrm{K}_1$ \ra\ $\pi\pi\gamma_\mathrm{E}$ \ and \
$\mathrm{K}_2$ \ra\ $\pi\pi\gamma_\mathrm{M}$
are allowed within \CPz\ invariance. A signal for \CPz\ violation is (e. g.) the simultaneous
occurence of E \ and M radiation from kaons, which have survived during times 
$t \gg \ts\ $.

The variety of radiations is due to scalar factors, which multiply E and M, which are not
determined by gauge invariance. They have to be measured or calculated from models.

The experiment \cite{taur} at CERN has identified the $\gamma$ radiation from
\ks\ \ra\ $\pi\pi\gamma$ as pure low energy bremsstrahlung. This determines the scalar factor for 
the E radiation to be the one from soft photon emission, and fixes the phase of the 
$\pi\pi\gamma$ amplitude to be the one of the $\pi\pi$ amplitude \cite{low}. This will become an
important ingredient for the model \cite{svl} below.

The experiment \cite{carr} at BNL has found two similarly strong components in the $\gamma$ radiation
of \kl\ \ra\ $\pi\pi\gamma$, (i) the bremsstrahlung, which is now \CPz\ suppressed, and (ii) the
M radiation, whose energy spectrum is compatible with a rise $\propto$ E$^3_{\gamma}$. We can thus
naturally expect that there is a value of the gamma ray energy E$_{\gamma}$\ , where the two
components have equal intensity, and where thus the radiation shows a marked polarization due to interference.

The model of \cite{svl} calculates this polarization, and finds that the correponding observable
asymmetry in the distribution of the angle between the planes (e$^+$e$^-$) and (\pip\pim) is of
the form
\begin{equation}\label{A}
A \propto \mitapm\ \sin(\fpm + \phi_\mathrm{FSI})\ ,
\end{equation}
where $\phi_\mathrm{FSI}$\ is determined by the final state interaction theorem,
and where
\fpm\ , as mentioned above, is fixed by the soft photon emission law.

In order for (\ref{A}) to be a direct manifestation of \Tz\ violation, we would like to see
$A$ disappear, if \Tz\ violation is switched off, while \CPz\ violation remains present.\\
Doing this, following \cite{wolfe} or (\ref{bellapp}), we see that $A$ persists, if we set
\begin{equation*}
\mitapm\ \neq 0\ , \ \fpm\ = \ \fsw\ + \frac{\pi}{2}\ ,\ \ \mathrm{and}
\ \ \phi_\mathrm{FSI} = 0 . \notag
\end{equation*}\\
The model presents thus a \Tz -odd observable which, in the absence of final state interactions,
takes still a finite value when \Tz\ invariance holds.

\subsection{Pure and mixed states}\label{pure}

Until now we have implicitly assumed that a single neutral
kaon represents a {\it pure} state, described by a state vector 
whose components develop in time coherently according to 
Eq. (\ref{eq:2.04}).\\ 
The {\it ensemble} of kaons in a beam is formed most often in
individual reactions, and the kaons develop in time 
independently of each other. This ensemble represents a
{\it mixed} state, and its description needs two state vectors
and the knowledge of their relative intensity.

It is a deeply rooted property of quantum mechanics that the pure state of an 
isolated particle does not develop into a mixed state. Such a (hypothetical) 
transition would entail a loss of phase coherence of the amplitudes, and thus 
become detectable by the weakening of the interference patterns. It would also 
violate \CPTz\ invariance \cite{wald,page}, but in a different way than 
described in previous sections.

The various interference phenomena shown by neutral kaons 
have already been used as a sensitive detector in the search 
for coherence losses. As analysis tool in this search the density-matrix
formalism used to describe mixed states seems appropriate.

\subsubsection{Density matrix description}\label{dens}

The time development of mixed states, and the results of measurements can be compactly
described by the positive definite (hermitian) density matrix $\rho (t)$
\cite{vneu,pauli,fano}.

All density matrices (in Quantum Mechanics, QM) develop in time in the same way, i. e. like those of pure states. A pure
state $\psi$ (with components $\psi_\kappa (t), \ \kappa = $ \kn, \knb) has the density matrix
$\rho (t) = \psi \psi^\dagger$ (with components $\psi_\kappa (t) \psi^\ast_{\kappa'} (t)$).\\
Density matrices thus develop according to 
$\rho (t) = \e^{-\i\Lz t} \psi_0 \ (\e^{-\i\Lz t} \psi_0)^\dagger$, or, denoting
$U(t) = \e^{-\i\Lz t}$ and $\rho (0) = \psi_0 \psi^\dagger_0$, like
\begin{equation}\label{ruru}
\rho (t) = U(t)\rho (0) U^\dagger (t)\ , \ \ t \ge 0\ , \ \ \ \ \ \mathrm{(QM)}.
\end{equation}
The form of (\ref{ruru}) grants the conservation of \\
(i)  \  the \ $rank$,\ \ \ \ and\\
(ii)  the \ $positivity$ of $\rho (t)$.\\\\
Since the pure states have (by construction) density matrices of \ $rank = 1$, the development (\ref{ruru})
keeps pure states pure.\\
Since a matrix  $\rho$\ of $rank = 1$ can always be written as a tensor product of two vectors,
$\rho = \psi {\psi}^\dagger$, the developments by Eqs. (\ref{eq:2.04}) and (\ref{ruru}) become
equivalent for pure states.

Eq. (\ref{ruru}) does not automatically conserve the trace, tr\{$\rho\}$, since $U(t)$ is not unitary. In order to avoid that the probability of existence of a neutral
kaon does exceed the value one, we separately require, as a property of $U(t)$, that
\begin{equation}\label{trac}
1 \ge \mathrm{tr}\{\rho (0)\} \ge \mathrm{tr}\{\rho (t \ge 0)\}\ .
\end{equation}

The outcome of measurements can be summarized as follows:\\
The probability $W$ for a neutral kaon with the density matrix $\rho$\ , to be detected by an
apparatus, tuned to be sensitive to neutral kaons with the density matrix $\rho_f$ , is 
\begin{equation}\label{Wtrr}
W = \mathrm{tr}\{\rho_f\ \rho\} \ .
\end{equation}

\subsubsection{Transitions from pure states to mixed states ?}\label{mixd}

It has been suggested \cite{hawk} that gravitation might influence the coherence of wave functions and
thereby create transitions from pure states to mixed states. These could look like a violation
of Quantum Mechanics (QMV). In order to quantify observable effects due to such transitions, the authors of
\cite{ehns} have supplemented the Liouville equation of motion by a QM-violating term, linear in the density
matrix. They have provided relations of the QMV parameters to a set of observables, to which the
CPLEAR experiment has determined upper limits.\\
Our description includes extensions, specifications, and generalizations of the formalism.

In order to characterize the effects of QMV, it has been successful to introduce the Pauli
matrices $\sigma^\mu , \mu = 0 , \cdots , 3$ (with $\sigma^0 =$ unit matrix) as a basis for the
density matrices $\rho (t) \equiv \rho = R^\mu \sigma^\mu$, and
$\rho (0) \equiv \rho_0 = R_0^\mu \sigma^\mu$. $R^\mu$ and  $R_0^\mu$ are reals.
We note that the determinant equals det$(\rho ) = R^\mu R_\mu$, and find from (\ref{ruru}) that
\begin{equation}\label{lore}
R^\mu R_\mu = \|U\|^2 R_0^\mu R_{0\mu}\ .
\end{equation}
$\|U\|$ is the absolute value of det$(U)$. Indices are lowered with the $4 \times 4$ matrix \
$g =(g_{\mu \nu})=(g^{\mu \nu})$, with
$g^{00} = - g^{11} = - g^{22} = - g^{33} = 1 \ ,\ 
g^{\alpha \beta} = 0$ for $\alpha \neq \beta $\ . \\\\
Eq. (\ref{lore}) is a multiple of a Lorentz transformation~\cite{geprl} between the
four-vectors $ R \equiv (R^\mu)$ and $ R_0 \equiv (R_0^\mu)$. We write its matrix as the
exponential $\e^{Tt}$, and the transformation as
\begin{equation}\label{eTt}
R= \e^{Tt}R_0
\end{equation}
 where $T = (T^{\mu\nu}) = T^{00} {\bf 1}_{4 \times 4} + L$ , and where $L$ is an element of 
the Lie algebra of the Lorentz transformations, and therefore satisfies 
\begin{equation}\label{glgml}
g L g = - L^\mathrm{T} \ .
\end{equation}
$()^\mathrm{T}$ denotes the transpose of ().
\\
This all is not too surprising. It expresses the homomorphism of the unimodular group SL(2,C) onto the proper 
Lorentz group\cite{rich}.
 
Eq. (\ref{eTt}) characterizes the quantum mechanical time evolution, which \it conserves the
purity of the states\rm\ . The purity is now expressed by
\begin{equation}\label{pu}
R^\mu R_\mu =0 \ \ \ \ (R \ \ \mathrm{lightlike \ \ \hat{=} \ \ pure \ state} ).
\end{equation}
An obvious way to let the formalism create transitions from pure states to mixed states is, to
supplement the matrix $T$ above with a matrix $X$, which is $not$ an element
of the Lie algebra of the Lorentz transformations, e. g. which satisfies
\begin{equation}\label{gxgpx}
g X g = + X^\mathrm{T} .
\end{equation}
We will explicitly use
\begin{equation}\label{X}
X = \left(
\begin{array}{rrrr}
0 & S^1 &	S^2	& S^3 \\		 
- S^1	& - J^1	& D^3	& D^2 \\		
- S^2 & D^3	& - J^2	& D^1	\\	 
- S^3	& D^2	& D^1	& - J^3 
\end{array}
\right) \ .
\end{equation}
The time evolution is now generated by the matrix
\begin{equation}\label{T}
T = T^{00} {\bf 1}_{4 \times 4} + L + X ,
\end{equation}
which is just a general \ $4 \times 4$ matrix. From (\ref{glgml}) and (\ref{T}) we see that QM
has 7 parameters, and (\ref{X}) shows that QMV has 9 parameters.

The probability for a neutral kaon, characterized by the four-vector
$R_0$ at time $t=0$ , to be detected, by an apparatus set to be sensitive to $R_f$ , is
\begin{equation}\label{W}
W(t) = \mathrm{tr}(\rho_f \rho (t)) = 2 R^\mu_f (\e^{Tt})^{\mu\nu} R_0^\nu \ \equiv
2\ \e^{T^{00}t} R_f^\mathrm{T} \e^{(L+X) t} R_0\ .
\end{equation}
Using~\cite{fano,feyn}
\begin{equation}\label{elxt}
\e^{(L+X) t}  = \e^{Lt}\ \e^{{\bf D}(t, - L, X)} = \e^{{\bf D}(t, L, X)}\ \e^{L t} =
({\bf 1}+{\bf D}+ \cdots + \frac{1}{n!}{\bf D}^n + \cdots)\ \e^{Lt}
\end{equation}
with
\begin{equation}\label{D}
{\bf D} \equiv {\bf{D}} (t, L, X) = \int\limits^t_0 d\tau\ \e^{L\tau} X \e^{-L\tau} =
-{\bf{D}} (-t, -L, X)\ , 
\end{equation} 
we obtain to first order in $X$
\begin{eqnarray}
W(t) &=& W_{QM}(t)+W^{(1)}_{QMV}(t)\label{Wapp} \\
W_{QM}(t) &=& 2\ \e^{T^{00}t}\ R_f^\mathrm{T}\ \e^{Lt}\ R_0 \label{WQM} \\
W^{(1)}_{QMV}(t) &=&
              2\ \e^{T^{00}t}\ R_f^\mathrm{T}\ {\bf D}\ \e^{Lt}\ R_0 \ .\label{WQMV}
\end{eqnarray}

These equations have an evident interpretation:  $R_0$ describes the kaon beam, $R_f$ describes
the detector, $\e^{Lt}$ describes the regular time evolution, and ${\bf D}$ describes
the decoherence.
$W_{QM}(t)$ represents the result within QM.\\
In order to calculate the expressions (\ref{WQM}), (\ref{WQMV}), we need $T^{00}$ and $L$.
\\
Details are given in the Appendix.
\\
For \ $T = (T^{\mu\nu}) = T^{00} {\bf 1}_{4 \times 4} + L$ \ we obtain, in terms of \Lz\ ,
\begin{equation}\label{TL}
T^{\mu\nu} = \im (\mathrm{tr}\{\sigma^\mu \Lz\ \sigma^\nu  \})\ ,
\end{equation}
especially
\begin{equation}\label{T00}
T^{00} = -(\gs + \gl)/2 \equiv -\gm \ ,
\end{equation}
and

\begin{eqnarray}\label{L}
L = 
\left(
\begin{array}{cccc}
0                &\ -\dg/2                          & 0    & -\Delta\Gzaa /2   \\
\noalign{\medskip}
-\dg/2           &\ 0                               & -\Delta\Mzaa \ \   & 2 \modulus{\epn}\modulus{\dlz} \\
\noalign{\medskip}
0                &\ \Delta\Mzaa                               & 0    & \dm \\
\noalign{\medskip}
-\Delta\Gzaa /2 \ \  &\ -2 \modulus{\epn}\modulus{\dlz} \ \ & -\dm & 0
\end{array}
\right) \ ,
\\ \notag
\end{eqnarray}
and with (\ref{eq:2.12}) 
{\small
\begin{eqnarray}\label{eLcpt}
&&\e^{Lt}=    \notag
\\            \notag
\\
&&\left( \begin {array}{cccc} \notag
C & -S & 0 & 0
\\\noalign{\medskip}
-S & C & 0 & 0 
\\\noalign{\medskip}
0 & 0 & c & s
\\\noalign{\medskip}
0 & 0 & -s & c
\end {array} \right)+ \\ \noalign{\bigskip} 
&2&\left( \begin {array}{cccc} \notag \label{eltcpt}
 2(C-c)(\re\ \epn )^2                 & 
-[S\ \re\ \epn^2 +s\ \im\ \epn^2 ]    & 
 [S\ \im\ \epn -s\ \re\ \epn ]        & 
-(C-c)\ \re\ \epn 
\\\noalign{\medskip}
-[S\ \re\ \epn^2 +s\ \im\ \epn^2 ]     & 
-2(C-c)(\im\ \epn )^2                  & 
-(C-c)\ \im\ \epn                      & 
\phantom{-}[S\ \re\ \epn+s\ \im\ \epn]
\\\noalign{\medskip}
\phantom{-}[S\ \im\ \epn -s\ \re\ \epn ] &
-(C-c)\ \im\ \epn                        & 
 2(C-c)(\im\ \epn ) ^2                   & 
-[S\ \im\ \epn^2 -s\ \re\ \epn^2] 
\\\noalign{\medskip}
\phantom{-}(C-c)\ \re\ \epn              & 
-[S\ \re\ \epn + s\ \im\ \epn]           & 
\phantom{-}[S\ \im\ \epn^2 -s\ \re\ \epn^2]& 
-2(C-c)(\re\ \epn )^2 
\end {array}  \right)+ \\ \noalign{\bigskip}
&&{\cal O  }(\modulus{\epn}^3) \  . 
\end{eqnarray}
} 
We have used
\begin{eqnarray}
s &=& \sin(\dm\ t)\ , \ \ \ \ \ S = \sinh((\dg/2)\ t)  \\
c &=& \cos(\dm\ t)\ , \ \ \ \ C = \cosh((\dg/2)\ t)\ .
\end{eqnarray}
Eq. (\ref{eLcpt}) corresponds to \CPTz\ invariance.
\\
We see that the second order terms in $\epn$ with factors $S$ or $C$ in Eq. (\ref{eLcpt}) 
may become dominating for large times $t$, and can thus not be neglected for a nonvanishing \epn\ .
\\
The first term is also \Tz\ invariant, and may be written as 
\begin{equation}\label{eLap}
\e^{Lt} = \left(									
\begin{array}{cc}
\sigma^0 \cosh (\frac{\dg}{2} t) - \sigma^1 \sinh (\frac{\dg}{2} t) & (0) \\
(0) & \sigma^0 \cos (\dm t) + i \sigma^2 \sin (\dm t)
\end{array}
\right) \ .
\end{equation}
It will be used in Eqs. (\ref{elxt}), (\ref{D}) as the starting point for the calculation of $W(t)$
in the presence of nonvanishing QMV parameters in $X$.
\\

We are now able to list all possible experiments to search for QMV \cite{geprl}. The four dimensions make $R_0$
and $R_f$ capable to define four independent beams and four independent measurements, to give a total of 16 experiments.

It is a fortunate fact that ${\bf D}$ in Eq. (\ref{WQMV}) introduces a sufficiently rich time
dependence, which just enables the existence of a specific set of four experiments \cite{geprl,geEPJCent},
that allows one to determine all 9 QMV parameters of (\ref{X}).\\

As an example we study the influence of QMV on the decay of an initially
pure \kn\ into two pions.
The expression for $R^{\pi\pi}_{+1}(t )$ in (\ref{eq:2.65}) corresponds to $W_{QM}(t)$.
We calculate the modification $W^{(1)}_{QMV}(t)$ due to QMV.\\\\
First, we verify that $R_f^\mathrm{T} = \frac{1}{2} (1\ 1\ 0\ 0)$ represents the state
$\bra{\mathrm{K}_1}$. We note that $\frac{1}{2} (1\ 1\ 0\ 0)\hat{=}\
\frac{1}{2}(\sigma^0+\sigma^1)
=\frac{1}{\sqrt{2}} \left ( \begin{array}{c} 1 \\ 1 \end{array} \right )
\frac{1}{\sqrt{2}} (1\ 1)$, and recognize that the last term is just the tensor product of
$\psi_{\mathrm{K}_1} = 
\frac{1}{\sqrt{2}} \left ( \begin{array}{c} 1 \\ 1 \end{array} \right )$
with $\psi^\dagger_{\mathrm{K}_1}$.\\
In the same way we find: 
$\mathrm{K}_2 \hat{=}\ \frac{1}{2} (1\ -1\ 0\ 0),\ $
$\kn          \hat{=}\ \frac{1}{2} (1\  0\ 0\ 1),\ $ and
$\knb         \hat{=}\ \frac{1}{2} (1\  0\ 0\ -1). $ \\\\
With \ $R_0 = \frac{1}{2}(1\ 0\ 0\ 1)^\mathrm{T} \hat{=} \ket{\kn}$ and \
${\bf D} = J^1 \ {\bf D} (t, L, \frac{\partial}{\partial J^1}X)$, we obtain from (\ref{WQMV})
\begin{equation}\label{WQMVpi}
W^{(1)}_{QMV}(t) = \frac{J^1}{4\dg} \ \e^{-\gl t} \ \ \ \ \ \mathrm{for} \ \ t \gg \ts\ .
\end{equation}
Eqs. (\ref{Wapp}) to (\ref{WQMV}) together with (\ref{WQMVpi}) and (\ref{eq:2.65}) yield the modified expression for the
decay rate
\begin{equation}\label{RQMVpipi}
R^{\pi\pi}_{+1\ QMV}\ (t ) \propto
\bigl [a_{QMV}\ \e^{-\gs t } + {\modulus{\eta_{QMV}} }^{2}\ \e^{-\gl t } 
            + \ 2{\modulus{\eta} }\ \e^{-\frac{1}{2}(\gs +\gl )t }\cos (\dm t - \phi )
            \bigr ]
\end{equation}
where
\begin{equation}\label{etaQMV}
{\modulus{\eta} }^{2} \ra\ {\modulus{\eta_{QMV}} }^{2}={\modulus{\eta} }^{2}+\frac{J^1}{2\dg}\ \ .
\end{equation}
The modification $1 \ra a_{QMV}$ of the short-lived term in (\ref{RQMVpipi}) will not be
considered further.\\
The outstanding features are the modification of ${\modulus{\eta} }^{2}$ in the long-lived term,
in contrast to ${\modulus{\eta}}$ in the interference term, and the fact that the $first$ order term
of QMV, $J^1$, combines with the $second$ order term of the \CPz\ violation, as seen in (\ref{etaQMV}).
This will allow one to determine an especially strict limit for $J^1$ \cite{ehns,huet}. \\
The parameters $J^2, J^3$, and $D^1$ are not present in (\ref{RQMVpipi}). Contributions from further
parameters in $X$ are presently ignored, and discussed below.\\

Eq. (\ref{W}), when $X \neq (0)$, does not guarantee positive values for $W(t)$, unless the
values of the parameters of $X$ fall into definite physical regions, since the time evolution
generated by a general $T$ of (\ref{T}) does not satisfy (\ref{ruru}).

We describe now the general law of time evolution of the density matrix and the ensuing
physical region for the values in $X$.\\
The intriguing mathematical fact \cite{choi,gori} is, that Eq. (\ref{ruru}) does guarantee the positivity of $W(t)$, not only 
for the evolution of a single kaon, but also for a (suitably defined) system of many kaons.
On the other hand, the precautions for the positivity (when $X \neq (0)$), tailored to the single-particle evolution, do,
in general, $not$ entail the positivity for the many-particle evolution, unless the single-particle evolution
has the property of $complete \ positivity$. For the application to neutral kaons, see
\cite{Ben97,Ben02,geEPJCent}.\\
We summarize three results.\\
(i) Complete positivity is a necessary condition \cite{geEPJCent}
for the consistent description of entangled neutral kaon pairs in a symmetric 
state, as
produced in the \PPb\ annihilation \cite{Ad97,pr}. A general law of evolution therefore
has to have this property.\\
The time evolution is completely positive if (and only if) it is given by (see \cite{choi})
\begin{equation}\label{rogen}
\rho (t) = U_i (t) \rho (0) U_i^{\dagger} (t) \ ,
\end{equation}
where the right-hand side is a sum over four terms, with suitably normalized $2 \times 2$
matrices $U_i$ \ \cite{alic}.\\\\
(ii) The physical region for the values of the QMV parameters in $X$ follows from (\ref{rogen})
\cite{Ben98,geEPJCent}. The most important conditions are:
\begin{equation}\label{phy1}
(D^i)^2 + (S^i)^2 \leq ( (J^i)^2 - (J^j - J^k)^2 )/4 \ ,
\end{equation}
\begin{equation}\label{phy2}
J^i \ge 0 \ \ \ \ \forall i = 1, \cdots ,\ 3 \ ,	
\end{equation}
\begin{equation}\label{phy3}
J^i \le J^j + J^k \ ,			
\end{equation}
\begin{center} $(ijk =$ permutation of 123)\ . \end{center}
We note:\\\\
If any one of the three diagonal elements, $J^i$, vanishes, then the other two ones are equal, and all 
off-diagonal elements, 
$D^i$ and $S^i$, vanish, and\\
if any two of the three diagonal elements, $J^i$, vanish, then all elements of $X$ vanish.\\\\
(iii) The condition $(1 \ge \mathrm{tr}\{\rho (0)\} \ge \mathrm{tr}\{\rho (t \ge 0)\})$ demands in 
addition
\begin{equation}\label{Tle0}
T^{00} \le 0 \ ,
\end{equation}\label{Tgeq}
\begin{equation}
(T^{00})^2 \geq ( {L}^{01} + S^1)^2 + ( {L}^{02} + S^2)^2 + ( {L}^{03} + S^3)^2 \ .
\end{equation}
This shows that, with the properties of neutral kaons, especially since \gs\ $\gg\ $\gl\ ,  there 
is little room for the values of the small parameters.\\\\

\subsection{Entangled kaon pairs}\label{enta}

Particles with a common origin may show a causal behaviour, still when they have become
far apart, that is unfamiliar in a classical description.

Pairs of neutral kaons \cite{gold,lipk} from the decay $\phi \ra \kn \knb $, with the kaons 
flying away in opposite directions in the $\phi$'s rest system, have a number of remarkable 
properties \cite{huet,enz}:\\\\
They are created in the entangled, antisymmetric $J^{PC} = 1^{--}$ state
\begin{eqnarray}
\ket{\psi_-} &=& \frac{1}{\sqrt{2}}(\ket{\kn}\ket{\knb}-\ket{\knb}\ket{\kn})\label{str}\\
             &=& \frac{1}{\sqrt{2}}\ (\ket{\kk_2}\ket{\kk_1}-\ket{\kk_1}\ket{\kk_2})\label{cp}\\
      &\approx & \frac{1}{\sqrt{2}}\ (\ket{\kl}\ket{\ks}-\ket{\ks}\ket{\kl})\label{mass}.
\end{eqnarray}

In each of these representations the two particles have opposite properties: opposite
strangeness (\ref{str}), opposite CP parity (\ref{cp}), or opposite shifts of the eigenvalues 
of \Lz\ (\ref{mass}).\\
This allows the experimenter to define an ensemble of neutral kaons which has one of these
properties with high purity \cite{dafn,buch}.\\
The identification of the particular quantum number shown by the
particle which decays first, assures the opposite value for the surviving one.\\
This intriguing feature is not merely a consequence of conservation laws, since, at the moment
of the pair's birth, there is nothing which determines, which particle is going to show what
value of what observable, and when.\\

From (\ref{mass}), we see that $\ket{\psi_-}$ develops in time just by a multiplicative factor,
$\ket{\psi_-} \ra \ket{\psi_- (t)} = \e^{-\i (\ml + \ms)\gamma t}\
\e^{-\frac{1}{2}(\gs +\gl)\gamma t}\ket{\psi_-}$ which is independent of the symmetry violations
($\gamma t$ = eigentime of the kaons).
$\ket{\psi_- (t)}$ and $ \ket{\psi_-}$ have thus the same decay properties, e. g. the kaon pair cannot
decay simultaneously into two \pipa\ pairs or into the same \pen\ triplets, at all times. It 
is due to this simplicity of the time evolution, and due to the antisymmetry of $\ket{\psi_-}$,
that the kaons from $\phi\ $ decay are so well suited to explore symmetry violations in the
decay processes, or to search for QMV, which ignores the states' antisymmetry.
\\\\
The Eqs. (\ref{str}) to (\ref{mass}) represent special cases of linear combinations.
\\
Let us express the states \ $\ket\kn$, $\ket\knb$ by some two new, arbitrary, independent states $\ket{\kk_\alpha},\ket{\kk_\beta}$ as follows
\begin{eqnarray}
\ket\kn &=& \braket{\kk_\alpha}{\kn}\ \ket{\kk_\alpha}+\braket{\kk_\beta}{\kn}\ \ket{\kk_\beta}\phantom{\ .} \\
\ket\knb &=& \braket{\kk_\alpha}{\knb}\ \ket{\kk_\alpha}+\braket{\kk_\beta}{\knb}\ \ket{\kk_\beta}
\end{eqnarray}
with the coefficient matrix
\begin{equation}
N=
\left(
\begin{array}{cc}
\braket{\kk_\alpha}{\kn}  & \braket{\kk_\beta}{\kn}\\\noalign{\medskip}
\braket{\kk_\alpha}{\knb} & \braket{\kk_\beta}{\knb}
\end{array}
\right)\ .
\end{equation}
Expressed in these new states, (\ref{str}) becomes (simply)
\begin{equation}\label{str1}
\ket{\psi_-} = \frac{\det (N)}{\sqrt{2}}\ (\ket{\kk_\alpha}\ket{\kk_\beta}-\ket{\kk_\beta}\ket{\kk_\alpha})\ .
\end{equation}
This says, that Eqs. (\ref{str}) and (\ref{str1}) are, apart from an irrelevant numerical factor, identical in their form.\\
Especially the \it antisymmetry\rm\ of $\ket{\psi_-}$ is inert against all kinds of (nonsingular) linear transformations of the basis vectors.
\\

For the formal description of a pair of neutral kaons in a general (\it mixed\rm) state, we use the positive
definite $4\times 4$ density matrix $\rho (t_1 ,t_2)$. The times $t_1, t_2$ indicate the moments when
later measurements on the individual particles will be performed. As a basis we use
$(\kn \kn, \ \kn \knb, \ \knb \kn, \ \knb \knb)$, and we assume (with \cite{enz}), that $\rho (t_1 ,t_2)$
evolves like $\rho (t_1) \otimes \rho (t_2)$. The two-particle evolution is thus uniquely determined by the
one-particle evolutions, and thus the introduction of QMV becomes obvious.\\
Similar to the one-particle $2 \times 2$ density matrices, we develop the $4 \times 4$ density
matrix $\rho (t_1 ,t_2)$ in terms of the products $(\sigma^\mu \otimes \sigma^\nu)$ with
coefficients $R^{\mu\nu} \equiv R^{\mu\nu} (t_1 , t_2) , \  \ R_0^{\mu\nu} \equiv R^{\mu\nu} (0,0)$,
as 
\begin{equation}
\rho (t_1 ,t_2) = R^{\mu\nu} (\sigma^\mu \otimes \sigma^\nu)
\label{eq:rho2}
\end{equation}
and obtain
\begin{equation}
R^{\mu\nu} =  (\e^{T t_1})^{\mu\alpha} R_0^{\alpha\beta} (\e^{T t_2})^{\nu\beta}
\label{Rmunu}.
\end{equation}
The generator \ $T$ may, or may not, contain QMV terms.
\\\\
The probability density that an apparatus, tuned to $\rho_f = R_f^{\mu\nu} (\sigma^\mu \otimes \sigma^\nu)$
detects the particles at the times $t_1, t_2$ is
\begin{eqnarray}\label{Wt1t2}
W(t_1 ,t_2) & = & \mathrm{tr}( \rho_f  \rho (t_1 ,t_2) ) \ = \ 4 R_f^{\mu\nu} R^{\mu\nu} = \ 4 R_f^{\mu\nu} (\e^{T t_1})^{\mu\alpha}
R_0^{\alpha\beta} (\e^{T t_2})^{\nu\beta}  \nonumber \\
& = & 4 R_f^{\mu\nu} (\e^{(L+X) t_1})^{\mu\alpha} R_0^{\alpha\beta} (g \ \e^{( -L+X) t_2}\ g)^{\beta \nu} \e^{T^{00}(t_1 + t_2)}
\nonumber \\ 
& \equiv & 4\ \mathrm{Tr} \{\ {\bf{R}}_f^\mathrm{T}\ \e^{(L+X) t_1}\ {\bf{R}}_0 \ \ g \ \e^{( - L+X) t_2}\ g\ \}\ \e^{T^{00}(t_1 + t_2)}\nonumber \\
&=& 4\ \mathrm{Tr} \{\ {\bf{R}}_f^\mathrm{T}\ \e^{{\bf D}(t_1, L, X)}\ \e^{L t_1}\ {\bf{R}}_0 \ \ g \ \e^{-Lt_2}\ \e^{{\bf D}(t_2, L, X)}\ g\ \}\ \e^{T^{00}(t_1 + t_2)} \ .
\end{eqnarray}
The elements of the
$4 \times 4$ matrices $\bf{R}_0$ and $\bf{R}_f$~, e.g., are respectively,
$R_0^{\alpha\beta}$ and $R_f^{\nu\mu}$. Tr acts on their superscripts.
\\\\
Eq. (\ref{Wt1t2}) is the general expression for the measurements of all the parameters of 
the (general) \kn \knb pair. It allows one to draw a complete list of all independent possible experiments.
\\\\
In the deduction of (\ref{Wt1t2}), the equations (\ref{glgml}) and (\ref{gxgpx}) have been used.
The different relative signs of $L$ and $X$ in the two exponents in (\ref{Wt1t2}) mark the
difference of the effects due to QM or to QMV.\\
As a benefit of complete positivity, $W(t_1 ,t_2)$, for $t_1, t_2 \ge 0$, will be positive, if $X$ satisfies the 
correspondig single-particle criteria.\\\\
If the two measurements at the times $t_1, \ t_2$ \ yield \it independent events in pure states\rm , then the matrix $\bf{R}\rm\it_f\rm$ factorizes into two arrays 
\begin{equation}
{\bf{R}}_f = R_1 R_2^T\ ,
\end{equation}
and (\ref{Wt1t2}) becomes 
\begin{equation}\label{W12s1}
W(t_1 ,t_2) 
= 4 \ R_1^\mathrm{T}\ \e^{{\bf D}_1}\ \e^{L t_1}\ {\bf{R}}_0 \ \ g \ \e^{-Lt_2}\ \e^{{\bf D}_2}\ g \ R_2 \ \ \e^{T^{00}(t_1 + t_2)} \ .
\end{equation}
We have denoted \ $ {\bf D}_i \equiv {\bf D}(t_i, L, X). $
\\
The assumed purity of the measured states is expressed by (\ref{pu}).
\\
 
It is convenient to have (\ref{W12s1}) approximated to first order in the QMV parameters using \mbox{$\e^{\bf D} \approx \bf 1+D\rm$.}\\
We obtain
\begin{eqnarray}
W(t_1 ,t_2) &\approx & W_{QM}(t_1 ,t_2)+W^{(1)}_{QMV}(t_1 ,t_2)\label{W12s2}
\\
W_{QM}(t_1 ,t_2)
&=& 4 \ R_1^\mathrm{T}\ \ \{\phantom{AC}\e^{L t_1}\ {\bf{R}}_0 \ \ g \ \e^{-Lt_2}\phantom{AC} \} \ \ g \ R_2 \ \ \e^{T^{00}(t_1 + t_2)}\label{Wqm}
\\
W^{(1)}_{QMV}(t_1 ,t_2)
&=& 4 \ R_1^\mathrm{T}\ \ \{{\bf D}_1\ \e^{L t_1}\ {\bf{R}}_0 \ \ g \ \e^{-Lt_2}\phantom{AC}\} \ \ g \ R_2 \ \ \e^{T^{00}(t_1 + t_2)}\notag
\\
&+& 4 \ R_1^\mathrm{T}\ \ \{\phantom{AC}\e^{L t_1}\ {\bf{R}}_0 \ \ g \ \e^{-Lt_2}\ {\bf D}_2\} \ \ g \ R_2 \ \ \e^{T^{00}(t_1 + t_2)} \ . \label{Wqmv}
\end{eqnarray}

We now give the general expression for the results of measurements on the pair (\ref{str}).
Its density matrix is
\begin{eqnarray}\label{ro00}
\rho (0 ,0)_-\
=&\frac{1}{\sqrt{2}} \left ( \begin{array}{c} \phantom{-}0 \\ \phantom{-}1 \\ -1 \\ \phantom{-}0 \end{array} \right )
\frac{1}{\sqrt{2}} (0\ 1 -1\ 0) 
= \frac{1}{2}
\left ( \begin{array}{cccc} 0 & \phantom{-}0 &\phantom{-} 0 &\phantom{-} 0 \\
                            0 & \phantom{-}1 &           -1 &\phantom{-} 0 \\ 
                            0 &           -1 &\phantom{-} 1 &\phantom{-} 0 \\ 
                            0 & \phantom{-}0 &\phantom{-} 0 &\phantom{-} 0 
\end{array} \right )\notag\\
& &  \notag \\
=& \frac{1}{4}(\sigma^0 \otimes \sigma^0 - \sigma^m \otimes \sigma^m)\ \ \
=\ \ (g^{\mu \nu}/4)\ (\sigma^\mu \otimes \sigma^\nu)\ ,
\end{eqnarray}
and thus
\begin{equation}\label{R0SI}
{\bf{R}}_0={\bf{R}}_0^{-} = g/4 \ .
\end{equation}
Inserting this interesting result into (\ref{W12s1}), we obtain
\begin{equation}\label{Wkkb}
W(t_1 ,t_2) =
R_1^\mathrm{T}\ \e^{{\bf D}_1}\ \e^{L (t_1-t_2)}\ \e^{{\bf D}_2}\ g \ R_2 \ \ \e^{T^{00}(t_1 + t_2)} \ .
\end{equation}
Again, all the terms in this expression, apart perhaps of $g$, have an obvious physical
interpretation.\\
Developping the exponentials $\e^{\bf{D}}$, Eq. (\ref{Wkkb}) allows one to calculate the
frequency of occurence of the events detected, by the apparatus tuned to ${\bf{R}}_f= R_1 R_2^T\ ,$ as a
function of all the 16 parameters to any order in the small ones.\\
Explicit expressions have been
published for 3 QMV parameters \cite{huet,ellis96}, for 6 ones \cite{Ben98,Ben971}, and
for 9 ones \cite{geEPJCent}.\\
The experimental study of these 9 parameters using the antisymmetric state 
$\ket{\psi_-}$, as suggested in \cite{geEPJCent}, does not require a high purity of this 
state with respect to the symmetric state  $\ket{\psi_+}$, considered below.
\\

Now, we consider neutral-kaon pairs in the symmetric state
\begin{equation}
\ket{\psi_+} = \frac{1}{\sqrt{2}}(\ket{\kn}\ket{\knb}+\ket{\knb}\ket{\kn})\ .\label{psip}
\end{equation}
They have the density matrix \cite{enz}
\begin{equation}\label{rotri}
\rho (0 ,0)_{+}\
= \frac{1}{2}
\left ( \begin{array}{cccc} 0 & 0 & 0& 0 \\ 0& 1& 1& 0 \\ 0& 1& 1&  0 \\ 0& 0& 0& 0 
\end{array} \right )
=\ \frac{1}{4}(\sigma^0 \otimes \sigma^0 + \sigma^m \otimes \sigma^m
-2\ \sigma^3 \otimes \sigma^3)\ ,
\end{equation}
and
\begin{equation}\label{R0TR}
{\bf{R}}_0 =
{\bf{R}}_0^{+}
= \frac{1}{4}
\left ( \begin{array}{cccc}\phantom{} 1 &\phantom{-} 0 &\phantom{-} 0& \phantom{-} 0\\
                           \phantom{} 0 &\phantom{-} 1 &\phantom{-} 0& \phantom{-} 0\\ 
                           \phantom{} 0 &\phantom{-} 0 &\phantom{-} 1& \phantom{-} 0\\
                           \phantom{} 0 &\phantom{-} 0 &\phantom{-} 0&            -1\\
\end{array} \right ) \ ,
\end{equation}
to be inserted into (\ref{W12s1}).\\
The explicit expression for $W(t_1 ,t_2)$ has been given, for the special case of QM, in
\cite{enz}.\\
$\ket{\psi_+}$, in contrast to $\ket{\psi_-}$, is allowed under QM, to evolve into
$\ket{\kn}\ket{\kn}$ and into $\ket{\knb}\ket{\knb}$.\\
\CPTz\ forbids that \ $\ket{\psi_+}$ evolves into \ \
$\sim (\ket{\ks}\ket{\kl}+\ket{\kl}\ket{\ks})$.\\
\\

For fundamental or practical reasons, we might be interested to measure how exactly, a given sample of kaon pairs, is in the entangled antisymmetric state $\ket{\psi_-}$.
\\
Be
\begin{equation}
\ket{\psi}=\ket{\psi_-}+\omega\ket{\psi_+}, \ \ \ \ \omega\ \mathrm{complex} \label{psi}
\end{equation}
the state of the \kn\knb\ pairs.

The authors of \cite{bebe} point out that $\omega \neq 0$ for the kaons in \ $\phi \ra\ \kn\knb\ $  represents a novel kind of \CPTz\ violation.
\\

We explain now, that $\omega$ can experimentally be distinguished from the 7 parameters, which describe the quantum mechanical time evolution, as well as from the additional 9 ones, which are quantum mechanics violating.
\\

In order to apply Eqs. (\ref{W12s2}) to (\ref{Wqmv}) we insert \ {\bf{R}}$_0$ , which we worked out to be
\begin{equation}\label{R0}
{\bf{R}}_0=\frac{1}{4}\ g\ +\ \frac{1}{2}\ {\bf Y}_\omega\ + {\cal{O}}(\modulus{\omega}^2)
\end{equation}
where
\begin{eqnarray} 
{\bf Y}_\omega=
\left ( \begin{array}{cccc} 0         & 0 & 0 &           -{\cal R} \\
                            0         & 0 & -{\cal I} &\phantom{-} 0 \\ 
                            0         & {\cal I} & 0 &\phantom{-} 0 \\ 
                            {\cal R}  & 0 & 0 &\phantom{-} 0 
\end{array} \right )
\end{eqnarray}
with \ ${\cal R}= \re\ \omega\ , \ \ {\cal I}=\im\ \omega $ ,
\\\\
and obtain
\begin{eqnarray}\label{Wallg}
W(t_1 ,t_2)&=& R_1^\mathrm{T}\ \e^{L (t_1-t_2)}\times \notag \\
&&\{\
g + 2\ \e^{L t_2}\ {\bf Y}_\omega (\e^{Lt_2})^\mathrm{T}
+{\bf D}_2\ g
+\e^{-L (t_1-t_2)}\ {\bf D}_1\ g\ (\e^{-L (t_1-t_2)})^\mathrm{T}\
\}\times \notag \\
&&R_2 \ \ \e^{T^{00}(t_1 + t_2)}\ .
\end{eqnarray}
For a data set consisting of identical pure states, \ $R_1=R_2\equiv R$,
and for equal times \ $t_1=t_2\equiv t$, this becomes independent of $\omega$ :
\begin{equation}\label{Widtt}
W_\mathrm{id}(t,t)= 2\ R^\mathrm{T}\ {\bf D}\ g
R \ \ \e^{2T^{00}t}\ ,
\end{equation}
and allows one to determine the QMV parameters separately from $\omega$ \cite{geEPJCent}. When these are known to be or not to be nonvanishing, then a suitable data set with $R_1 \neq R_2$ introduced into (\ref{Wallg}) provides \ $\omega$.
\\

The treatment presented here is based on the description of the time evolution of the density
matrix, generated by a general $4 \times 4$ matrix. An important difference to the regular
quantum-mechanical time evolution is, that conservation laws do not follow anymore from
symmetry properties, and that their existence is no more compulsory \cite{ehns,huet,ellis96}.
The question has to be left to the models, which enable QMV, whether the 
creator of QMV may
also be the supplier of the otherwise missing conserved quantities.

\section{Measuring neutral kaons}\label{sec:exp_1}
\subsection{Basic considerations}

Many measurements concerning the neutral-kaon
system have been carried out with beams containing a mixture
of \kn\ and \knb\ \cite{blois}. Neutral kaons are
identified, among other neutral particles, by their masses, 
as obtained  from measurements of their decay products.\\
The relative proportion of the numbers of \kn\ and \knb\  
particles at the instant of production is measured separately, 
and taken into account in the analysis.
If the beam crosses  matter, regeneration effects take place,
and the measured ratio has to be corrected \cite{good1,klkn}.\\
Alternatively, \ks\ are separated from \kl\
taking advantage of the fact that a beam containing
\kn\ and \knb\ decays as a nearly pure \ks\ or \kl\ beam
depending on whether it decays very near or far away
from the source. This property was exploited at  CERN
in a precision measurement of the double ratio
\begin{center}
$\dfrac{
\Gamma\ (\kl \ra 2\pin)/
\Gamma\ (\kl \ra \pip\pim)
}{
\Gamma\ (\ks \ra 2\pin)/
\Gamma\ (\ks \ra \pip\pim)
}
=\modulus{\itaoo/\itapm}^2
\approx\ 1-6\re(\epsp/\epn)\approx\ 1-6\epsp/\epn
$~\ ,
\end{center}
which led to the discovery \cite{bu} of $\epsp\ \ne 0$  and to its precise
measurement \cite{bat}.

In a different approach neutral kaons have been identified and measured 
at their birth through the accompanying particles, in a 
convenient exclusive reaction. Conservation of energy and momentum allows 
neutral particles with the neutral-kaon mass to be selected. Differentiation 
between \kn\ and \knb\ is achieved taking advantage of the conservation of 
strangeness in strong and electromagnetic interactions, through which kaons 
are produced. This dictates that the strangeness of the final state is equal 
to that of the initial state.

Thus, opposite-sign kaon beams have been used to produce \kn\ and \knb\ by 
elastic 
charge-exchange in carbon, in order to compare  \kn\ and \knb\ decay 
rates to \pipi\ \cite{bann73}. \\
Similarly, Ref.~\cite{nieb74} reports 
measurements on \pen\ decays from \kn s, which have been obtained by inelastic 
charge-exchange of positive kaons in hydrogen.\\ 
CPLEAR \cite{pr} produced concurrently \kn\ and \knb\ starting 
from \PPb\ annihilations, by selecting two charge-conjugate annihilation
channels, \kn\km\pip\  and \knb\kp\pim . 

Of fundamental interest are the \kn\knb\ pairs, created in the annihilation
\PPb\ $\rightarrow $ \kn\knb , or in the decay $\phi \rightarrow $ \kn\knb\ .
Their speciality is that they are in an \it entangled state\rm. This property 
has been exploited with different aims. In the CPLEAR experiment (see 
\cite{pr}), the EPR-type long-distance correlation given by the two-particle 
wave function, according to Quantum Mechanics, has been verified \cite{EPR}.  
In experiments at $\phi$ factories like the KLOE experiment at DA$\Phi$NE 
\cite{misc}, the long-distance correlation is used to define a neutral kaon 
as a \ks\ or a \kl\ according to the decay mode, or to the interaction, of 
the other neutral kaon of the pair.
   
In the course of the neutral-kaon time evolution, pionic and semileptonic 
decays may be used to define a fixed time $t$ subsequent to the production 
time ($t=0$). \\
Pionic (\pipa\ and \pipb ) final states (which are \CPz\ 
eigenstates or a known superposition of them) are suitable for \CPz\ 
studies.\\ 
Semileptonic (\pen\ and \pmn ) final states allow \kn\ to be 
differentiated from \knb\ at the decay time, and  are convenient for \Tz\ and 
\CPTz\ studies. 
 
Alternatively, in order to identify the strangeness at a time $t$, 
neutral kaons could be observed to interact in a thin slab of matter (in 
most cases bound nucleons), in a two-body reaction like $\kn \p  \ra \kp\nn $
and  $\knb\nn \ra \km\p $ or $\knb\nn \ra \pin\Lz (\ra \pim\p )$, where the
charged products reveal the strangeness of the neutral kaon.

As a case study we shall focus on the CPLEAR measurements, which yield 
results on \Tz\ violation and on \CPTz\ invariance. Our presentation 
follows closely the description given by the CPLEAR group, summarized in Ref. 
\cite{pr}.

\subsection{CPLEAR experiment}\label{sub:cpl} 
\subsubsection*{\hspace{1cm}Experimental method}\label{sub:meth}

\noindent
Developping the ideas discussed in Ref. \cite{erice}, CPLEAR chose 
to study the neutral-kaon time evolution by labelling (tagging) the 
state with its strangeness, at two subsequent times, see \cite{pr}. 

Initially-pure \kn\ and \knb\ states were produced concurrently by
antiproton annihilation at rest
in a hydrogen target, via the {\em golden} 
channels:
\begin{equation}\label{eq:3_1}
\PPb  \ra  \begin{array}{c}\km \pip \kn \\
\kp \pim \knb \end{array}\,,
\end{equation}
each having a branching ratio of $\approx  2  \times  10^{-3}$. The 
conservation of strangeness in the strong interaction dictates that a 
\kn\ is accompanied by a \km , and a \knb\ by a \kp . Hence, the 
strangeness of the neutral kaon at production was tagged by measuring
the charge sign of the accompanying charged kaon, and was therefore 
known event by event. The momentum of the produced \kn (\knb ) was 
obtained from the measurement of the $\pi^{\pm} {\rm K}^{\mp}$ pair 
kinematics. If the neutral kaon subsequently decayed to \pen , its
strangeness could also be tagged at the decay time by the charge of 
the decay electron: in the limit that only transitions with 
$\DS = \DQ$ take place, neutral kaons decay to \elp\ if the strangeness 
is positive at the decay time and to \elm\ if it is negative. This clearly
was not possible for neutral-kaon decays to two or three pions.

For each initial strangeness, the number of neutral-kaon decays was 
measured as a function of the decay time $t$. These numbers
were combined to form asymmetries -- 
thus dealing mainly with  ratios between measured quantities.
However, the translation of measured numbers of events into decay rates
requires (a) acceptance factors which do not cancel in the asymmetry, (b)
residual background, and (c) regeneration effects to be taken into account.
These experimental complications were handled essentially with the same 
procedure in the different asymmetries. Here we exemplify the procedure
referring to \pen\ decays.

\renewcommand{\theenumi}{(\alph{enumi})}
\renewcommand{\labelenumi}{\theenumi}
\medskip
\begin{enumerate}
\item
Detecting and strangeness-tagging neutral kaons at production and decay
relied on measuring, at the production (primary) vertex, a \kpm\pimp\ 
track-pair and the corresponding  momenta $\pvec_{\kpm}$ and $\pvec_{\pimp}$,
and, at the decay (secondary) vertex, an \elmp\pipm\ track-pair and the 
corresponding momenta $\pvec_{\elmp}$ and $\pvec_{\pipm}$. The detection 
(tagging) efficiencies of the \kpm\pimp\  track-pairs depend on the pair 
charge configuration and momenta, and are denoted by $\eps(\pvec_{\kpm },
\pvec_{\pimp })$. A similar dependence exists
for the detection efficiencies of the \elmp\pipm\ track-pairs, 
$\eps(\pvec_{\elmp },\pvec_{\pipm })$.
Since the detection efficiencies of primary and secondary track-pairs were
mostly uncorrelated, the acceptance of a signal (\pen ) event was factorized 
as $\varrho_S\times \eps(\pvec_{\kpm }, \pvec_{\pimp })\times 
\eps(\pvec_{\elmp },\pvec_{\pipm })$. The factor $\varrho_S$ represents
the portion of the acceptance which does not depend on the charge
configuration of either primary or secondary particles.  
The acceptances of the events corresponding to different
charge configurations were then equalized (or normalized) by introducing 
two functions:
\begin{subequations}\label{eq:norm1}
\begin{eqnarray}
\xi(\pvec_{\kk }, \pvec_{\pi }) & \equiv &
                        \frac{\eps(\pvec_{\kp}, \pvec_{\pim})}
                        {\eps(\pvec_{\km}, \pvec_{\pip})},\label{eq:norm1a}\\
\eta(\pvec_{\e }, \pvec_{\pi}) & \equiv &
                        \frac{\eps(\pvec_{\elm}, \pvec_{\pip})}
                        {\eps(\pvec_{\elp}, \pvec_{\pim})}.\label{eq:norm1b}
\end{eqnarray}
\end{subequations}
These functions, referred to as  {\em primary-vertex normalization factor} and
{\em secondary-vertex normalization factor}, respectively, are weights
applied event by event, $\xi$ to \kn\ events and  $\eta$ to the events
with a neutral kaon  decaying to \elp \pim .\\
\item
The background events mainly consist of  neutral-kaon decays to final states
other than the signal. Their number depends on the decay time $t$. To a
high degree of accuracy the amount of background is the same for initial
\kn\ and \knb\ and hence cancels in the numerator but not in the denominator
of any asymmetry: thus it is a dilution factor of the asymmetry.
To account for these events, the analytic expressions of the asymmetries
were modified by adding to the signal rates \rr\ and \rrb\ the corresponding
background rates \brr\ and \brrb :
\begin{eqnarray}\label{eq:3_2} 
\brr  (t) = \sum_{i} R_{Bi} \times \varrho_{Bi}/\varrho_S \,,&\quad &
\brrb (t) = \sum_{i} \overline{R}_{Bi} \times \varrho_{Bi}/\varrho_S\,,
\end{eqnarray}
where $ R_{Bi}, \overline{R}_{Bi} $ are the rates of the background
source $i$ for initial \kn\ and \knb , respectively, $\varrho_S$ is
defined above and $\varrho_{Bi}$ is the corresponding term for the 
acceptance of events from the  background source $i$. The quantities
$\varrho_{Bi}$
and $\varrho_S $ were obtained by Monte Carlo simulation.
Experimental asymmetries were formed from event rates including signal
and background: $\rr ^* =\rr +\brr $ and $\rrb^* =\rrb +\brrb $.
These asymmetries were then fitted to the asymmetries of the {\em 
measured rates} (see below), which included residual background.\\
\item
The regeneration probabilities of \kn\  and \knb\ propagating through the 
detector material are not the same, thus making  the measured ratio of
initial \knb\ to \kn\  decay events at time $t$  different from that
expected in vacuum \cite{regen1}. A correction was performed by giving 
each \kn\ (\knb ) event a weight \wra\ (\wrb ) equal to the ratio of the 
decay probabilities for an initial \kn\ (\knb ) propagating in vacuum and 
through the detector.\\
\end{enumerate}
Finally, when \pen\ decays were considered, each initial-\kn\ event was 
given a total weight $\wt _+ =\xi\times\eta\times\wra $ or $\wt _-=\xi
\times\wra $ if the final state was $\elp\pim\net$ or $\elm\pip\netb$, 
respectively. The summed weights in a decay-time
bin are \nwp ($t$) and \nwm ($t$). In the same way, each initial-\knb\ event
was given a total weight $\wtb _+=\eta\times\wrb $ or $\wtb _-=\wrb $ if 
the final state was $\elp\pim\net$ or $\elm\pip\netb$. The corresponding
summed weights are \nwpb ($t$) and \nwmb ($t$). In the case 
of decays to two or three pions, each initial-\kn\ event was given a total
weight $\wt = \xi\times\wra $, and each initial-\knb\ event a total weight
$\wtb = \wrb $. The corresponding summed weights are \nw ($t$) and \nwb ($t$).
In the following the summed weights are referred to as the {\em measured
decay rates}. With these quantities are formed the {\em measured asymmetries}.

The measured asymmetries of interest here are
\begin{subequations}\label{eq:asym_1}
\begin{eqnarray}
\ATexp  (t) & = & 
       \frac{\nwpb (t) - \nwm (t)}{\nwpb (t) + \nwm (t)}\,,
       \label{eq:asym_1a} \\
\Adexp  (t)& = &
       \frac{\nwpb (t) - \alpha\nwm (t)}{\nwpb (t) +\alpha\nwm (t)} +  
       \frac{\nwmb (t) - \alpha\nwp (t)}{\nwmb (t) +\alpha\nwp (t)}\,,
       \label{eq:asym_1b} \\
\Apmexp (t)  &=&
       \frac{\nwb (t) - \alpha\nw (t)}{\nwb (t) + \alpha\nw (t)}\,,
       \label{eq:asym_1c} 
\end{eqnarray}
\end{subequations}

The quantity $\alpha = 1 + 4\reel$ is related to the primary vertex 
normalization procedure, see below.
The phenomenological asymmetries to be fitted to each of the above 
expressions include background rates. Explicit expressions of the 
phenomenological asymmetries, in the limit of negligible background, 
can be written using (\ref{eq:2.50}), (\ref{eq:2.52}) and (\ref{eq:2.67}).
For \ATexp , as we shall see, Eq. (\ref{eq:2.56}) is also used.

(Two points are worth mentioning with regard to this method. Effects
related to a possible violation of charge asymmetry in the reactions of 
Eq.~(\ref{eq:3_1}) are taken into account by the weighting procedure at 
the primary vertex. When comparing the measured asymmetries with the 
phenomenological ones we take advantage of the fact that those reactions 
are strangeness conserving. A small strangeness violation (not expected 
at a level to be relevant in the CPLEAR experiment) would result in a 
dilution of the asymmetry and affect only some of the parameters.)

\subsubsection*{\hspace{1cm}The detector}\label{sub:detec1}

\noindent 
The layout of the CPLEAR experiment is shown in Fig. \ref{fig:detec}; a
comprehensive description of the detector is given in Ref.~\cite{det}.

The detector had a typical near-4$\pi$ geometry and was embedded in a (3.6 m
long, 2 m diameter) warm solenoidal magnet with a 0.44 T uniform
field (stable in a few parts in $10^4$). The 200 MeV/$c$ \Pb\ provided 
at CERN by the Low Energy Antiproton Ring (LEAR) \cite{ba} were stopped in a 
pressurized hydrogen gas target, at first a sphere of 7 cm radius at 16 bar  
pressure, later a 1.1 cm radius cylindrical target at 27 bar pressure.

A series of cylindrical tracking detectors provided information about the
trajectories of charged particles. The spatial  resolution $\sigma  \approx
300~\mu \mathrm{m}$ was sufficient to locate the annihilation vertex, as well
as the decay vertex if \kn\  decays  to charged particles, with a precision 
of a few millimetres in the transverse plane. Together with the momentum
resolution $\sigma_p/p \approx 5 ~\mathrm{to}~10$\% this enabled a lifetime
resolution of $\sigma \approx (5-10)\times 10^{-12}$~s. 
 
The tracking detectors were followed  by the particle identification detector
(PID), which comprised
a threshold Cherenkov detector, mainly effective for K/$\pi$
separation above 350 MeV/$c$ momentum ($> 4\sigma$), and scintillators
which measured the energy loss (d$E$/d$x$) and the time of flight of 
charged particles. The PID was also used to separate \e\ from
$\pi$ below 350~MeV/$c$.
 
The outermost detector was a  lead/gas sampling calorimeter
designed to detect the photons of the $\kn  \to 2\pi^0$ or $3\pi^0$ decays.
It also provided e/$\pi$ separation at higher momenta ($p> 300$ MeV/$c$). 
To cope with the branching ratio for reaction (3)
and the high annihilation rate (1 MHz), a set of hardwired
processors (HWP) was specially designed to  provide full event
reconstruction and selection in a few microseconds.

\subsubsection*{\hspace{1cm}Selection of \pen\ events}\label{sub:pen_sel}

\noindent
The \PPb\ annihilations followed by the decay of the neutral
kaon into \pen\  are first selected by topological criteria and by
identifying one of the decay tracks as an electron or a positron, 
from a Neural Network algorithm containing the PID information. The electron 
spectrum and identification efficiency are shown in Fig. \ref{fig:eleff}a. 

The method of kinematic constrained fits was used to further reduce
the background and also determine the neutral-kaon lifetime with an
improved  precision (0.05 \ts\ and 0.2--0.3 \ts\ for short and long
lifetime, respectively). The decay-time resolution was known to
better than  $\pm 10\%$. In total $1.3\times 10^6$ events were selected,
and  one-half of these entered the \ATexp asymmetry.    

The residual background is shown in Fig.~\ref{fig:eleff}b. The 
simulation was controled by relaxing some of the selection cuts to increase 
the background contribution by a large factor. Data and simulation agree 
well and a conservative estimate of 10\% uncertainty was made. The background 
asymmetry arising from different probabilities of misidentifying \pip\ 
and \pim , was determined to be $0.03\pm 0.01$ by using \PPb\  multipion 
annihilations. 

Each event selected, labeled  by the initial kaon strangeness and the 
decay electron charge, was then properly weighted before forming 
the numbers $N$  of events entering the asymmetries \ATexp\ and \Adexp ,
see Eqs. (\ref{eq:asym_1a}) and (\ref{eq:asym_1b}).

\subsubsection*{\hspace{1cm}Weighting \pen\ events and building measured
                 asymmetries}\label{sub:pen_wei}

\noindent
Regeneration was corrected on an event-by-event basis using the amplitudes 
measured by CPLEAR \cite{regen2}, depending on the momentum of the neutral 
kaon and on the positions of its production and decay vertices. Typically, 
this correction amounts to a positive shift of the asymmetry \ATexp  of 
$0.3 \times 10^{-3}$ with an error dominated by the amplitude measurement.
                                                            
The detection efficiencies common to the classes of events being compared 
in the asymmetries cancel; some differences in the geometrical acceptances are 
compensated to first order since data were taken with a frequently reversed  
magnetic field. 

No cancellation takes place for the detection probabilities of the charged 
(K$\pi$) and (\e$\pi$) pairs used for strangeness tagging, thus the two 
normalization factors $\xi$ and $\eta$ of Eqs. (\ref{eq:norm1a}) and 
(\ref{eq:norm1b}) were measured as a function of the kinematic configuration.

The factor $\xi$, which does not depend on the decay mode, was obtained from 
the data set of \pipi\ decays between 1 and 4~\ts\ , where the number of events
is high and where the background is very small, see Ref. \cite{pipm}.
At any time $t$ in this interval, after correcting for regeneration, and 
depending on the phase space configuration, the ratio between the numbers 
of decays of old \kn\ and old \knb , weighted by $\xi$, is compared to the
phenomenological ratio obtained from (\ref{eq:2.67}):  
\begin{eqnarray}\label{eq:xi}
\frac{\xi N(\kn \rightarrow \pi^+\pi^-)}
         {N(\knb\rightarrow \pi^+\pi^-)} = 
(1 -  4\reel)\times \biggr (1 + 4|\eta_{+-}|\cos(\dm\ t -\phi_{+-})
\e^{\frac{1}{2}\gs\ t } \biggl ) \; .
\end{eqnarray}

Thus, the product $\xi\times (1+4\reel)$ can be evaluated. 
The oscillating term on the right-hand side is known with a precision
of $\approx 1 \times 10^{-4}$ (with the parameter values from Ref. 
\cite{pdg}), and remains $< 4\times 10^{-2}$. The statistical error 
resulting from the size of the \pipi\ sample is $\ \pm 4.3\times 10^{-4}$. 
 
The effectiveness of the method is illustrated in Fig. \ref{fig:norm}.
For the order of magnitude of $\xi$, as given by its average $\mean{\xi}$,
CPLEAR quotes $\mean{\xi} = 1.12023 \pm 0.00043$, with $2\reel \approx \dell 
= (3.27 \pm 0.12) \times 10^{-3}$ \cite{pdg}. 

Some of the measured asymmetries formed by CPLEAR, (\ref{eq:asym_1b}) and  
(\ref{eq:asym_1c}), contain just the product $\xi\times [1+4\reel]$, 
which is the quantity measured. However, for \ATexp , (\ref{eq:asym_1b}),
$\xi$ alone was needed.  The analysis was then performed taking \reel\ 
from the measured \kl\ charge asymmetry, $\dell = 2\reel - (\rexm + \rey )$. 
As a counterpart, the possible contribution to \dell\ of direct \CPTz\ 
violating terms had to be taken into account. 

The factor $\eta$ was measured as a function of the pion momentum, using \pip\ 
and \pim\ from  \PPb\ multipion annihilations. The dependence on  the electron 
momentum was determined using $\e^+ \e^-$ pairs from $\gamma$ conversions, 
selected from decays $\kn (\knb )\rightarrow 2\pin$, with a $\pin \rightarrow 
2\gamma$. 

The value of $\eta$, averaged over the particle momenta, is $\mean{\eta} = 
1.014\pm 0.002$, with an error dominated by the number of events in the \elpm\ 
sample.
 
The factors $\xi$ and $\eta$ are the weights applied event by event, which 
together with the regeneration weights, allowed CPLEAR to calculate the 
summed weights, in view of forming the measured asymmetries. The power of 
this procedure when comparing \kn\ and \knb\ time evolution is illustrated 
in Figs. \ref{fig:twopi_a} and \ref{fig:twopi_b} for the \pipi\ decay case. 

The comparison of the measured asymmetries with their phenomenological 
expressions allows the extraction of the physics parameters, as reported 
in Section \ref{meas}.
     
\section{Measurements}\label{meas}
\subsection{Overview}\label{oview}

The results to be presented concern the time evolution and the decays.
\\

Those about the time evolution correspond to the entries of the matrix $L$, which we repeat here for convenience.
\begin{eqnarray}
L = 
\left(
\begin{array}{cccc}
0                &\ -\dg/2                          & 0    & -\Delta\Gzaa /2   \\
\noalign{\medskip}
-\dg/2           &\ 0                               & -\Delta\Mzaa \ \   & 2 \modulus{\epn}\modulus{\dlz} \\
\noalign{\medskip}
0                &\ \Delta\Mzaa                               & 0    & \dm \\
\noalign{\medskip}
-\Delta\Gzaa /2 \ \  &\ -2 \modulus{\epn}\modulus{\dlz} \ \ & -\dm & 0
\end{array}\notag
\right) \ .
\\ \notag
\end{eqnarray}
The elements are real numbers which range in three groups of widely different magnitudes, approximately given by:
\begin{equation}\notag
\begin{array}{rlllll} 
\hline \\
1 & dominant^* &\phantom{All} \dg/2= \dm/0.95 &= 0.56\times 10^{10}\ & \hbar/s\  & \hat{=} \ \mathrm{decay, oscillation}\\\\
2 & 160 \times smaller&\phantom{All}2\modulus{\epn}\modulus{\dlz} &= 0.35\times 10^{8\phantom{0}}\ &\hbar/s  & \hat{=} \ \Tz\ \mathrm{violation}       
\\  &                   & \approx\ \dm \cdot\  6.7\times 10^{-3}\\\\
3 & vanishing  & \phantom{All} \Delta\Mzaa  & < 0.5\times 10^7 & \hbar/s\ & \hat{=} \ \CPTz\ \mathrm{invariance}  \\
&                   & \phantom{All} \Delta\Gzaa  & < \ \ \ 2\times 10^7 \ & \hbar/s\               \\\\
\hline\\
\phantom{}^*   & \mathrm{related \ to}& \phantom{lll}\modulus{\dlz}^2 = (\dm)^2+(\dg/2)^2  \\\\
\end{array}
\end{equation}
Transitions from pure states to mixed states have not been found.
\\

Those about the decay processes confirm \CPTz\ invariance and the $\Delta S = \Delta Q$ rule to a level, which is sufficient, to be harmless to the conclusions regarding the time evolution.

\subsection{\CPTz\ invariance in the time evolution}

What one measures is the parameter \den\ defined in (\ref{eq:2.28a}).

\noindent As for \red ,
exploring the fact that $\re (y+x_{-})$, which expresses \CPTz\ violation in the semileptonic
decay process, cancels out in the sum \AT ($t$)\ + \ACPT ($t$)\ , the CPLEAR group has formed
a data set $A_\delta^{\mathrm{exp}}(t)$ which measures this sum \cite{pen3}. 
Using Eqs. (\ref{eq:2.50}), (\ref{eq:2.52}), and (\ref{eq:2.39}), for $\ACPT (t)$, $\AT (t)$
and \AT , respectively, the measured quantity is shown to become
\begin{eqnarray}
A_\delta^{\mathrm{exp}}(t) &=& \ACPT (t) + \AT (t) - 4\re (\epn -\den )\label{adexp}\\
                           &=& 8\red\ + f(t,\imxp,\rexm,\red,\imd)\ .\label{adexp2}
\end{eqnarray}
The term $4\re (\epn -\den )$ follows from the normalization procedure, and the use of 
the decay rates to two pions (\ref{eq:2.65}). It does not require, however, a measurement 
of \dell\ (\ref{eq:2.56}). \\ 
The function $f$ is given in \cite{pen3}. It is negligible for \ $t\ \widetilde{>}\ 5\tau_S $\ .\\

Fig. \ref{fig:ad_mod} shows the data, together with the fitted curve
$A_\delta^{\mathrm{exp}}(t)$,
calculated from the corresponding parameter values.\\
The main result is \cite{pen3}
\begin{eqnarray*}
\red   = (0.30 \pm 0.33_{\mathrm{stat}}  
                          \pm 0.06_{\mathrm{syst}})\times 10^{-3}~.
\end{eqnarray*}
The global analysis \cite{phen2} gives a slightly smaller error
\begin{equation}\label{Redelta}
\red\ = (0.24 \pm 0.28)\times 10^{-3}\ .
\end{equation}
It confirms \CPTz\ invariance in the kaon's time evolution, free of assumptions on the semileptonic
decay process (such as \CPTz\ invariance, or the \ $\Delta S = \Delta Q$ rule ).\\\\
As for \imd , the most precise value 
\begin{eqnarray*}
\imd &=& (0.000 \pm 0.020)\times 10^{-3}
\end{eqnarray*}
is  obtained by inserting
\ $\mitapp\ =(2.284 \pm 0.014)\times 10^{-3},\  \fsw\ = (43.51 \pm 0.05)^\circ$,
and \ $\phi_\eta\ =(43.5 \pm 0.7)^\circ$, all from \cite{pdg4},
into (\ref{imappr}).\\
A more detailed analysis \cite{NA} yields (within the statistical error) the same result.\\
The formula (\ref{imappr}) also shows that the uncertainty of \imd\ is, at present, just a multiple of the one of $\phi_\eta$.\\\\
Using the Eqs. (\ref{phisw}) and (\ref{dpardsen}) with the values of \imd\ given above, and of
\red\ in (\ref{Redelta}),
we obtain
\begin{eqnarray*}
\dpar = ( 0.17 \pm 0.20)\times 10^{-3} ~,\quad
\dper = (- 0.17 \pm 0.19)\times 10^{-3}~.
\end{eqnarray*}
The mass and decay-width differences then follow from Eqs. (\ref{eq:2.32}) and (\ref{eq:2.33}). \\
With
$\dm = (3.48  \pm 0.01 )\times10^{-15}\ \geve$ and \
$\dg = (7.335 \pm 0.005)\times10^{-15}\ \geve$,
we obtain
\begin{equation}\label{dGdM}
\begin{split}
\Delta\Gzaa\ \equiv\  \Gamma_{\kn\kn}\ -\  \Gamma_{\knb\knb} \ =& \phantom -(\phantom - 3.5 \pm 4.1)
                                      \times 10^{-18}\; \geve ~,\\ 
\Delta\Mzaa\ \equiv\ {\mathrm M}_{\kn\kn}-{\mathrm M}_{\knb\knb} \ =& \phantom -(- 1.7 \pm 2.0)
                                      \times 10^{-18}\; \geve ~.
\end{split}
\end{equation}
\imd\ is constrained to a much smaller value than is \red\ . \imd\ could thus well be neglected.
The results (\ref{dGdM}) \ are then, to a good approximation, just a multiple of \red\ .

\subsection{\CPTz\ invariance in the semileptonic decay process}

We can combine the result on \red\ with the measured values for \dell\ ,
Eq.\ (\ref{eq:2.56}), and \ree , Eq.~\ (\ref{reappr}), and obtain
\begin{equation}\label{ryx}
\re (y+x_{-}) = -\red\ + \ree\ - \dell/2\ .
\end{equation}
For \dell\ the value 
\begin{equation}
\dell = \ (3.27 \pm 0.12)\times 10^{-3}
\end{equation}
has been used \cite{dorf, benn}.\\
For \ree , when entering Eq.\ (\ref{reappr}) with the values of \mitapp\ and with \fsw\  
given above, we have
\begin{equation}
\ree = \ (1.656 \pm 0.010)\times 10^{-3}.
\end{equation}
The value of $\re (y+x_{-})$  thus obtained
is in agreement with the
one reached  
in a more sophisticated procedure by the CPLEAR group
\cite{phen2}:
\begin{equation}\label{ryxval}
\re (y+x_{-})   = (-0.2 \pm 0.3)\times 10^{-3} .
\end{equation}
\\
This result confirms the validity of \CPTz\ invariance in the semileptonic decay
process, as defined in \ref{semilep}. (The new, more precise values of
\dell \cite{ah2,na48ru} do not change this conclusion).\\
We note in passing \cite{phen2}
\begin{eqnarray*}
\rey   = (0.5 \pm 3.0)\times 10^{-3} .
\end{eqnarray*}

\subsection{\Tz\ violation in the kaon's time evolution}

The measured asymmetry between the rates of \ \knb\ \ra\ $\e^+\pi^-\nu$ \ and
of \ \kn\ \ra\ $\e^-\pi^+\nu$
shall now be identified as an asymmetry between the rates of the mutually inverse
processes \ \knb\ \ra\ \kn \ and \ \kn \ra \knb, and thus be a demonstration of time reversal
violation in the kaon's time evolution, revealing a violation of
\ $\Tz ^{-1} \ \Hwk\ \ \Tz\ =\ \Hwk $\ .\\\\
Based on Eq. (\ref{eq:2.50}), whose time independent part (\ref{eq:2.51})\ becomes slightly
modified by the normalization procedure \cite{pen2}, the CPLEAR data is expected to follow
\begin{equation}\label{atexp}
\AT^{exp} (t) = \AT - 4\re (y+x_{-})\ + g(t,\rexm,\imxp)\ .
\end{equation}
The function $g$ is given in \cite{pen2}. It is negligible for \ $t\ \widetilde{>}\ 5\tau_S $\ .\\
Fig. \ref{fig:at_mod} shows \ $\langle A_\mathrm{T}^{\mathrm{exp}}\rangle
=6.6\times10^{-3}$, $-\ 4\re (y+x_{-})$, and \ $g(t,\rexm,\imxp)$, calculated
with the values for \rexm\ and \imxp\ , given below, but increased by one standard deviation.\\\\
The main result is
\begin{equation*}
\langle A_\mathrm{T}^{\mathrm{exp}}\rangle   = (6.6 \pm 1.3_{\mathrm{stat}}  
                          \pm 1.0_{\mathrm{syst}})\times 10^{-3},
\end{equation*}
in agreement with its theoretical prediction (\ref{eq:2.39}). \\\\
This is the only occasion in physics where a different transition rate of a subatomic process,
with respect to its inverse, has been observed.

\subsection{Symmetry in the semileptonic decay process ($\Delta S = \Delta Q$ rule)}

The measurements discussed above allow one also to confirm the $\Delta S = \Delta Q$ rule
for the semileptonic decay processes.
\\
Much of the information is contained in the time
dependent parts $f(t,\imxp,\rexm,\red,\imd)$ and $g(t,\rexm,\imxp)$, and in the decays to \pen\ .
\\
CPLEAR has found \cite{pr},\cite{phen2},\cite{pen1},\cite{pen4}
\begin{align*}
\rexm  &= (          - 0.5 \pm  3.0_{{\rm stat}} \pm 0.3_{{\rm syst}})
\times 10^{-3}~, \notag\\
\imxm  &= (          - 0.8 \pm  3.5 \phantom{AAAAAAA})
\times 10^{-3}~, \notag\\
\imxp  &= (          - 2.0 \pm  2.6_{{\rm stat}} \pm 0.5_{{\rm syst}})
\times 10^{-3}~. \notag 
\end{align*}

\subsection{\CPTz\ invariance in the decay process to two pions}

A contribution to the study of \CPTz\ invariance in the two-pion decay
is the measurement \cite{pdg4}\\
\begin{equation*}
\foo\ - \fpm = (0.2 \pm\ 0.4)^\circ\ ,\\
\end{equation*}
which is
in agreement with the request (\ref{fpmmoo}) that  
\begin{equation*}
\modulus{\fpm\ -\ \foo } \approx\ (\frac{1}{50})^\circ\ .
\end{equation*}
The following \CPTz-violating
quantities have been given values,
using Eqs. (\ref{rbdra}) and (\ref{B2A2}),
\begin{eqnarray}
\frac{\reba }{\reaa } &=& (0.24 \pm 0.28)\times 10^{-3} ~,\quad
\frac{\rebc }{\reac } = (0.32 \pm 0.32)\times 10^{-3} ~.\label{bodao}
\end{eqnarray} 
In addition to the above value for $(\fpm\ -\foo\ )$, we have entered 
\red\ and \imd\ from Section 4.1, $|A_2/A_0| = 0.0448 \pm 0.0002$ from 
\cite{dedi},
\mitaoo\ and \fsw\ from \cite{pdg4}. See also \cite{pr}.

All together \CPTz\ invariance is confirmed. The measurements \cite{pdg4,ah3} 
below fulfil (\ref{fecpt}):
\begin{eqnarray*}
\fpm &=& (43.4 \phantom{1}\pm 0.7\phantom{9})^{\circ},\\
         \phi_{\ita} = 2/3 \ \fpm + 1/3 \ \foo &=& (43.5\phantom{1} \pm
                                            0.7\phantom{9})^{\circ} ,\\
 \fsw  &=& (43.51 \pm 0.09)^{\circ}.\\
& &
\end{eqnarray*}

In terms of the \ \kn \knb\  mass difference we note that all the terms on 
the rhs of (\ref{mmmb}) are negligible with respect to \red\ and we regain, 
in good approximation, Eq. (\ref{dGdM})
\begin{equation*}
\Mzaa\ -\ \Mzbb\ \approx -2\modulus{\dlz}\red\ \sin(\fsw) = (-1.7 \pm 2.0)
                                                        \times 10^{-18}\ \geve.
\end{equation*}
Authors \cite{pdg4wt,NA} have considered the model, which assumes $\dpar =0 $ \
($\hat{=}\  \Gzaa -\Gzbb = 0$), entailing \
$\den = -\den'=\imd\ (\tan{(\fsw) - \i})$, and thus leading to a roughly ten 
times more precise constraint. See also \cite{pr}.\\

\subsection{Further results on \CPTz\ invariance}

Each of the two terms on the rhs of Eq. (\ref{etdl}), $\reba/\reaa$ and 
$\re{(y+x_-)}$,   vanishes under \CPTz\ invariance. This is confirmed by the
experimental results (\ref{ryxval}) and (\ref{bodao}).\\
The experiment \cite{ah2} has allowed one, in addition, to confirm the 
vanishing of their sum by the experimental determination of the lhs of 
Eq. (\ref{etdl})
\begin{equation}\label{etdllhs}
\re{\ (\frac{2}{3}\itapm\ +\frac{1}{3}\itaoo)}\ -\frac{\dell}{2}
= (-0.003 \pm 0.035)\times 10^{-3}=\frac{\reba}{\reaa}+\re{(y+x_-)} \ .
\end{equation}
Although these two terms represent two hypothetical \CPTz\ violations of very 
unlike origins, we can see from Eqs. (\ref{rbdra}) and (\ref{ryx}), that, with 
to-day's uncertainties on the
values of \imd, \ree, and \dell\ , their possible sizes are roughly equal to 
\red:
\begin{equation}\label{bravo}
\frac{\reba}{\reaa}\approx -\re{(y+x_-)}\approx \red \ .
\end{equation}

\subsection{Transitions from pure states of neutral kaons to mixed states ?}

The authors of \cite{ehns} assume, for theoretical reasons, $J^3=0$. Taking complete 
positivity into account, we obtain $J^1=J^2$, and all other elements of $X$ vanish. 
This allows one to use (\ref{WQMVpi}) to (\ref{etaQMV}).
For $\modulus{\eta_{QMV}}$, the measurement  of $R^{\pi \pi}_s (t)$ by the 
CERN-HEIDELBERG Collaboration \cite{CHDG}, 
with the result $\modulus{\eta } = (2.30 \pm 0.035) \times 10^{-3}$, is well suited, 
since it would include effects of QMV. For $\modulus{\eta}$ in (\ref{etaQMV}) we take 
the value of $\modulus{\epn}$ reported in \cite{Ad95} from a first fit to CPLEAR data
(mainly to the asymmetry $\ACPf (t )$ of the decay rates to $f=\pipi$), with three of
the QMV parameters left free. One could then evaluate that this result $\modulus{\epn} 
= (2.34 \pm 0.08) \times 10^{-3}$ corresponds to the quantity $\modulus{\eta }$, 
which is free of QMV effects.
Inserting the values above into (\ref{etaQMV}), we obtain \ $\frac{1}{2}J^1 = (-1.4 
\pm 2.9) \times 10^{-21}\ \geve$. The analysis by \cite{Ad95} (for $\frac{1}{2}J^1 
\equiv \gamma$, $\frac{1}{2}J^2 \equiv \alpha$, $\frac{1}{2}D^3 \equiv \beta$, all 
other QMV parameters $= 0$, and without the constraint of complete positivity)
has yielded an upper limit (with 90 \% CL) of
\begin{equation}
\frac{1}{2}J^1 <  3.7 \times 10^{-21}\ \geve.
\end{equation}
We note that this value is in the range of ${\cal O}(m_\mathrm{K}^2/m_\mathrm{Planck})
= 2\times10^{-20}$ \geve . \\\\ 

\section{Conclusions}\label{conc}

Measurements of interactions and decays of neutral kaons, which have been produced in well
defined initial states, have provided new and detailed information on \Tz\ violation and
on \CPTz\ invariance in the time evolution and in the decay.

\Tz\ violation in the kaon evolution has been demonstrated by measuring that \ \knb\ \ra\ \kn\
is faster than  \ \kn\ \ra\ \knb\ , and by proving that this result is in straight conflict
with the assumption, that \ \Tz\ and \Hwk\ would commute. \\
Complementary measurements have confirmed that hypothetical violations of the
$\Delta$S=\ $\Delta$Q rule or of \CPTz\ invariance in the semileptonic decays, could not have
simulated this result. See Fig. \ref{fig:at_mod}.

\CPTz\ invariance is found intact. The combination of measurements on the decays to \pipa\ and
to \pen\ yields constraints on parameters, which have to vanish under \CPTz\ invariance, as well
concerning the evolution as the decay processes.

The interplay of results from experiments at very high energies (CERN, FNAL) and at medium ones
(CERN) has been displayed. A typical constraint on a hypothetical \kn\knb\ mass or decay width
difference is a
few times $10^{-18}$ \geve, resulting from the uncertainty of (the time evolution parameter)
\red\ .\\

In the future, more experiments with entangled neutral kaon-antikaon
pairs, in an antisymmetric
(\ref{str}, \ref{cp}, \ref{mass}) \ or in a symmetric (\ref{psip}) state,
will be performed.\\
The $\phi$ decay is a source of pairs in an antisymmetric state, which
allows one to select a
set of particles with precisely known properties. We wish to remind that
pairs in the symmetric
state have a complementary variety of phenomena, and also allow for a
particular \ \CPTz\ test.

The experiments have achieved precisions which may open the capability to explore the validity
of some of the often tacitly assumed hypotheses.\\
Some examples are: 
(i) the quantum-mechanical result for the correlation among two distant particles in an 
entangled state, rather than a weaker correlation limited by the assumption of 
Local Realism \cite{be,dido,da}, 
(ii) the perturbation expansion of the Schr\"odinger equation \cite{dagr,urba} leading to 
the two dimensional spinor representation with the exponential decay law, 
(iii) the unitarity relations \cite{tanner}, 
(iv) the complementarity of \CPz\ parity and strangeness \cite{bjo,bramoPR,bramo}, 
(v) that the \kn\ and the \knb\ from $\phi$ decay are exact \CPTz\ conjugates \cite{bebe},
(vi) the conservation of the purity of states of isolated particles, manifested by the long time 
coherence of the kaon matter wave \cite{hawk,ehns}.\\
On the last subject, data from 
the CPLEAR Collaboration in combination with earlier data from the CERN-HEIDELBERG 
Collaboration achieve a sensitivity of \ $\approx 10^{-21}$ \geve.\\ 
The long time coherence has not only been studied for kaons, but also for neutrons 
\cite{ehns}, and lately for neutrinos \cite{bm}.\\

Neutral kaons might well bring even more surprises. Probably the best probes in the 
world.\\\\

\section*{Acknowledgements}\label{memo}

We thank the many colleagues, students and staff we have met at CERN 
and at ETH, who, with their work, their questions and their advice, 
helped us to  gain the insights about which we report in this article.

\section{Appendix \\ The relationship between neutral kaons, spin polarization, and optical interferometry}

\medskip
We often encounter physical phenomena with widely different roots but strikingly similar appearances. For instance, the   neutral kaon and its antiparticle form a two-state system, and so does the spin-1/2 particle, with respect to a direction in space. Besides this obvious similarity, namely to constitute a doublet in some physical variable, there is a much deeper relationship, which we now shall study.

Both systems show the quantum-mechanical property of \it complementarity\rm\ in a similar fashion.\\
Complementarity means that it is not possible to measure the values of certain observables simultaneously with optional precision.
\\
For most neutral kaons this entails (e.g.) our 
unability to measure whether they are kaons or antikaons, i.e. to measure their strangeness. The result comes out ambiguous. Or, in a two-way optical interferometer, the interference pattern is washed out, when we enable the device to reveal the actual path the light has taken.

The description which succeeds to exhibit the doublet property as well as the observed complementarities, is achieved by the use of two-component wave functions in the two-dimensional space, where the observables are represented by the Pauli matrices.
\\

The analogy of the spin precession in a magnetic field with the neutral kaon's strangeness oscillation is outlined in Table \ref{ana}.

\subsection{Physics with wave functions in two dimensions, a universal tool box }
\subsubsection{All the measurements}

The two-component wave functions \ $ \psi =\left(\begin{array}{c} a\\ b\end{array}\right)$ constitute the space where the time development and all the measurements concerning the particle are being formulated. 
\\
The observables are represented by the Pauli matrices $\sigma^i,\ i=1, .., 3$. \ $\sigma^0$ is the $2\times 2$ unit matrix.
\\
The results of the measurements are the expectation values of these matrices
\begin{equation}\label{siexp}
\langle \sigma^i\rangle = \frac{ \langle\psi | \sigma^i | \psi\rangle}{\langle\psi | \psi\rangle}\ .
\end{equation}

The general observable is represented by a linear combination of the matrices \ $\sigma^i$. The corresponding measured value is then a weighted sum of the expectation values $\langle\sigma^i \rangle$.
\\
When we describe \it unstable\rm\ particles, such as kaons, then we also have to keep track of the time-dependent normalization $\langle \psi|\psi \rangle$.
\\

Due to the relations 
\begin{eqnarray}\label{cr}
[\sigma^i,\sigma^j] &=&  2\i\ \eps^{ijk}\sigma^k \\
       (\sigma^i)^2 &=&  \sigma^0\ , \label{s0}
\end{eqnarray}
the expectation values are located on the unit sphere
\begin{equation}\label{one}
\langle \sigma^1\rangle^2+\langle \sigma^2\rangle^2+\langle \sigma^3\rangle^2
=1
\end{equation}
in the three-dimensional space with the unit vectors ($\hat{x}^1,\hat{x}^2,\hat{x}^3$) \ .
\\
(Summation over multiple indices, when missed on the lhs, is understood).
\\

Since all the Pauli matrices $\sigma^i$ have \ $\pm 1$ as their eigenvalues, the expectation values $\langle\sigma^i \rangle$ are readily \it measurable\rm\ as an \it asymmetry\rm , as follows.
\\
Be
\begin{equation}
w^{(i)}_\pm \propto \modulus{ \langle\psi^{(\sigma^i)}_\pm|\psi \rangle}^2
\end{equation}
the probability density that a particle in the state $|\psi\rangle$ will be detected, when it is offered to a detector, which is tuned to the eigenstate of \ $\sigma^i$, \ $|\psi^{(\sigma^i)}_\pm\rangle$ .
\\
Then, we can determine $\langle\sigma^i \rangle$ as
\begin{equation}\label{asy}
\langle\sigma^i \rangle
=
\frac{w^{(i)}_+ - w^{(i)}_-}{w^{(i)}_+ + w^{(i)}_-} \ .
\end{equation}
We note in passing
\begin{equation}
-1 \le \langle\sigma^i \rangle \le +1 \ .
\end{equation}
%


\begin{table}
\begin{center}
\caption{\textbf{Analogy: \ Spin precession - Neutral kaon oscillation. A comparison of languages.
}}
The spin precesses around the $\vec{B}$ field which is directed along $\hat{x}^1$ .
The spin components along  
\\
$\hat{x}^2$ and $\hat{x}^3$ oscillate. \ For neutral kaons, the variables \ $\langle\sigma^2\rangle$ \ and $\langle\sigma^3\rangle$
($\hat{=}$strangeness) \ oscillate.
\\
The kaon eigenstates of \ $\sigma^2$, K$_\mathrm{y}$ and K$_\mathrm{-y}$\ , have not been considered before, and are not in the
\\
literature (yet).
\medskip
{
\small   
\begin{tabular}{p{1.5cm}p{1cm}|p{1.6cm}p{1.2cm}p{1.8cm}|p{2.0cm}p{1.0cm}p{2.4cm}}
\hline\hline 
& & & & & & & \\
Observable \newline $\sigma^m$ & Eigen\newline value    &  Spin \newline polarization & along & Magnetic \newline energy & Neutral kaon \newline eigenvalue &   Name & Masses \\[0.7cm]
\hline\hline 
& & & & & & & \\
$\sigma^1$  &
$+1$ \newline $-1$ & 
Spin \ \it forth\rm \newline \phantom{Spin\ } \it back\rm &
$\hat{x}^1\ \parallel \vec{B}$ &
\it different\rm\newline in \ $\pm\ \hat{x}^1$ &
$CP=+1$ \newline \phantom{CP=+1}$-1$ &
$\kk_1$ \newline $\kk_2$ &
$\kk_1$ and $\kk_2$ have \newline \it different masses\rm
    \\[0.4cm]
$\sigma^2$   &
$+1$ \newline $-1$ & 
Spin \ \it east\rm \newline \phantom{Spin\ } \it west\rm  &
$\hat{x}^2$ &
\it same\rm &
& K$_\mathrm{y}$ \newline K$_\mathrm{-y}$ &
\it same\rm\ as \newline \kn\ and \knb
\\[0.4cm]
$\sigma^3$  &
$+1$ \newline $-1$ &
Spin \ \it north\rm \newline \phantom{Spin\ } \it south\rm &
$\hat{x}^3$ &
\it same\rm &
$S\ =\ +1$ \newline \phantom{$S\ \;=\ $}$-1$  &
\kn  \newline \knb &
\kn\ and \knb\ have \newline \it the same masses\rm\
  \\[0.4cm]
\hline\hline
\end{tabular}
}
\label{ana}
\end{center}
\end{table} 

\subsubsection{The meaningful and the meaningless measurements}

An illustration of the formula (\ref{asy}) is an ensemble of electrons, all having their spins, say, in the $\hat{x}^1$ direction, in which also the detector is tuned.
\\
We then have \ $i=1,\ w^{(1)}_+ = 1, w^{(1)}_- = 0 $, and thus $\langle\sigma^1 \rangle = 1$.
\\
Conversely, if we have prepared a state with \ $\langle\sigma^1 \rangle = 1$, then we conclude from (\ref{asy}), that $w^{(1)}_+ = 1$, i.e. the electron spins look along the $\hat{x}^1$ direction \it with certainty\rm\ . All the measured values of $\sigma^1$ are equal to +1. Their average is +1, and their variance vanishes.
\\\\
On the other hand, if we prepare a state with \ $\langle\sigma^3 \rangle = 0$, then we conclude that \ $w^{(3)}_+ = w^{(3)}_-$, and we have no indication, whether the spin looks \it along\rm\ or \it opposite\rm\ to the $\hat{x}^3$ direction. Both cases are equally probable. We measure the values +1 and -1 equally often. Their average value vanishes, and their variance is equal to \it one\rm\ .
\\

These examples suggest us, to use the \it variance\rm , as a measure of the relevance of a particular measurement, for the determination of the state, in which the particle happens to be.
\\
Let $X$ be an observable. We define the variance, as usual, by
\begin{equation}
(\Delta X)^2 \equiv \langle \ (X-\langle X \rangle)^2 \ \rangle = \langle X^2 \rangle -\langle X \rangle^2.
\end{equation}
Applied to $\sigma^i$, we find, simply by using (\ref{s0}),
\begin{equation}\label{unc1}
(\Delta\sigma^i)^2=1-\langle\sigma^i\rangle^2.
\end{equation}
We note in passing
\begin{equation}
0 \le (\Delta \sigma^i )^2 \le 1 \ .
\end{equation}

Eq. (\ref{unc1}) says, that the variance of $\sigma^i$ vanishes, when $\langle \sigma^i\rangle^2$ takes its maximum value of $\langle\sigma^i\rangle^2=1$. 
\\
This is the case if (and only if) the particle is in one of the eigenstates of \ $\sigma^i$\ : \ \ $\psi=\psi^{(\sigma^i)}_+$ or  \ \ $\psi=\psi^{(\sigma^i)}_-$.
\\

We can derive a statement of \it complementarity\rm\ by evaluating Eq. (\ref{unc1}) with an eigenstate of $\sigma^j$, $j\neq i$\ . Then, $(\Delta\sigma^i)^2=1$, \ i.e. $\sigma^i$ is predicted to have its maximum uncertainty.
\\
We shall obtain this result again, on more general grounds, in Section \ref{uncer}.
\\

From (\ref{unc1}) and (\ref{one}), we see that only one of the $\sigma^i$ can have a vanishing variance at one time. The other two ones then have maximum variance ($=1$).
\\

In order to measure the variance, we combine (\ref{unc1}) with (\ref{asy}) to obtain
\begin{equation}
(\Delta\sigma^i)^2
=
\frac{4\ w^{(i)}_+  w^{(i)}_-}
{(w^{(i)}_+ + w^{(i)}_-)^2} \ \ .
\end{equation}
The maximum uncertainty, \ $(\Delta\sigma^i)^2=1$, is measured, when \ $w^{(i)}_+=  w^{(i)}_-$,
\\ 
i. e. for vanishing asymmetry (\ref{asy}).

\subsubsection{The uncertainty relation}\label{uncer}

The general uncertainty inequality for two observables $X$ and $Y$\ is
\begin{equation}\label{unc}
(\Delta X)^2 (\Delta Y)^2 \ge\ \textstyle\frac{1}{4}\ \modulus{\ \langle [X,Y]\rangle\ }^2 \ .
\end{equation}

For $\sigma^i$ and $\sigma^j$, using Eqs. (\ref{unc})\ and (\ref{cr})\ , we obtain
\begin{equation}\label{u}
(\Delta\sigma^i)^2\ (\Delta\sigma^j)^2
\ge
\textstyle\modulus{\eps^{ijk}}\langle\sigma^k\rangle ^2
\end{equation}
We note a remarkable property of (\ref{u}):
\\\\
Given an arbitrary orthogonal coordinate system in space, ($\hat{x}^1,\hat{x}^2,\hat{x}^3$), to which the Pauli matrices refer.
If the particle is in an eigenstate of $\sigma^k$, which entails $\langle\sigma^k\rangle ^2=1$, then both of the variances, \ $(\Delta\sigma^i)^2 $ and $(\Delta\sigma^j)^2$, \ $i \neq j \neq k$, \ which thus refer to directions perpendicular to $\hat{x}^k$, have to have their maximum value of 1.\ Single measurements of \ $\sigma^i$ and $\sigma^j$ then have maximum uncertainty.
\\
In short: complementary variables belong to orthogonal directions of $\vec{P}=(\langle\sigma^1\rangle,\langle\sigma^2\rangle,\langle\sigma^3\rangle)^\mathrm{T}$.
\\

We shall discuss further below some principles of experimentation on this subject with neutral kaons.

\subsubsection{The polarization and the density matrix four-vector}

The \it polarization four-vector\rm\ $P$ and the \it density matrix four-vector\rm\ $R$ allow us to gain a very useful visualization of the time development of the particle's observables and thus to guide the intuition for the invention of experiments. The formal notions are reminiscent to the classical Stokes parameters which characterize polarized light \cite{jr}.
\\\\
The \it polarization vector\rm\ $\vec{P}$, whose components are the expectation values of the Pauli matrices, is a unit vector along (or opposed to) the $\hat{x}^i$ axis, when it is calculated with the eigenvector $\psi^{(\sigma^i)}_+$ (or $\psi^{(\sigma^i)}_-$) \ of $\sigma^i$.
\\

For practical calculations, the following parametrization of the wave function $\psi$ is often useful.
\begin{equation}\label{psi1}
\psi =
\left(
\begin{array}{c}
a\\\noalign{\medskip}
b
\end{array} \right)
=
\left(
\begin{array}{rl}
\modulus{a}&\e^{-\i \phi/2}
\\\noalign{\medskip}
\modulus{b}&\e^{\i \phi/2}
\end{array} \right)
=
\sqrt{\modulus{a}^2+\modulus{b}^2} \
\left(
\begin{array}{rl}
\cos\frac{\theta}{2} &\e^{-\i \phi/2}
\\\noalign{\medskip}
\sin\frac{\theta}{2} &\e^{\i \phi/2}
\end{array} \right)\ ,
\phantom{AA} \theta, \phi\ \ \mathrm{real}\ .
\end{equation}
%
Inserting (\ref{psi1}) into (\ref{siexp}), we obtain the \it polarization four-vector\rm\
\begin{equation}\label{P}
P =
\left(
\begin{array}{c}  
P^0\\  \\ \vec{P}\\  \\
\end{array}
\right)
=
\left(
\begin{array}{c}  
\langle\sigma^0\rangle\\ \langle\sigma^1\rangle\\ \langle\sigma^2\rangle\\ \langle\sigma^3\rangle 
\end{array}
\right)
=
\left(
\begin{array}{c}  
1\\
2\re(a^*b)/(\modulus{a}^2+\modulus{b}^2)\\
2\im(a^*b)/(\modulus{a}^2+\modulus{b}^2)\\
(\modulus{a}^2-\modulus{b}^2) /(\modulus{a}^2+\modulus{b}^2) \ \ \ \
\end{array}
\right)
=
\left(
\begin{array}{c}  
1\\
\sin\theta \cos\phi\\
\sin\theta \sin\phi\\
\cos\theta 
\end{array}
\right)
\end{equation}
We see here that \ $\theta , \phi\ $ of Eq. (\ref{psi1}) are the polar and the azimuthal angles of $\vec{P}$, and that
\begin{eqnarray}
\cos\theta=\frac{\modulus{a}^2-\modulus{b}^2        }{\modulus{a}^2+\modulus{b}^2} \ ,\ \
\sin\theta=\frac{2\modulus{a} \modulus{b}        }{\modulus{a}^2+\modulus{b}^2} \ ,\ \
\phi = \arg (b/a)\ .
\end{eqnarray}
\medskip
\\
As $\psi$ evolves in time, the end point of $\vec{P}$ (applied at the origin of
the $\hat{x}^i$ reference frame describes a curve on the surface of the unit 
sphere, as shown for neutral kaons in Figs. \ref{fig:kbfaster} and 
\ref{fig:Regs}) .
\\

The \it density matrix four-vector\rm\ $R$ is simply a multiple of the polarization four-vector
\begin{eqnarray}
R &\equiv& \textstyle\frac{1}{2} \langle \psi|\psi \rangle P=\textstyle\frac{1}{2} (\modulus{a}^2+\modulus{b}^2) P\ , \ \ \ \ \ \ \mathrm{or} \label{R}           \\
P &=& R/R^0 \ .         \label{Pp}
\end{eqnarray}
The density matrix \ $\rho=\psi\psi^\dagger$ is then found to be
\begin{equation}\label{roh}
\rho=\psi\psi^\dagger =
\left(
\begin{array}{ll}
aa^* & ab^* \\
ba^* & bb^*
\end{array}
\right)
=
\left(
\begin{array}{ll}
R^0+R^3 & R^1-\i R^2 \\
R^1+\i R^2 & R^0-R^3
\end{array}
\right)
\equiv
R^\mu \sigma^\mu \ .
\end{equation}
We verify the important property that
\begin{equation}\label{detr}
\mathrm{det}(\rho) = 0  = (R^0)^2-{\vec{R}}^2 \equiv R^\mu R_\mu \ , \ \ \ \forall\ \psi\ .
\end{equation}
\\
This indicates that $R$ describes a \ \it pure state\rm , i.e. a state described by a wave function, because, whenever we have a wave function $\psi$, we obtain \ $\det(\rho)=0$. Conversely, if, for given $\rho$ , we have $\mathrm{det}(\rho) =0$, then  $\rho$ can be factorized as \ $\rho=\psi\psi^\dagger$.  The state then is described by $\psi$. It is thus a pure state.
\\
If $\mathrm{det}(\rho) \neq0$, then there is no wave function to describe the state. We then genuinely need $\rho$ to do the calculations. Such a state is called a \it mixed \rm\ state.
\\
Mixed states are treated in Section \ \ref{pure}. In this Appendix, we shall consider pure states only.
\\
Table \ref{coor} \ summarizes the values of \ $a, b, R^\mu , P^\mu\ $ of the most important neutral kaon states.


\begin{table}
\begin{center}
\caption{\bf Components of neutral kaon states. Basis is \ \kn \knb\ . \rm}
\epn\ and \den\ are defined in Section \ref{sub:evol}. \    \epn\ signals \Tz\ violation, \den\ signals \CPTz\ violation.
\\
$\modulus{\epn}^2 \ll 1$ and $\modulus{\den}^2 \ll 1$ are neglected.
\\
\medskip
{
\small   
\begin{tabular}{p{1.2cm}|p{2.5cm}p{5.3cm}p{4.8cm}}
\hline\hline \\[-0.0cm]
%
Neutral \newline kaon  & $\vec{P}$ in Figs. \newline \ref{fig:kbfaster} and \ref{fig:Regs} & Wave function $\psi^\mathrm{T}$\newline (a , b) & Density matrix four-vector \newline  $R^\mathrm{T}=P^\mathrm{T}/2$
\\[0.5cm]
\hline\hline \\
$\kk_1$ \newline $\kk_2$ &
$\phantom{\approx\ }$$+\hat{x}^1$ \newline $\phantom{\approx\ }$$-\hat{x}^1$  &
$\left(1 ,\ \phantom{-}1 \right)/\sqrt{2}$ \newline
$\left(1 ,\ -1 \right)/\sqrt{2}$ &
[1 \ $\phantom{-}$1 \ 0 \ 0]/2  \newline [1 \ $-1$ \ 0 \ 0]/2 
\\[0.4cm]
\ks \newline \kl &
$\approx+\hat{x}^1$ \newline $\approx-\hat{x}^1$  &
$\left(1+\epn +\den \ ,\ \phantom{-}1-\epn -\den \right)/\sqrt{2}$ \newline
$\left(1+\epn -\den \ ,\ -1+\epn -\den \right)/\sqrt{2}$ &
[1 \ $\phantom{-}$1 \ $-2$\im(\epn$+$\den) \ 2\re(\epn$+$\den)]/2  \newline [1 \ $-1$ \ $\phantom{-}2$\im(\epn$-$\den) \ 2\re(\epn$-$\den)]/2
\\[0.8cm]
K$_\mathrm{y}$ \newline K$_\mathrm{-y}$ &
$\phantom{\approx\ }$$+\hat{x}^2$ \newline $\phantom{\approx\ }$$-\hat{x}^2$ &
$(\phantom{-}1-\i, \ \phantom{-}1+\i)/2$ \newline $(-1-\i, \ -1+\i)/2$ &
[1 \ 0 \ $\phantom{-}$1 \ 0]/2  \newline [1 \ 0 \ $-1$ \ 0]/2 
\\[0.8cm]
\kn  \newline \knb &
$\phantom{\approx\ }$$+\hat{x}^3$\newline $\phantom{\approx\ }$$-\hat{x}^3$ &
(1, \ 0) \newline (0, \ 1)  &
[1 \ 0 \ 0 \ $\phantom{-}$1]/2  \newline [1 \ 0 \ 0 \ $-1$]/2 
  \\[0.4cm]
\hline\hline
\end{tabular}
}
\label{coor}
\end{center}
\end{table} 

\subsubsection{The time evolution is (almost) a Lorentz transformation}

If $\psi$ evolves in time, then $R$ develops according to Eq. (\ref{R}) with the constraint of Eq. (\ref{detr}), which we read as \ $R^\mu R_\mu \ = 0$, and which is fulfilled (e.g.), when $R$ undergoes a \it multiple\rm\ of a Lorentz transformation. We shall recognize that this is a considerable restriction of the actions a two-state system can perform.
\\

Let us assume that 
\begin{equation}\label{psit}
\psi(t)=\ U(t)\ \psi(0)
\end{equation}
is the linear transformation which allows one to calculate the time evolution of an initial state $\psi(0)$. The $2\times 2$ matrix $U(t)$ has to be justified by the equation of motion. For the neutral kaons, Eq. (\ref{psit}) is the result of a perturbation calculation.
\\

We formulate Eq. (\ref{psit}) now to make it obey the \it invariance\rm\ with respect to \Tz\ \it translation\rm\ . This is the requirement that the outcome of our experiment does not depend on the moment $t_0$, when we start it, as long as we use the same initial states ($\psi(t_0)$ or $\psi(0)$) and the same procedures.
I.e. we want \ $\psi(t+t_0)=U(t)\psi(t_0)$ to be implied by \ $\psi(t)=U(t)\psi(0)$ . This is accomplished by
the requirement that \ $U(t+t_0)=U(t)U(t_0)$,
or that the transformations $U(t)$ form a group with this composition law, or equivalently by
\begin{equation}
U(t)=\e^{-\i\Lz t}\ ,
\end{equation}
which is nonsingular, and where $\Lz$ is a constant \ $2\times 2$ matrix.
\\
We can thus factorize $U(t)$ as a complex number times a unimodular matrix
\begin{equation}
U(t)=\sqrt{\det(U(t))} \ \breve{U}(t) \ \ \ \ \ \mathrm{with} \ \ \ \det(\breve{U}(t))=1 \ .
\end{equation}
Let us call \lzls\ the eigenvalues of \Lz\ . Then
\begin{equation}
\mathrm{det}(U(t))=\e^{-\i(\lzl+\lzs)t}
\end{equation}
and
\begin{equation}\label{duabs}
\modulus{\mathrm{det}(U(t))}=\e^{-\bar{\Gz}t}
\end{equation}
with \ $\bar{\Gz}=-\im(\lzl+\lzs) \equiv (\gl+\gs)/2$ .
\\
The eigenvalues are complex numbers, \Lz\ is not hermitian, and neither $U(t)$ nor $\breve{U}(t)$ are unitary.
\\
(For simplicity the variables of this section are named as in the specific case of neutral kaons in Section \ref{sub:evol} .)
\\

It is now straightforward to obtain the relations of the measurements to the parameters of the theory. We have to calculate the time evolution of the density matrix $\rho(t)$ in terms of the elements of \Lz\ .
\\\\
First, we note that \ $\rho(t)=\psi(t)\psi^\dagger(t)=U(t)\rho(0)U^\dagger(t)$.
\\
Then, using
\begin{equation}
\mathrm{tr}\{\sigma^\mu\sigma^\nu\}=2\ \delta^{\mu\nu}\ ,
\end{equation}
we invert and apply Eq. (\ref{roh}) in the following steps
\begin{eqnarray}
R^\mu(t)
&=& {\textstyle \frac{1}{2}} \mathrm{tr}\{\sigma^\mu\rho(t)\}
={\textstyle \frac{1}{2}} \mathrm{tr}\{\sigma^\mu\ U(t) \rho(0)U^\dagger(t)\}
={\textstyle \frac{1}{2}} \mathrm{tr}\{\sigma^\mu\ U(t) \sigma^\nu U^\dagger(t)\}\ R^\nu(0)\\
&=& \e^{-\bar{\Gz}t}\cdot
{\textstyle \frac{1}{2}} \mathrm{tr}\{\sigma^\mu\ \breve{U}(t) \sigma^\nu \breve{U}^\dagger(t)\}\ R^\nu(0)\ .
\end{eqnarray}
To summarize

\begin{equation}\label{pt}
\fbox{$ \displaystyle
\ \ R(t)=\e^{-\bar{\Gz}t}\ \e^{Lt}\ R(0) \ \
$}
\end{equation}
where \ $\e^{Lt}$ is the Lorentz transformation announced above, represented by the $4\times4$ matrix with elements
\begin{equation}\label{eLtbr}
(\e^{Lt})^{\mu\nu}
={\textstyle \frac{1}{2}} \mathrm{tr}\{\sigma^\mu\ \breve{U}(t) \sigma^\nu \breve{U}^\dagger(t)\}\ .
\end{equation}
We calculate \ $\breve{U}(t)$ from Eqs. (\ref{eq:2.11}) and (\ref{eq:2.12}) and obtain, with Eq. (\ref{eLtbr}), directly Eq. (\ref{eLcpt}).
\\\\
Eq. (\ref{pt}) allows one to calculate the time evolution (even if it is non unitary) of all observables of all those systems which are described by a $2\times 2$ density matrix, including also mixed states.
\\

For the following considerations, the use of \ $\e^{Lt}$ is unnecesssarily complicated, the knowledge of $L$ is sufficient. The relation between the matrices $L$ and \Lz\ is obtained by recognizing that $L$ is part of the time derivative of \ $\e^{Lt}$.
\\
From \ $L^{\mu\nu} = (\frac{d}{dt}\ \e^{Lt}    )^{\mu\nu}_{t=0}$ \ we find

\begin{equation}\label{lmn}
L^{\mu\nu} =\im(\mathrm{tr}\{\sigma^\mu \Lambda \sigma^\nu\}) - \delta^{\mu\nu}\ \im(\mathrm{tr}\{\Lambda\}),
\end{equation}
and we may verify the property of $L$, that
\begin{equation}\label{genL}
gLg=-L^T ,
\end{equation}
and that thus \ $\e^{Lt}$ is, as expected, a Lorentz transformation . ($g$ is defined in Section \ref{mixd} .)

We shall now exploit the property (\ref{lmn}) to simplify the calculation of \ $\e^{Lt}$.
\\
Be \ $R_\alpha $  , $\alpha\ = 0, ..., 3$ , the eigenvectors, and  be  \ $L_\alpha $ the corresponding eigenvalues of $L$, satisfying 
\begin{equation}\label{ewert}
\ \ L R_\alpha = L_\alpha \cdot R_\alpha \ .
\end{equation}
Then we have
\begin{equation}
\e^{Lt} R_\alpha = \e^{L_\alpha t} \cdot R_\alpha \ .
\end{equation}
Or, if we set
\begin{equation}
R(0)=r_\alpha R_\alpha \ , \ \ 
\end{equation}
then we find for (\ref{pt}) the obvious result, which we could almost have guessed,
\begin{equation}\label{ralf}
R(t) = r_\alpha\ \e^{-\bar{\Gz}t} \ \e^{L_\alpha t} \ R_\alpha \ .
\end{equation}
It is special for \ $L$ of (\ref{lmn}), that the eigenvalues form a real and an imaginary pair, which we call (arbitrarily, for the moment)
$\pm\ \dg/2$ and  $\pm\ \i\dm$ , and that two eigenvectors are real, the other two ones complex conjugate. In general, the eigenvectors $R_\alpha$ are not orthogonal.
\\
Specifically, we assign \ $L_1=-L_0=\dg/2>0$ and \ $L_2=-L_3=\i \dm\ , \ \dm >0$. Then, $R_0, R_1$ real, $R_2=R_3^*$ .
The relation of \dm\ and \dg\ to the eigenvalues of \Lz\ is given in Section \ \ref{sub:evol} .

\subsubsection{Visualization of the time evolution}

This section lets us view a graphical inventory of how neutral kaons are able to develop. The results are summarized in Figs. \ref{fig:kbfaster}--\ref{fig:Regs}, which we shall discuss in detail below.
\\

The expectation values of the Pauli matrices lie on the unit sphere, as seen from Eqs. (\ref{one}) and \ (\ref{P}). It thus becomes natural to use the sphere's surface for the visualization of the time evolution of observables, mainly of $\vec{P}$. Since \ $P^0 \equiv 1$ by definition and since \ $(P^0)^2 = \vec{P}^2, \ \ \forall\ t\ \ge 0$, we are assured that the endpoint of $\vec{P}(t)$ stays on the surface.

We study now the question of the general behavior of the curves $\vec{P}(t)$, their dependence on the starting points, and their destinations. \\\\
\it Complexity of the curves $\vec{P}(t)$.\rm\ Equations (\ref{pt}) and (\ref{Pp}) tell us that the numerical factor $\e^{-\bar{\Gz}t}$, which causes \ $R(t)$ to shrink, does not influence $\vec{P}(t)$.
\\
The complexity of the time evolutions \ $\vec{P}(t)$ is determined by the complexity of the Lorentz transformations \ $\e^{Lt}$.
\\
We note from (\ref{lmn}) that the matrix $L^{\mu\nu}$ is \ \it symmetric\rm\ with respect to the first row and the first column, $L^{m 0}=L^{0 m}$, with $L^{0 0}=0$, but \it antisymmetric\rm\ with respect to the other ones, \ $L^{mn}=-L^{nm}$, $m\neq 0, n\neq 0$. The symmetric elements give rise to a time-dependent boost on $R(t)$ along a fixed direction given by $\vec{b}$, the antisymmetric ones cause $\vec{R}(t)$ (and $\vec{P}(t)$) to rotate around a fixed axis defined by $\vec{a}$.

This may be seen from the following little calculation. Let $R_{Lor}(t)=\e^{Lt} R(0)$ describe the evolution due to the Lorentz transformation alone, i.e. without the shrinking by $\e^{-\bar{\Gz}t}$.
\\
Then, $dR_{Lor}=L R_{Lor}\ dt$, or
\begin{equation}\label{dRlor}
dR_{Lor}=\left(
\begin{array}{r}
dR_{Lor}^0  \\\noalign{\medskip}
d\vec{R}_{Lor}  
\end{array}
\right)
=
\left(
\begin{array}{c}
\vec{b}\cdot\vec{R}_{Lor}  \\\noalign{\medskip}
\vec{b}\ R_{Lor}^0 + \vec{a}\times\vec{R}_{Lor}  
\end{array}
\right) dt
=
\left(
\begin{array}{c}
\vec{b}\cdot\vec{P}_{Lor}  \\\noalign{\medskip}
\vec{b} + \vec{a}\times\vec{P}_{Lor}  
\end{array}
\right) R_{Lor}^0\ dt
\end{equation}
where
\begin{eqnarray}
\vec{b}&=&(b^i)=(L^{i 0}) \notag \\
 \vec{a}&=&(a^i)=(-{\textstyle\frac{1}{2}}\ \epn^{ijk}L^{jk}), \ \ \ \mathrm{or} \notag \\
a^1&=&-L^{23},\ a^2=-L^{31},\ a^3=-L^{12}. \notag \\
& & \notag
\end{eqnarray}
The variety of movements of $R_{Lor}$ , \ and in turn of $\vec{P}$, is herewith obtained by a superposition of a boost along $\vec{b}$ with a rotation around 
$\vec{a}$ \ in Eq. (\ref{dRlor}).
\\\\
\it Frozen points\rm\ . If a four-vector \ $R_\alpha(t)$ changes with time \it in its own direction only\rm\ , then the corresponding point $\vec{P}_\alpha$ stays fixed on the sphere. The condition is thus \ $dR_\alpha \parallel R_\alpha$ , i.e. $LR_\alpha = L_\alpha \cdot R_\alpha$ . This is exactly Eq. (\ref{ewert}). Out of the four eigenvectors $R_\alpha$\ , the two real ones lead to fixed points on the sphere, called \ks\ ($\hat{=}-\dg/2$) and \kl\ ($\hat{=}+\dg/2$)  in Figs. \ref{fig:kbfaster}--\ref{fig:Regs} .
\\\\
\it Endpoint\rm\ . Consider any point $\vec{P}(0)$ on the sphere as an initial state. We know how the two points \ks\ and \kl\ develop with time: they stay where they are. What is the fate of all the other points ?
\\
Consult Eq. (\ref{ralf}) and assume $r_1\neq 0$. Then, for $t\rightarrow\infty$ , it is the term with $L_1=\dg/2$ ,  belonging to \kl\ , which dominates the others ones.
\\
Therefore all states, except \ks\ ,  become finally \kl\ .
\\

Figure \ref{fig:kbfaster} shows the trajectories of a \kn\ and of a \knb , calculated from \ Eq. (\ref{eLcpt}) with the numerical values of Section \ref{oview} (having set \ $\Delta\Mzaa\ =\Delta\Gzaa\ =\ 0$). The curves display a combination of a rotation around an axis, located in the $\hat{x}^1\hat{x}^2$ plane, close to the $-\hat{x}^1$axis, with a boost along $-\hat{x}^1$. They both tend towards the \kl\ situated in the close neighborhood of \kk$_2$ . Both dotted curves span the same total amount of time.
\\
We discover here that the original \knb\ crosses the equator earlier than does the original \kn\ .
\\
The \ \Tz\ violation is making the transition \ $\knb\ra\kn$  run \ \it faster\rm\ than, the reverse one, \ $\kn\ra\knb$.
\\
(For better visibility the effect (of \epn)\ is enhanced by a factor of 20.)
\\

The physical significance of this phenomenon is the main subject of Section \ref{tviol}, where a formula, which combines the elegance of Euler and of Pauli, is able to relate the effect directly to the general \it weak interaction hamiltonian\rm\ \Hwk\ of the neutral kaons.
\\

It might be surprising 
that the combination of a rotation with a boost along two (almost parallel) axes, both located in the horizontal $\hat{x}^1\hat{x}^2$ plane, can move the originally symmetric points \kn\ and \knb\ with different vertical speeds. In other words: how can the initially symmetric geometrical situation in Fig. \ref{fig:kbfaster} develop the \kn\ and the \knb\ into states which are no more symmetric with respect to the equatorial plane ?
\\
The deeper reason lies in a basic property of the Lorentz transformations, namely that its infinitesimal transformations, with respect to non-parallel axes, do not commute. From this follows unavoidably that the combination of a boost along $\hat{x}^1$ with a rotation around the axis \ $a\neq\ \hat{x}^1$ entails a boost along $\hat{x}^3$ !
\\
This is readily demonstrated as follows.
\\
Use Eqs. (\ref{elxt}) and (\ref{D}), or directly the Baker-Campbell-Hausdorff relation, 
\\
$
\e^{At}\ \e^{Bt} = \e^{\left( t(A+B)+ \frac{1}{2} t^2 [A,B] + \cdots \right)}
$
for the two operators A and B,
\\
with the assignments \  $A=-L_0,\ B=L=L_0+X$, to obtain
\begin{equation}\label{eltbch}
\e^{Lt}=\e^{(L_0+X)t}=\e^{L_0 t}\ \e^{\left(Xt\ +\ \frac{1}{2}t^2\ [X,L_0]\ +\ {\cal O}(t^3)\right)}.
\end{equation}
If we now identify $L_0$ with a boost along $\hat{x}^1$, and X with a rotation around \ $a$, then \ $e^{Lt}$ obtains a term 
$[X,L_0]\equiv XL_0-L_0X $. Let us introduce the symmetric $4\times 4$ matrices \ $c_i$ with elements $(c_i)^{\mu\nu}=\delta^{0\mu} \delta^{i\nu}+\delta^{0\nu} \delta^{i\mu}$, which represent the infinitesimal Lorentz boosts along \ $\hat{x}^i$, and see what $[X,L_0]$  is ! Assume \CPTz\ invariance, and write $L$ of Eq. (\ref{L}) with \ $L_0=-\frac{\dg}{2}\ c_1 $ and \ $X=L-L_0$.  Calculate \ $[X,L_0]$, and find
\begin{equation}\label{xlo}
[X,L_0]=[L,L_0]=-\frac{\dg}{2}[L,c_1]=
\dg \modulus{\epn}\modulus{\dlz}c_3=2\ree \modulus{\dlz}^2 c_3 \approx \frac{1}{2}A_T \modulus{\dlz}^2 c_3 \ .
\end{equation}
Thus, Eq. (\ref{eltbch}) contains a boost along \ $\hat{x}^3$. It is this boost, which creates the higher speed of \ \knb\ra\kn\ compared to \ \kn\ra\knb\ .
\\
Eq. (\ref{xlo}) states the basic ingredients for the detection of \ \Tz\ violation. 

\subsubsection{Complementary kaon states}

Let us consider the state, when the \kn\ crosses the equator in Fig. \ref{fig:kbfaster}, and call the particle K$_{eq}$ \ . This state is, according to Section (\ref{uncer}), complementary to the state of the original \kn . We must thus not be able, to find out by a measurement, whether the \ K$_{eq}$ is a \kn\ or a \knb .
\\
This means that a device, capable to measure strangeness, must randomly identify the K$_{eq}$ as often as a \kn\ as as a \knb\ . Repeated measurements of \ $\langle\sigma^3\rangle$ on a K$_{eq}$ must have maximum variance $(\Delta\sigma^3)^2=1$, and, according to Eq. (\ref{unc1}), yield \ $\langle\sigma^3\rangle=0$.
\\
Since the \kn\ decays (semileptonically) into a positron (plus others), and since the \knb\ decays into an electron (plus others), the \ K$_{eq}$ must decay into one or the other with equal frequency. I.e. the positron-electron asymmetry must vanish in the decay of K$_{eq}$. The same is true for the decay of the \ $\bar{\mathrm{K}}_{eq}$ , the original \ \knb\ crossing the equator (see Fig. \ref{fig:kbfaster}).
\\

This effect of complementarity has been observed in the CPLEAR experiment at CERN \cite{pen1}. It is displayed in Fig. \ref{fig:PR35}, where we see the electron-positron asymmetry vanish, at the moment, the kaons cross the equator, at time
\begin{equation}\label{teq}
t_{eq}\ \approx\ (\pi/2)/\dm\ =\ (\pi/2)\ \frac{\gs}{\dm}\ \tau_S\ = 3.32\ \tau_S\ .
\end{equation}
Needless to say that Eq. (\ref{teq}) also provides a way to measure \dm\ .
\\
The same effect of complementarity has been observed by checking whether a newborn \knb\ is 
capable, as time evolves, to produce a \Lz\ particle in a strong interaction, and thus to have 
a strangeness of  $\langle S\rangle\ = \langle \sigma^3\rangle\ =-1$ . The CPLEAR measurements 
\cite{dm_a} summarized in Figs. \ref{fig:PR41}a--b show that a \knb\ looses this 
capability to one half as its age approaches \ $t_{eq}$ , and completely, the closer it comes to \kl\ .
\\

However beautiful these confirmations of complementarity are, we may ask the question whether they constitute \it tests\rm\ , i.e. what would we conclude if they failed ? Certainly, questioning complementarity would be a very last resort. Before, we would reexamine other things, such as the \DS\ = \DQ\ rule, or the general description of neutral kaons with the wave functions in two dimensions, which rests on a perturbation expansion, as explained in ection \ref{sub:evol}.
\\

The uncertainty relation (\ref{u}) allows for a stronger prediction than the ones considered above. If a neutral kaon would cross the equator in the point K$_{\mathrm{y}}$ of Fig. \ref{fig:Regs}, then it would be in a state for which none of the two usual variables, $CP$ parity ($\hat{=}\sigma^1$) and strangeness($\hat{=}\sigma^3$) , would yield a definite, reproducible value.
\\\\
We shall study now, how such points like K$_{\mathrm{y}}$ can be reached by experiment.

\subsubsection{Coherent regeneration caused by a passage through matter\\
Settlement of the front side of the globe}

The curves $\vec{P}(t)$ which describe the evolution of \kn\ and \knb\ traverse regions which are located entirely on the back hemisphere, i.e. between longitudes of \ $\pi/2\le\phi\le3\pi/2$ . With the combination of the two effects of  \it coherent regeneration\rm\ and of \it entanglement\rm\ , we are able to produce a variety of states, including K$_{\mathrm{y}}$ ,  which are represented by points also on the front hemisphere $-\pi/2\le\phi\le+\pi/2$.

Coherent regeneration is the result of scattering processes of the wave function caused by the material when it is traversed by the neutral kaon. Since the strong interactions depend on strangeness, they influence the \kn\ part differently from the \knb\ part. The polarization vector undergoes a Lorentz transformation and an attenuation \cite{geprl}.
\\
Let $P_r$ describe the kaon after a traversal through a thin slab of matter, then
\begin{equation}
P_r=\e^{L_r} P
\end{equation}
with
\begin{equation}\label{r}
\e^{L_r} = \e^{-n(\bar{\sigma}+\sigma)/2}
\left(
\begin{array}{rrrr}
C_r & 0 & 0 & S_r\\\noalign{\medskip}
0 & c_r & s_r & 0\\\noalign{\medskip}
0 & -s_r& c_r & 0\\\noalign{\medskip}
S_r & 0 & 0 & C_r
\end{array}
\right) \ .
\end{equation}
We have \ $C_r=\cosh(n\Delta/2),\ S_r=\sinh(n\Delta/2),\ c_r= \cos(n\omega),\ s_r= \sin(n\omega),\ n=\ $areal density of nuclei,\ $\Delta=\bar{\sigma}-\sigma,\ \omega= \ 2\pi \hbar \ \re(f-\bar{f})/p,\ f =\ $forward scattering amplitude,\ $\sigma =$\ total cross section,\ $p =$ kaon momentum. The bar refers to antiparticles.
\\

The polarization vector $P$ of the kaon before the slab suffers, according to (\ref{r}), a boost along the $\hat{x}^3$ axis, and a rotation around it. 
\\\\
We anticipate that the coherent regeneration by 1 cm of carbon, acting on kaons of 500 MeV, causes a rotation of approximately \ $-0.5$\ degree (see Section \ref{carbon}).
\\\\
The kaons represented by the curves in Fig. \ref{fig:Regs} have been prepared by coherent regeneration starting with a \ks\ . We now explain, how a clean sample of \ks\ particles can be produced.

\subsubsection{The enigma of entanglement}

The \kn\knb\ pair, created as a product in the decay $\phi \rightarrow$\ \kn\knb, \  is expected \cite{gold,lipk,huet,enz} to be in the entangled, antisymmetric state  
\begin{eqnarray}
\ket{\psi_-}
      &=& \frac{1}{\sqrt{2}}(\ket{\kn}\ket{\knb}-\ket{\knb}\ket{\kn}) \label{strm} \\
&\approx& \frac{1}{\sqrt{2}}\ (\ket{\kk_\alpha}\ket{\kk_\beta}-\ket{\kk_\beta}\ket{\kk_\alpha})\ , \label{kalb}
\end{eqnarray}
where $\ket{\kk_\alpha}$ and  $\ket{\kk_\beta},\ \alpha \neq \beta$, are linearly independent combinations of the $\ket{\kn},\ket{\knb}$ .
\\
If the $\phi$ meson decays at rest, then the two neutral kaons fly apart, back to back, with momenta \ $\pm\ \vec{p}$, and Eq. (\ref{strm}) should, for clarity, be written, more inconveniently, as
\\
$\ket{\psi_-} = \frac{1}{\sqrt{2}}(\ket{\kn(\vec{p})}\ket{\knb(-\vec{p})}-\ket{\knb(\vec{p})}\ket{\kn(-\vec{p})})$. We keep this in mind.
\\

The property (\ref{kalb}) has astonishing physical consequences.
Examples of such linear transformations are the time evolution (!), or  \it simultaneous coherent regenerations\rm\ on both particles by material slabs placed in both particle paths at \it symmetric positions\rm\ with respect to the decay point of the $\phi$ meson. These two processes thus do not alter any properties of the state $\ket{\psi_-}$, except the intensity.
\\
Figure \ref{fig:regpx} illustrates this "regeneration paradox". While one slab of material does lead to regenerated \ks , indicated in the figure by \ \ks\ \ra\ $2\pi$, two symmetrically placed slabs do not, because the state $\ket{\ks}\ket{\ks}$ is not contained in any $\ket{\psi_-}$ of Eq. (\ref{kalb}). Where does the kaon inside the right hand side slab know from, whether there exists a corresponding slab on the left hand side ? 
\\\\
The Einstein Podolsky Rosen paradox has to do with the fact that a measurement on one of the kaons implies instant knowledge about the other one.
\\
Suppose the state \ $\psi_-$ has been prepared, and a subsequent measurement has determined that the kaon with momentum \ $\vec{p}$\ is in the state $\ket\kn$, say. What can be said about the kaon on the other side with momentum $-\vec{p}$ ? 
\\
Consider the complete set of two-particle states ($\ket\kn\ket\kn,\ \ket\kn\ket\knb,\ \ket\knb\ket\kn,\ \ket\knb\ket\knb$). Then, the state \ $\ket\kn\ket\knb$ is the only one which has both, a \kn\ with momentum $\vec{p}$, and also a nonvanishing component in $\ket{\psi_-}$.
\\
The state on the opposite side of $\ket\kn$ is thus, with certainty, the state $\ket\knb$.
\\
At $\phi$ factories, \ this effect is exploited to produce an ensemble of pure \ $\ket\ks$ states by monitoring the occurence of \ $\ket\kl$ states on the opposite side.
\\
More in Section \ref{enta}\ .

\subsubsection{The production of the state $\ket{\mathrm{K}_y}$ and other ones on the front hemisphere}

We now make use of the possibility to prepare a clean sample of \ks\ mesons. Left alone, a \ks\ would, before its decay, just stay in the \ks\ state. However, if it encounters some material, it undergoes coherent regeneration and, in turn, finally develops into a \kl\ . Figure \ref{fig:Regs} demonstrates that small variations of the regenerator thickness lead to considerably different trajectories of $\vec{P}(t)$. 

\subsubsection{Neutral kaons and optical interference}

Neutral kaons and optical interference have two fundamental aspects in common.
\\
First, they have two basic states (up, down, e.g.) whose linear combinations are sufficient to describe the observed phenomena.
\\
Second, some of these phenomena can not be observed simultaneously.
\\
The formal link among these widely different groups of phenomena stems from the successful description of the measurable quantities by the expectation values $P^\mu$ of the Pauli matrices, with respect to an underlying two-dimensional (Hilbert) space.
\\

Consider the interference pattern of two monochromatic light beams, which had been generated by a beam splitter. The intensity distribution is then of the form $I \propto 1+\nu_0\cos\ (ks)$, where $s$ is an appropriate length coordinate on the screen, and $k$ is a constant. For the following analogy, it will not be necessary to identify $ks$ with any specific property of the kaons. The value of $\modulus{\nu_0}$ is called \it fringe visibility\rm\ . Classically, the fringes disappear ($\nu_0=0$) when the light takes only one path; quantum mechanically, the fringes disappear, when the spectrometer is built to be capable to reveal the path, each photon has taken.
\\
The complementarity between fringe visibility and the \it which path knowledge\rm\ $\Pi\ $ is most easily implemented, if we associate the two paths with the two eigenstates of one of the Pauli matrices $\sigma^k$; i.e. $\Pi=\langle\sigma^k\rangle$. For the fringe visibility then remain the expectation values of $\sigma^l$ and of $\sigma^m$, which are, for $l\neq k$, $m\neq k$, complementary to $\sigma^k$, or some combination of the two. With the choice \ $\nu_0^2=\langle\sigma^l\rangle^2+\langle\sigma^m\rangle^2$ and Eq. (\ref{one}), we unavoidably obtain the celebrated relation
\begin{equation}\label{junk}
\nu_0^2+\Pi^2=1, \ \ \ \ \ \ \ \forall\ \ \psi (t)\ .
\end{equation}
Eq. (\ref{junk}) expresses a mathematical identity among the definitions of the variables. This implies the danger that an experimental falsification of it, is not so much a test of quantum mechanical principles, but possibly just a depreciation of the attributions of the observables of the Pauli matrix toolbox with the variables of the measuring apparatus in use.
\\\\
The analogy of optical interferences with neutral kaons can be seen from Fig. \ref{fig:kbfaster}. We attribute the which path knowledge with the strangeness \ $\langle\sigma^3\rangle$, and the fringe visibility with \ $\sqrt{\langle\sigma^1\rangle^2+\langle \sigma^2\rangle^2}$. This expression is just the length of the projection of \ $\vec{P}(t)$ onto the equatorial plane. Should a measurement of it yield the value one, then the state is represented by a point on the equator, and Eq. (\ref{junk}) tells us that we have zero knowledge of whether the particle came originally from the noth pole, or from the south pole.
\\
For more detailed considerations, see for instance
\cite{bjo,bramoPR,bramo}. 

\subsection{Numerical values for the regeneration in carbon}\label{carbon} 

\begin{tabular}{llll}
Kinetic energy of kaon         & $E$                             & \phantom{AAA} 500                    &MeV\\
Momentum                       & $p=\hbar k$                     & \phantom{AAA} 860                    &Mev/c\\
Wave length                    & $\lambda=2\pi/k$              & $\phantom{-}1.44\times 10^{-15}$   & m\\
Forward scattering amplitudes  & $\re(f-\bar{f})$              & $-4.82\times 10^{-15}$  & m\\
                               & $\im(f-\bar{f})$              & $-3.97\times 10^{-15}$  & m\\
                               & $\arg(f-\bar{f})$             & $-140.5$                & degree \\
Phase angle per atom density   & $\omega=\lambda\ \re(f-\bar{f})$& $-6.94\times 10^{-30}$  & m$^2$\\
Cross section difference       & $\Delta=-2\lambda\ \im(f-\bar{f})$&$\phantom{l}11.43\times 10^{-30}$     & m$^2$\\
Areal density, 1 cm carbon     & $n$                     & $\phantom{-}1.2\phantom{4} \times 10^{27}$   &atoms/m$^2$\\
Regeneration angle             & $ n \omega$                   & $-8.33\times 10^{-3} $  & rad/(1 cm carbon)\\
Regeneration boost             & $n\Delta/2$                   & $\phantom{-}6.86\times 10^{-3}$& for\ 1 cm carbon\\\\
1 cm carbon causes      & $s_r=\sin(n\omega)$           & $\hat{=} -0.5$\ degree & around $\hat{x}^3$.
\end{tabular}
%

\section*{Figure captions}\label{fcap}
\subsection*{}
\begin{figure}
\vspace{5cm}
\begin{center}

\includegraphics[angle=-90,bb=90 -175 -275 469,width=10cm,clip]
                                          {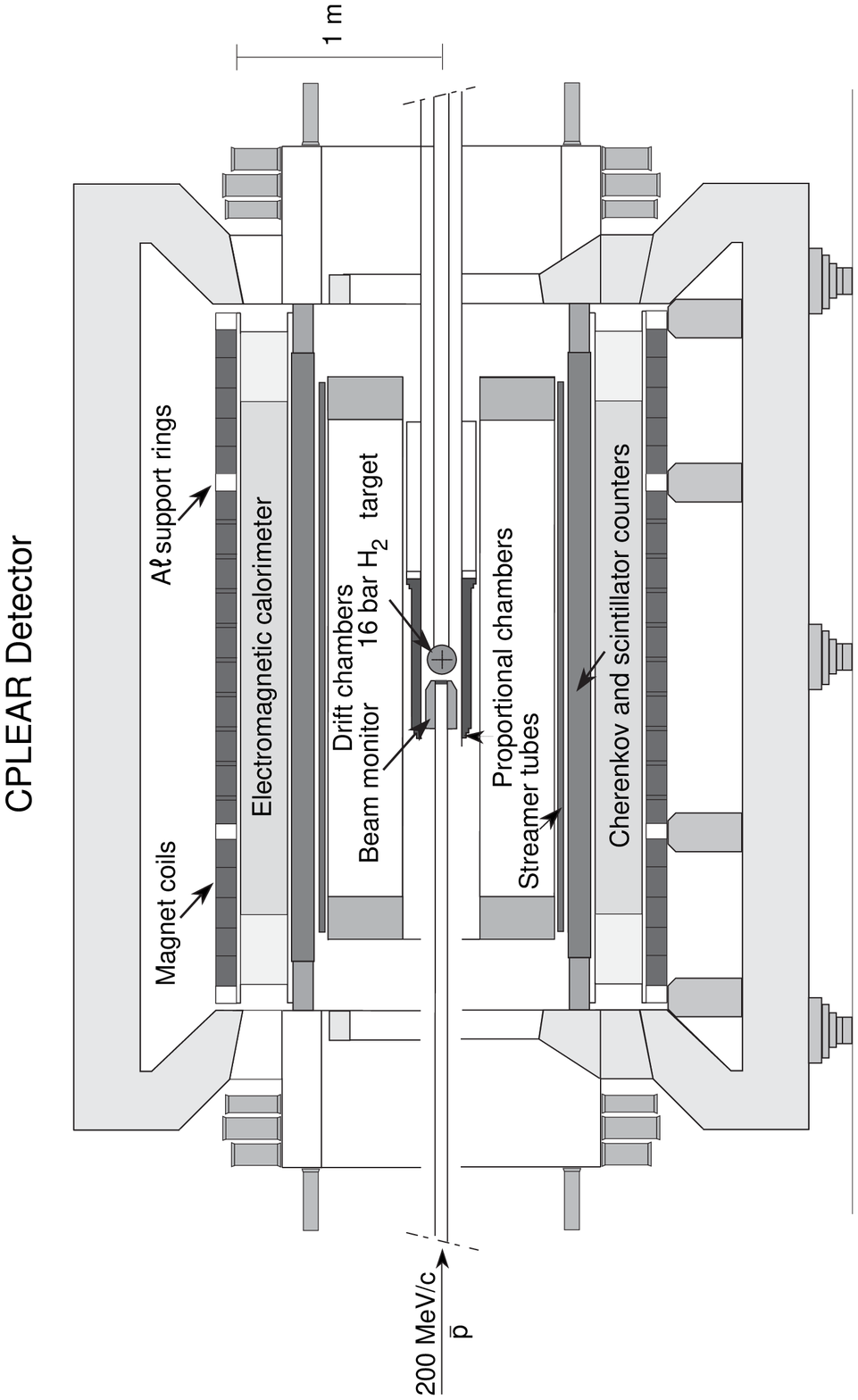}\\[5mm]
  (a) \\[15mm]
\includegraphics[bb=119 232 506 550,width=7.14cm,clip]{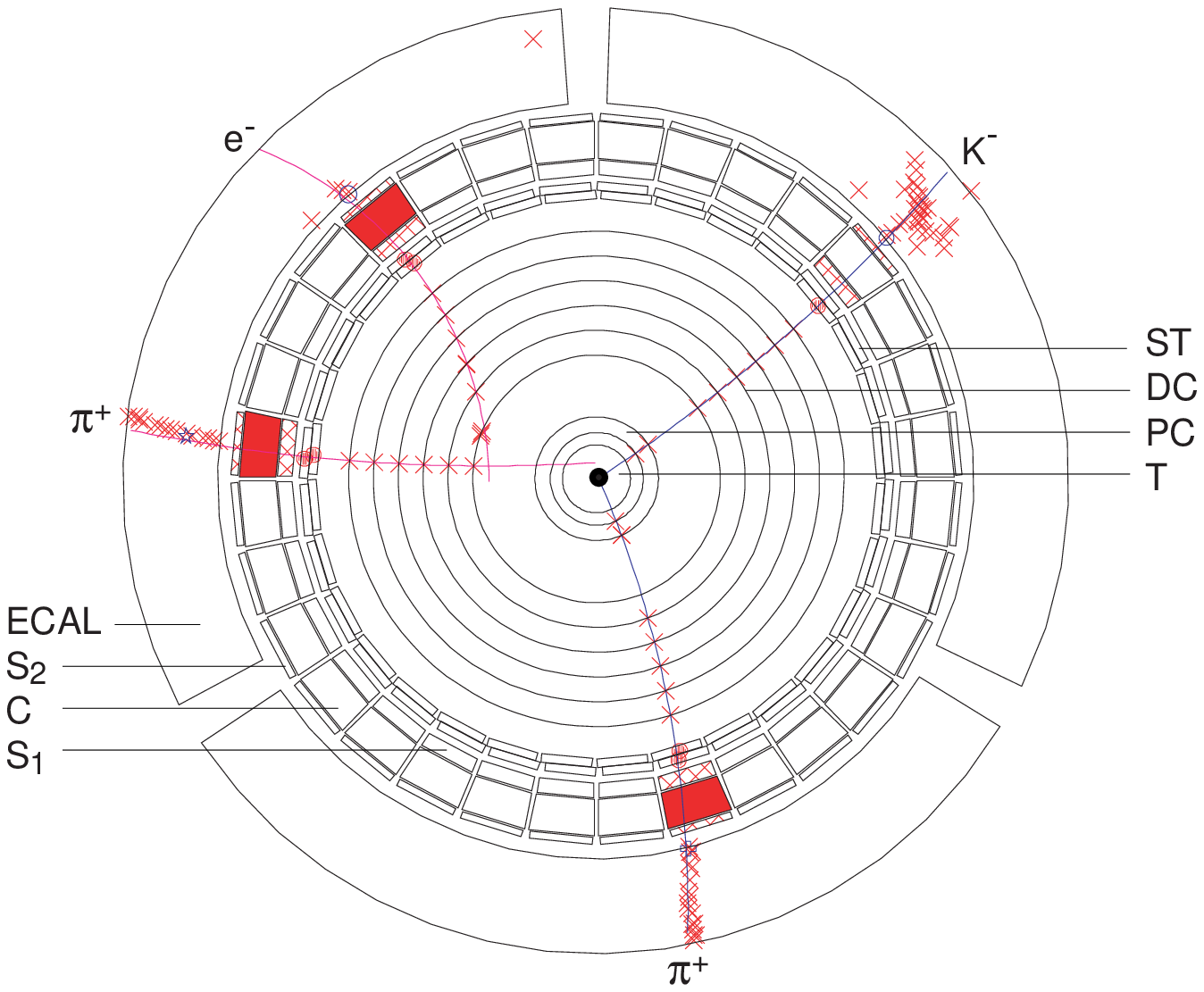}\\[5mm]
  (b) \\[15mm]
\caption{}
\label{fig:detec}
\end{center}
\end{figure}

\begin{figure}
  \begin{center}                                                      
    \begin{tabular}{cc}
  (a) &  (b) \\                
  \includegraphics[bb=188 291 396 494,width=0.45\linewidth,clip]   
                                                 {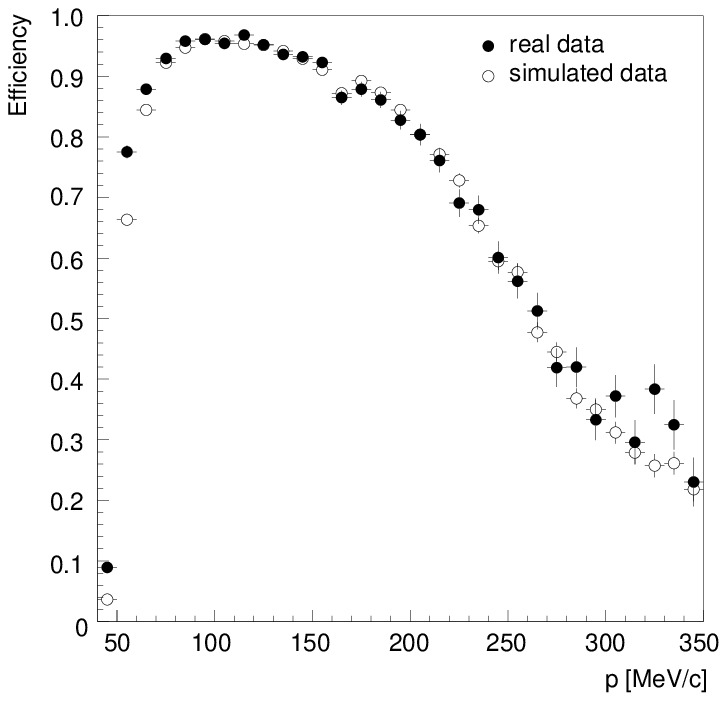} &    
  \includegraphics[bb=14 124 570 651,width=0.5\linewidth,clip]     
                                                  {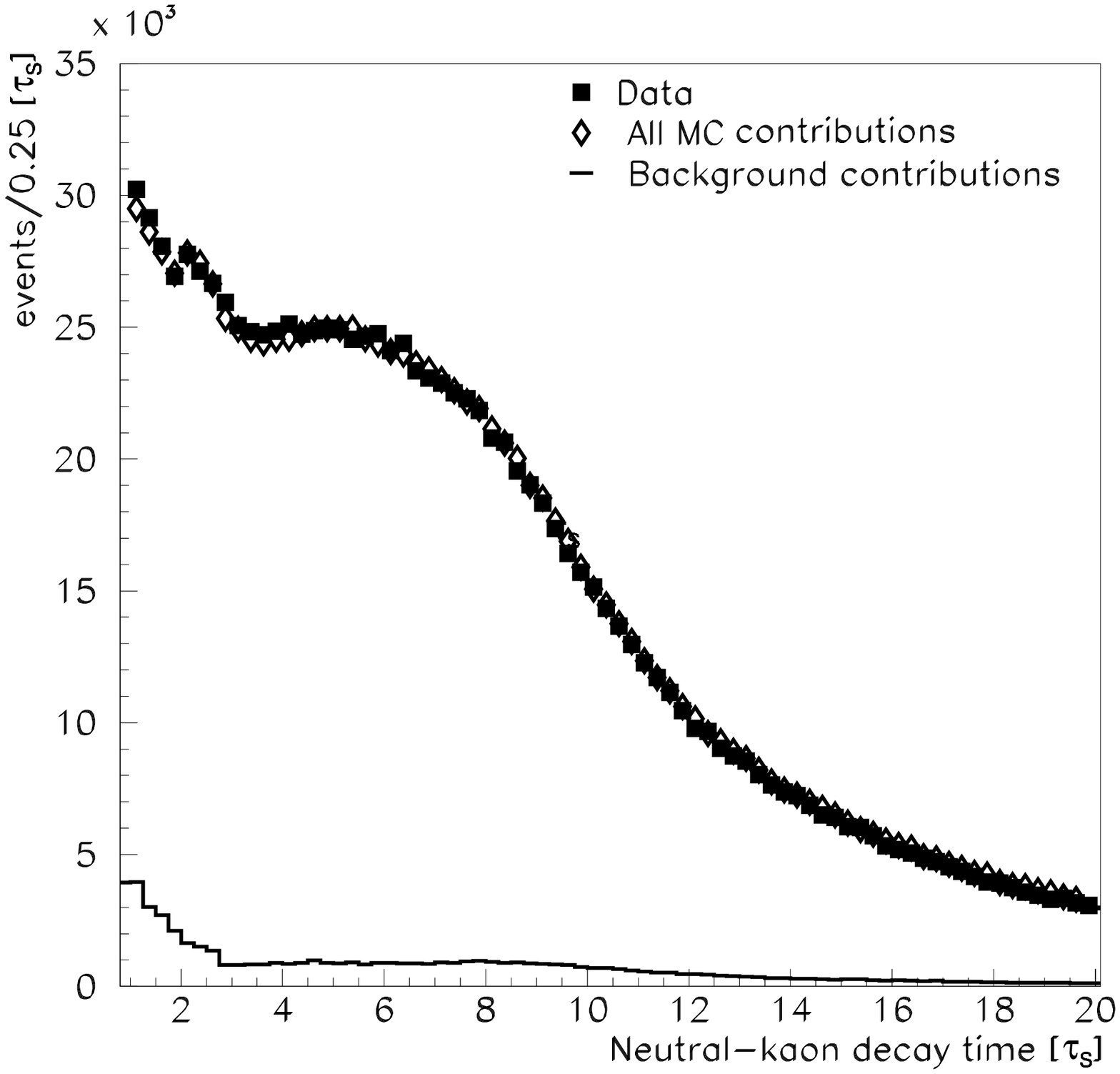}   
    \end{tabular}                                                      
  \end{center}                                                         
\caption{}
\label{fig:eleff}                                               
\end{figure}

   Fig.~ \ref{fig:detec} --  CPLEAR detector.\\ 
   (a) longitudinal view.\\ 
   From the left, the $200$ \mev\ \Pb\ beam delivered by LEAR  enters the 
   magnet along its axis, and through a thin scintillator (Beam monitor) 
   reaches a pressurized hydrogen gas target (T) where antiprotons stop 
   and annihilate. Cylindrical tracking detectors provide information about 
   the trajectories of charged particles in order to determine their charge 
   signs, momenta and positions: two proportional chambers (PC), six drift 
   chambers (DC) and two layers of streamer tubes (ST). The particle 
   identification detector (PID) comprising a threshold Cherenkov 
   detector (C) sandwiched between two layers of scintillators (S1 and S2)
   allows the charged-kaon identification, and also the separation of 
   electrons from pions. The outermost detector is a  lead/gas sampling 
   calorimeter (ECAL) to detect the photons from $\pi^0$ decays.\\
   (b) transverse view and display of an event.\\ 
   $\PPb$ (not shown) $\rightarrow$ \km\pip\kn\ with the neutral kaon 
   decaying to \elm\pip\netb . The view (b) is  magnified twice with respect 
   to (a) and does not show the magnet coils and outer detector components.\\ 
   From Ref. \cite{pr}.

\subsection*{}   
   Fig.~ \ref{fig:eleff} -- Electron detection.\\
   (a) Electron identification efficiency as a function of 
                   momentum when $<2\%$ of pions fake electrons for real
                   $(\bullet )$ and simulated $(\circ )$ calibration    
                   data; \\                                                
   (b) Decay time distribution for real $({\tiny \blacksquare })$ 
      and simulated  
      $(\diamond )$ data. The expected background contribution is shown 
      by the solid line. \\ 
    From Ref. \cite{pr}.

\subsection*{}   
\begin{figure}
\begin{center}
\includegraphics[bb=172 330 484 597,width=0.8\linewidth,clip]
                                                   {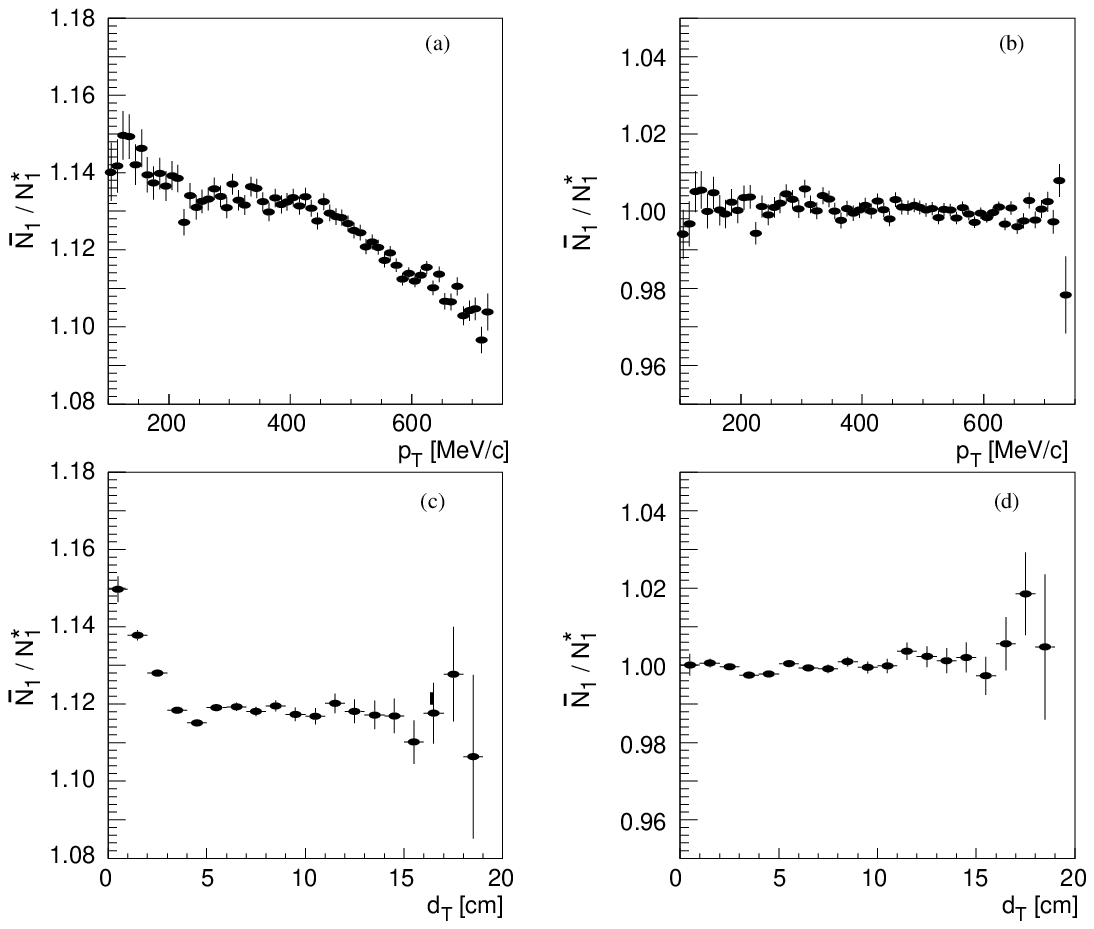}
\end{center}
\caption{}
\label{fig:norm}
\end{figure}

   Fig.~\ref{fig:norm} --  \knb\ to \kn\ normalization.\\
   The ratio between the numbers of \knb\ and \kn\ decays 
   to \pipi\ in the $(1 - 4)$ \ts\ interval, $\Nb_1 /N^*_1 $ 
   (corrected for regeneration and \CPz\  violating effects) as a 
   function of the neutral-kaon transverse momentum \pt , \\
   (a) before and\\ 
   (b) after giving each \kn\ event its normalization weight.\\
   The same ratio as a function of the decay vertex transverse separation 
   from the production vertex \dt ,\\
   (c) before and \\
   (d) after applying the normalization weights. \\
   From Ref. \cite{pipm}.

\subsection*{}
\begin{figure}
\begin{center}                          
\includegraphics[bb=404 -245 660 9,width=0.8\linewidth,clip]
                                                {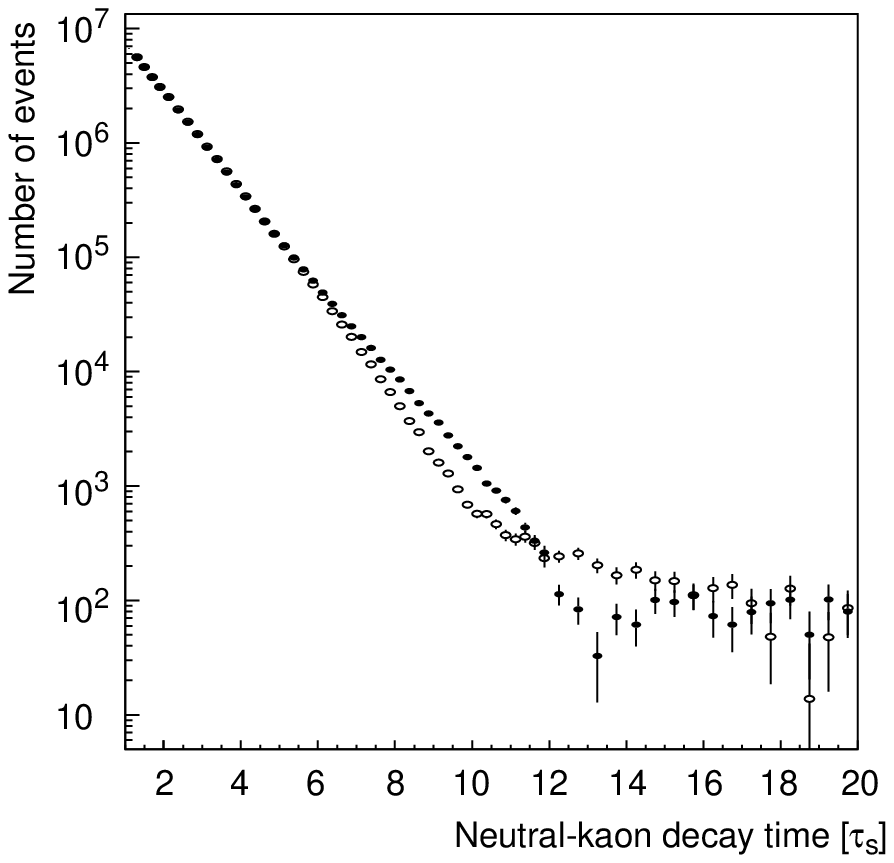}
\end{center}
\caption{}
\label{fig:twopi_a}
\end{figure}

   Fig.~\ref{fig:twopi_a} -- Neutral-kaon decay to \pipi\ : the different
                               decay rates indicate \CPz\ violation.\\ 
   The measured decay rates 
   for \kn\ ($\circ$) and \knb\ ($\bullet$) after acceptance correction
   and background subtraction.\\
   The spatial extension of this pattern is about 30 cm.
   \\
   From Ref. \cite{pipm}.

\subsection*{}
\begin{figure}
\begin{center}                          
\includegraphics[bb=0 19 232 235,width=0.8\linewidth,clip]
                                               {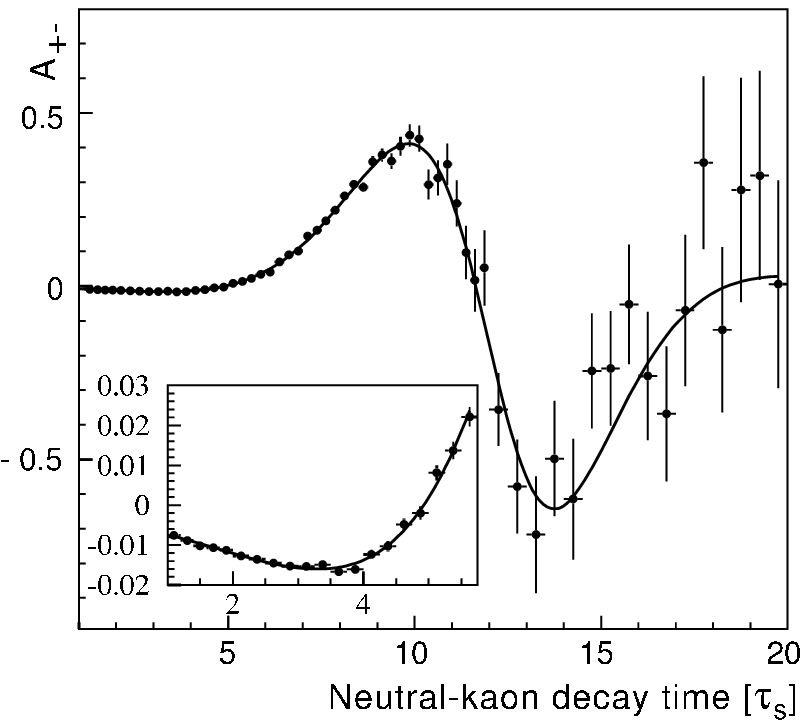}
\end{center}
\caption{}
\label{fig:twopi_b}
\end{figure}

   Fig.~\ref{fig:twopi_b} -- Neutral-kaon decay to \pipi\ : the rate
                         asymmetry \Apm ($t$) demonstrates \CPz\ violation. \\
   The measured asymmetry \Apmexp ($t$), Eq. (\ref{eq:asym_1c}), reduces to 
   \Apm ($t$) when the background is subtracted from the measured rates. The 
   dots are the data points. The curve is the result of the fit making use of 
   the rates (\ref{eq:2.65}). \\
   From Ref. \cite{pr}.

\subsection*{}   
\begin{figure}
  \begin{center}
   \includegraphics[bb=0 0 368 231,width=0.8\linewidth,clip]
                                                 {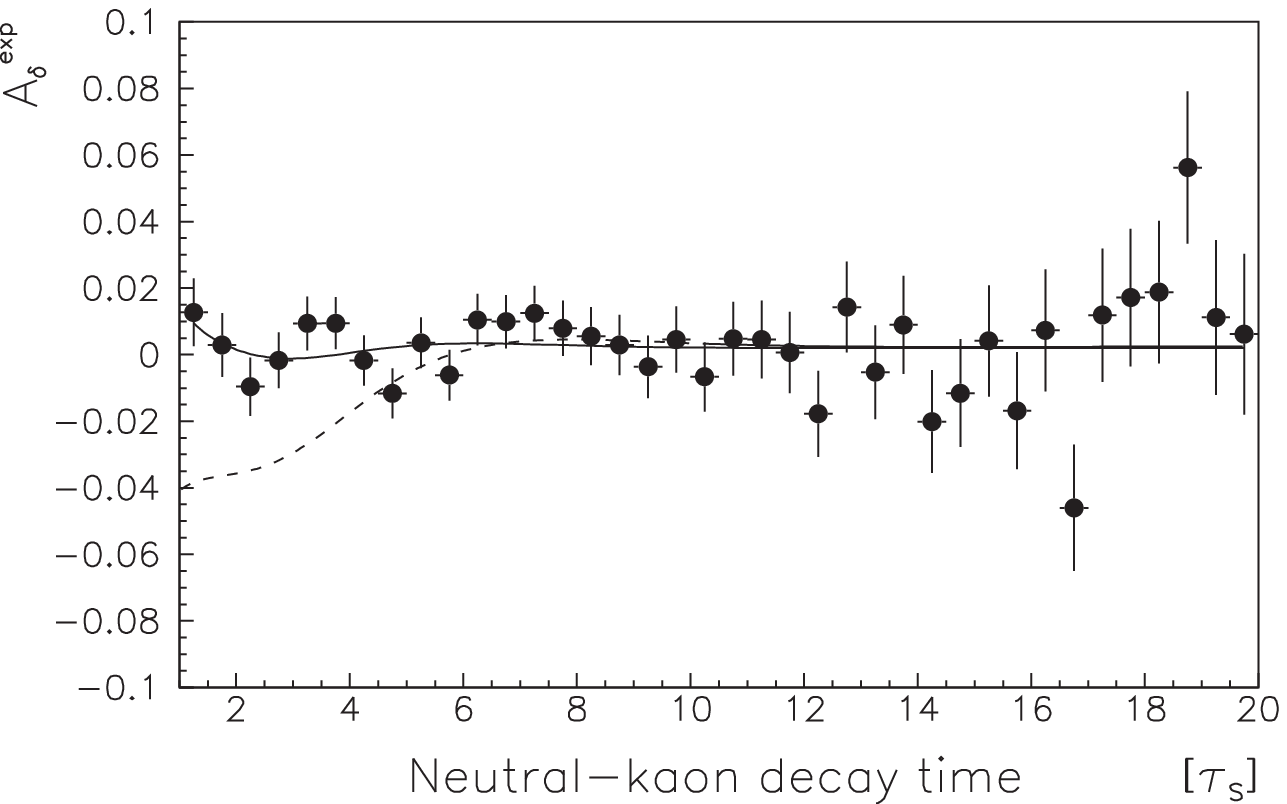}
  \end{center} 
\caption{}
\label{fig:ad_mod}
\end{figure}

   Fig.~\ref{fig:ad_mod} -- Experimental confirmation of \CPTz\ invariance 
   in the $time \ evolution$ of neutral kaons. \\ 
   The present result is the determinant input to~a measurement of the
   decay-width and mass differences between the neutral kaon
   and its antiparticle, the value of the latter being 
   $(- 1.7 \pm 2.0)\times 10^{-18}\; \geve ~.$ See Eqs. (\ref{dGdM}).\\ 
   Values for times $t\ \widetilde{>}\ 5\tau_S\ $ depend entirely on 
   a hypothetical \CPTz\ violation in the time evolution, independently 
   of further hypothetical violations of \CPTz\ invariance in the $decay$, 
   or of violations of the $\DS = \DQ$ rule.\\
   The points are the measured values of \Adexp .
   The solid line represents the fitted curve (\ref{adexp2}). The dashes 
   indicate a hypothetical violation of the $\DS = \DQ$ rule with the final 
   CPLEAR values \cite{phen2}, exagerated by one standard deviation for 
   $\imxp\ (\ra \ -4.6 \times 10^{-3})$, and for $\rexm\ 
   (\ra \ 2.5 \times 10^{-3})$.\\
   The validity of the $\DS = \DQ$ rule is confirmed.\\
   Data from Ref. \cite{pen3}.

\subsection*{}   
\begin{figure}                                              
  \begin{center}          
  \includegraphics[bb=108 325 490 545,width=0.8\linewidth,clip]
                                                 {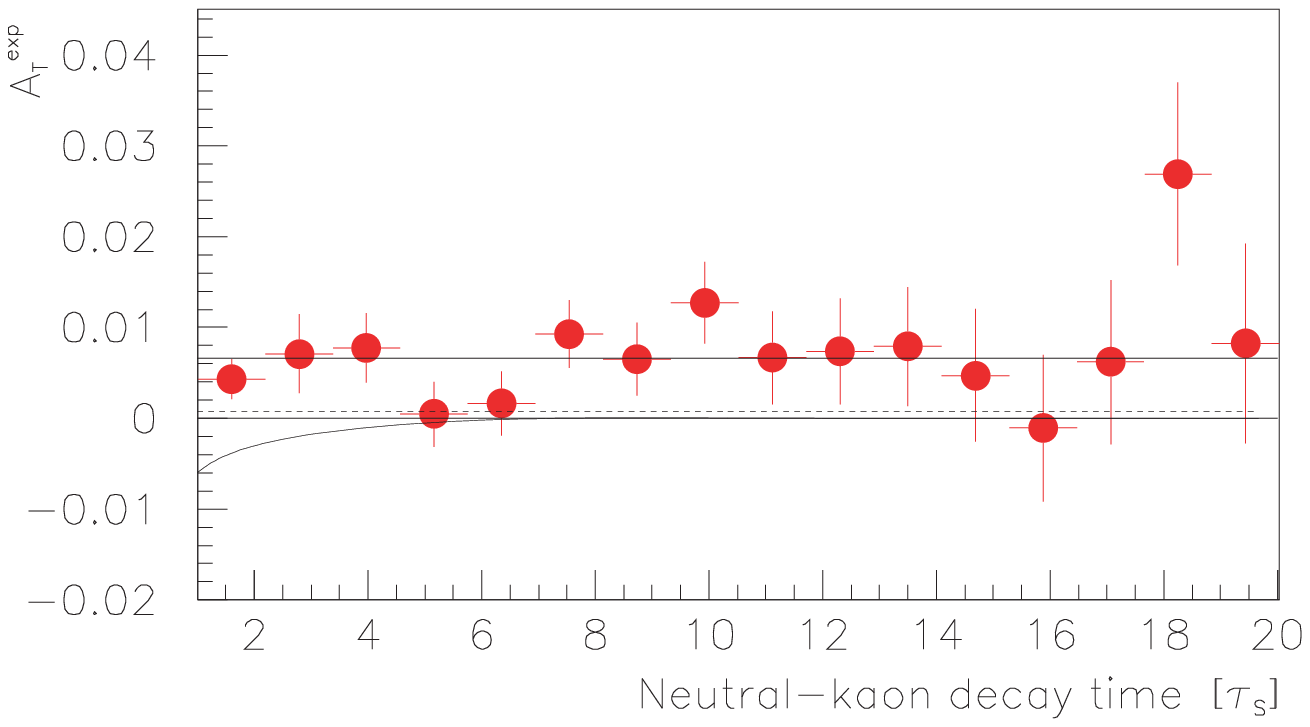} 
  \end{center}                                                    
\caption{}
\label{fig:at_mod}                                               
\end{figure}

\begin{figure}                                              
  \begin{center}          
   \includegraphics[angle=0,bb=1 0 498 347,width=0.8\linewidth,clip]
                                                    {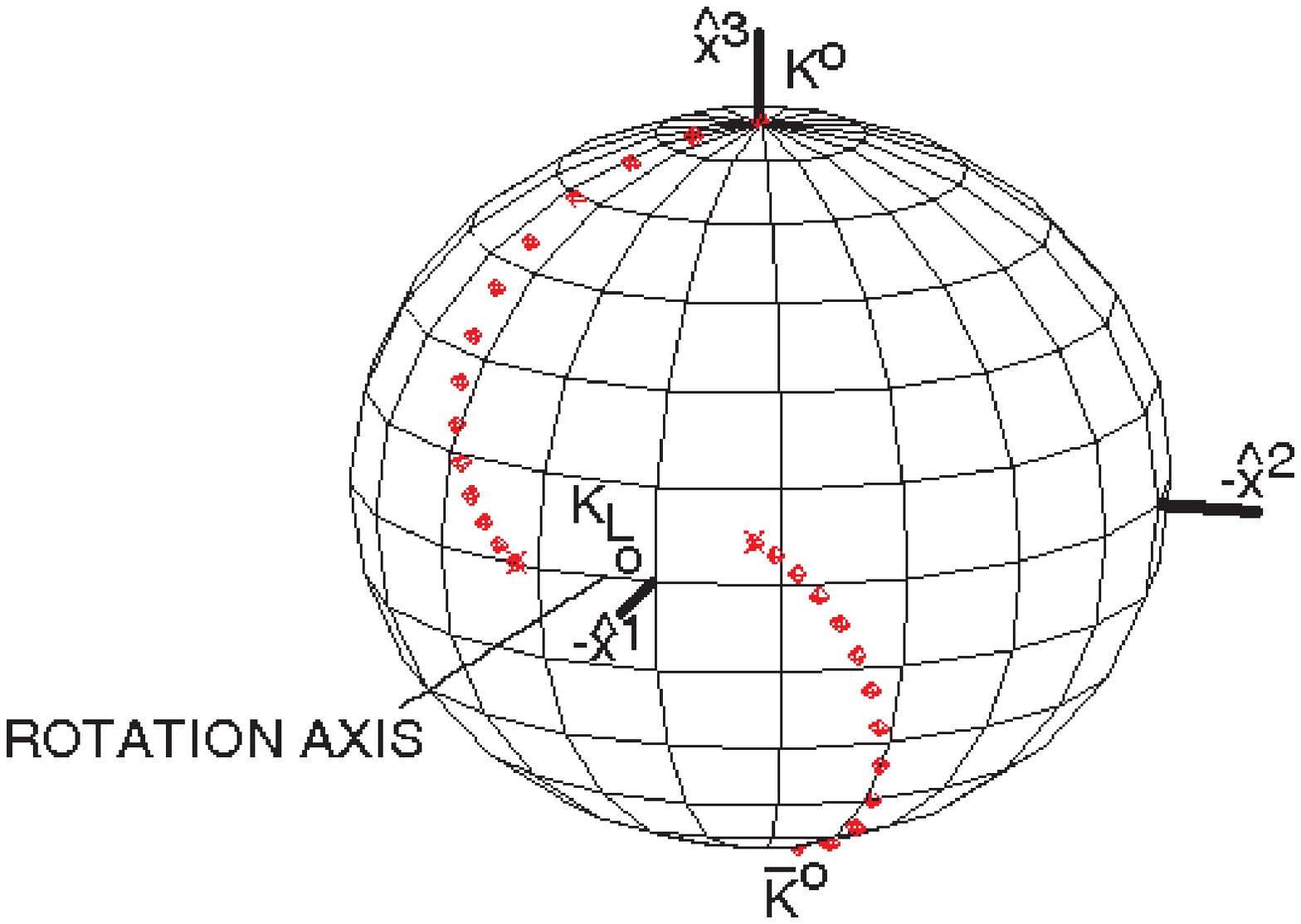}
  \caption{}
  \label{fig:kbfaster}
  \end{center}                                               
\end{figure}

\begin{figure}
  \begin{center}
  \includegraphics[angle=0,bb=13 1 376 361,width=0.6\linewidth,clip]
                                                   {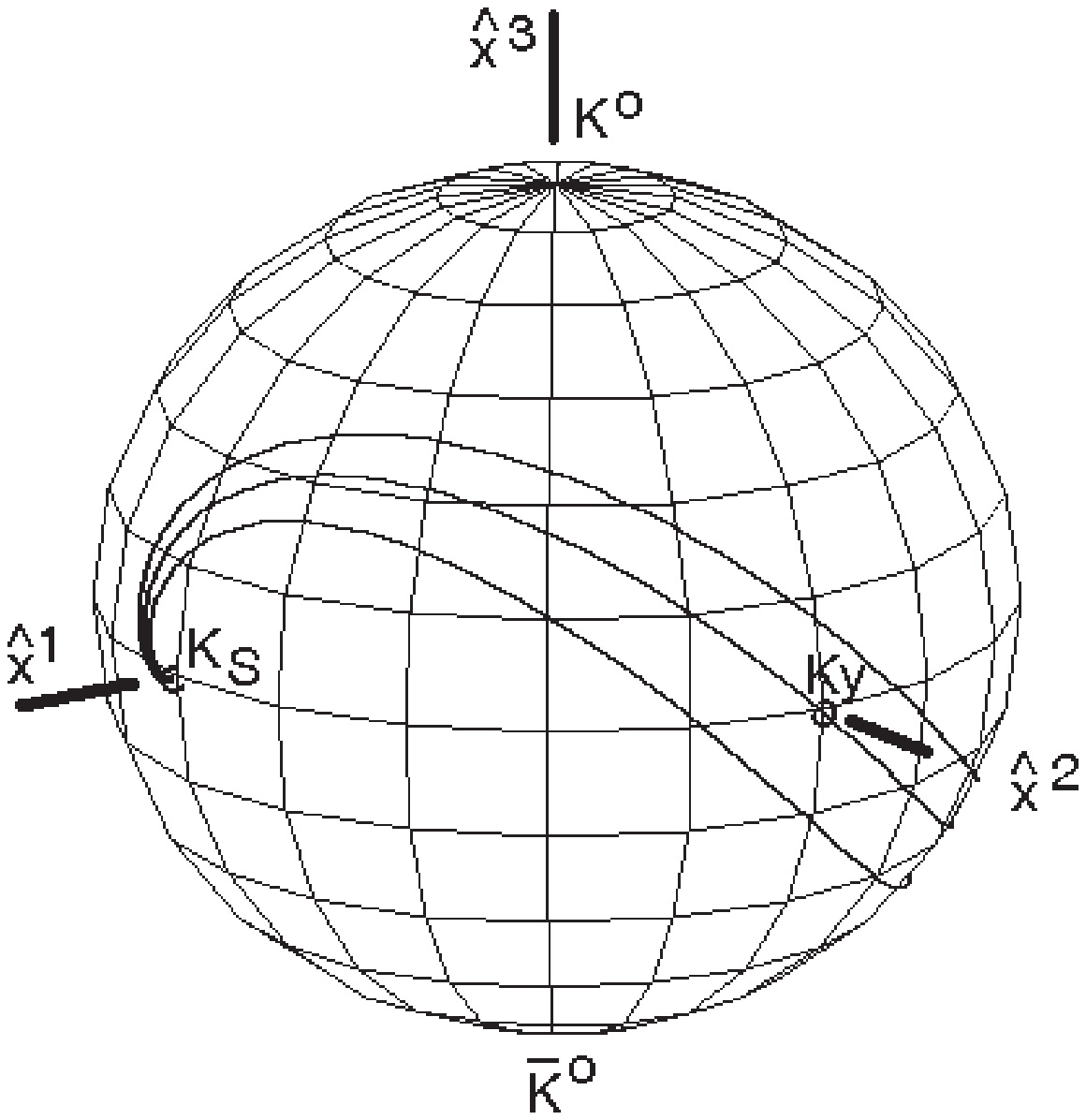}
  \caption{}
  \label{fig:Regs}
  \end{center}
\end{figure}

   Fig.~\ref{fig:at_mod} -- Experimental demonstration of \Tz\ invariance 
   violation in the $time \ evolution$ of neutral kaons. \\ 
   The positive values show that a \knb\ develops into a \kn\ with higher 
   probability than does a \kn\ into a \knb . This $contradicts 
   \ \Tz ^{-1} \Hwk\ \Tz = \Hwk$ \ for neutral kaons.\\
   The points are the measured values of $\ATexp (t)$.
   The solid line represents the fitted constant \avATexp.
   All symmetry violating parameters concerning the semileptonic 
   $decay \ process$ are compatible with zero.
   The dashes indicate $- 4\re (y+x_{-}) = 0.8 \times 10^{-3}$, 
   the contribution of the constrained hypothetical \CPTz~\ invariance 
   violation in the $decay$, expressed in Eq. (\ref{atexp}).
   The time dependent curve, seen at early times, indicates a further 
   constrained hypothetical violation of the $\DS = \DQ$ rule,
   expressed by the function g (with exagerated input values). \\
   Data from Ref. \cite{pen2}.

\subsection*{}
   Fig.~\ref{fig:kbfaster} --
   Insight into \Tz\ violation: \ the \knb\ runs faster than the \kn.\\
   The curves on the surface of the unit sphere display $\vec{P}(t)$ of \kn\ and \knb\ .
   The \kn\ starts at the north pole, the \knb\ starts at the south pole.
   Their movement is a superposition of a boost and a rotation, as Eq. (\ref{dRlor}) explains.
   The curves show that the trajectory of the original \knb\ crosses the 
   equator earlier, than does the one of the original \kn\ .
   Both curves represent the same total amount of running time.\\
   If \Tz\ invariance were not violated, then both curves would cross the equator at the same time $t_{eq}$ .\\
   (The \Tz\ violation effect is exagerated in this figure by a factor of 20 .)
   
\subsection*{}
   Fig.~\ref{fig:Regs} --
   High sensitivity of the time development, to the a regenerator acting on
   \ks\ particles .\\
   The curves 
   $\vec{P}(t)$ on the surface of the unit sphere show the time evolutions of initial \ks\ particles which have passed 
   regenerators of different thicknesses
   (0.9 mm, 1.2 mm, and 1.5 mm; values from Section \ref{carbon}).
   One of the particles evolves through a state, called $\mathrm{K}_y$ , which is
   located on the equator at 90 degrees of longitude.\\
   The state $\mathrm{K}_y$ is one of the two states which are complementary to
   all of \kn , \knb , $\mathrm{K}_1$\ , $\mathrm{K}_2$\ .
   
\subsection*{}
\begin{figure}
  \begin{center}
  \includegraphics[angle=0,bb=25 389 375 578,width=0.6\linewidth,clip]
                                                       {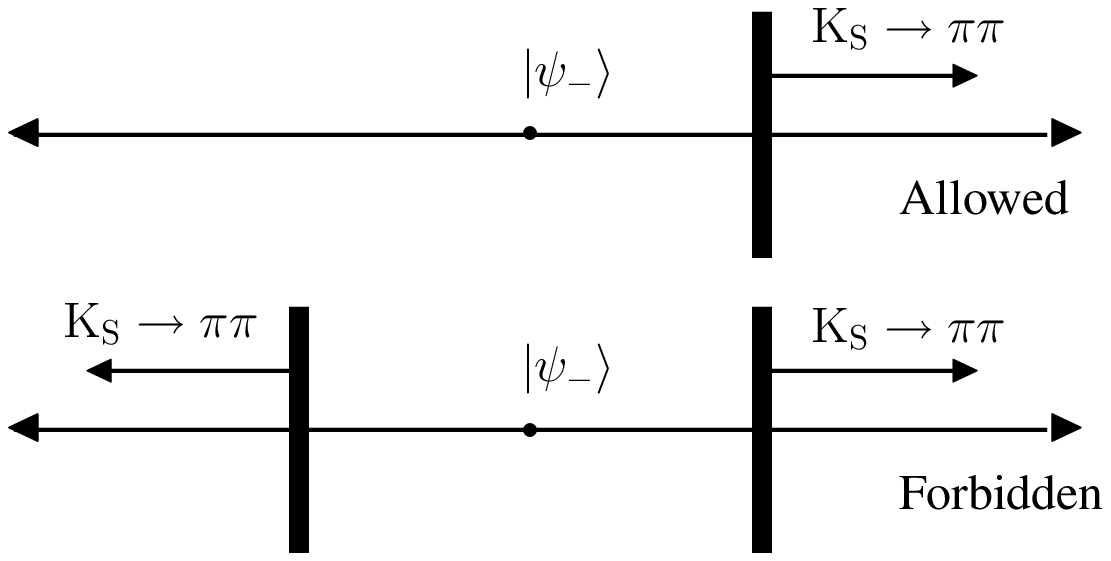}
  \caption{}
  \label{fig:regpx}
  \end{center}
\end{figure}

   Fig.~\ref{fig:regpx} --    
   Symmetrically placed regenerators in the paths of the two kaons, which form
   the entangled state \ $\ket{\psi_-}$ , do not produce identical 
   regeneration products, such as e. g. \ks\ks\ .

\subsection*{}
\begin{figure}
  \begin{center}
  \includegraphics[angle=0,bb=47 411 520 723,width=\linewidth,clip]
                                                     {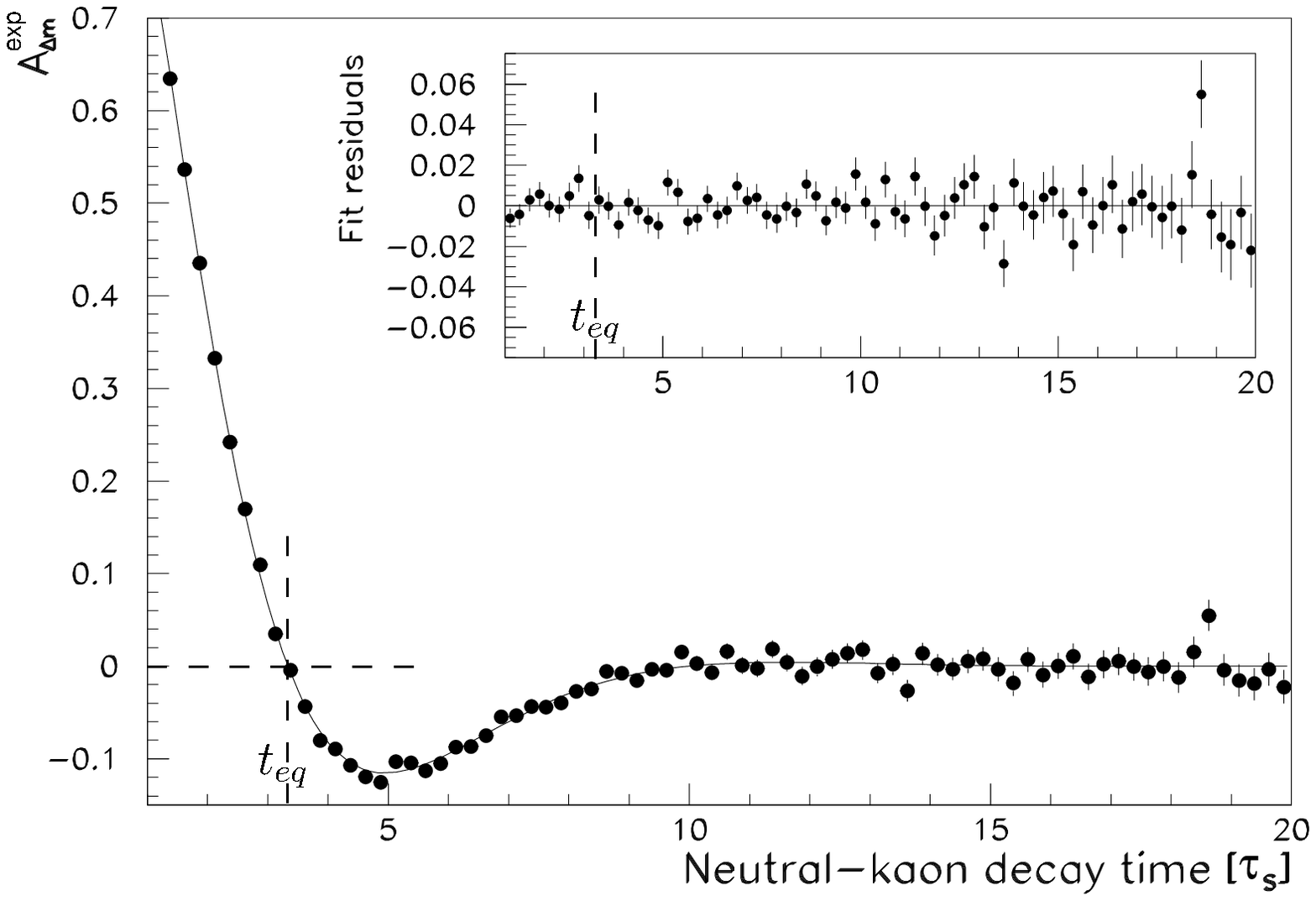}
  \caption{}
  \label{fig:PR35}
  \end{center}
\end{figure}

   Fig.~\ref{fig:PR35} --  
   At age \ $t_{eq}$ a \kn\ (or \knb ) decays as often into \ \ra\ \e$^+$ \pim\ $\nu$ \ as into
   \ \ra\ \e$^-$ \pip\ $\bar{\nu}$.
   \\
   At this time the kaon is in a state complementary to the ones of \ \kn\ and of \ \knb , and thus cannot
   decide whether it wants to decay as a \ \kn\ \ra\ \e$^+$ \pim\ $\nu$ \ or as a \ \knb\ \ra\ \e$^-$ \pip\ 
   $\bar {\nu}$.
   \\
   The fact that the values for $t_{eq}$ in this figure and in Fig. \ref{fig:PR41} coincide is a confirmation
   of the \DS\ = \DQ\ rule.
   \\
   Data from Ref. \cite{pen1,pen4}.

\subsection*{}
\begin{figure}
 \begin{center}
   \begin{tabular}{cc}
   (a)&(b)\\
   \includegraphics[bb=172 159 502 483,width=7cm,clip]{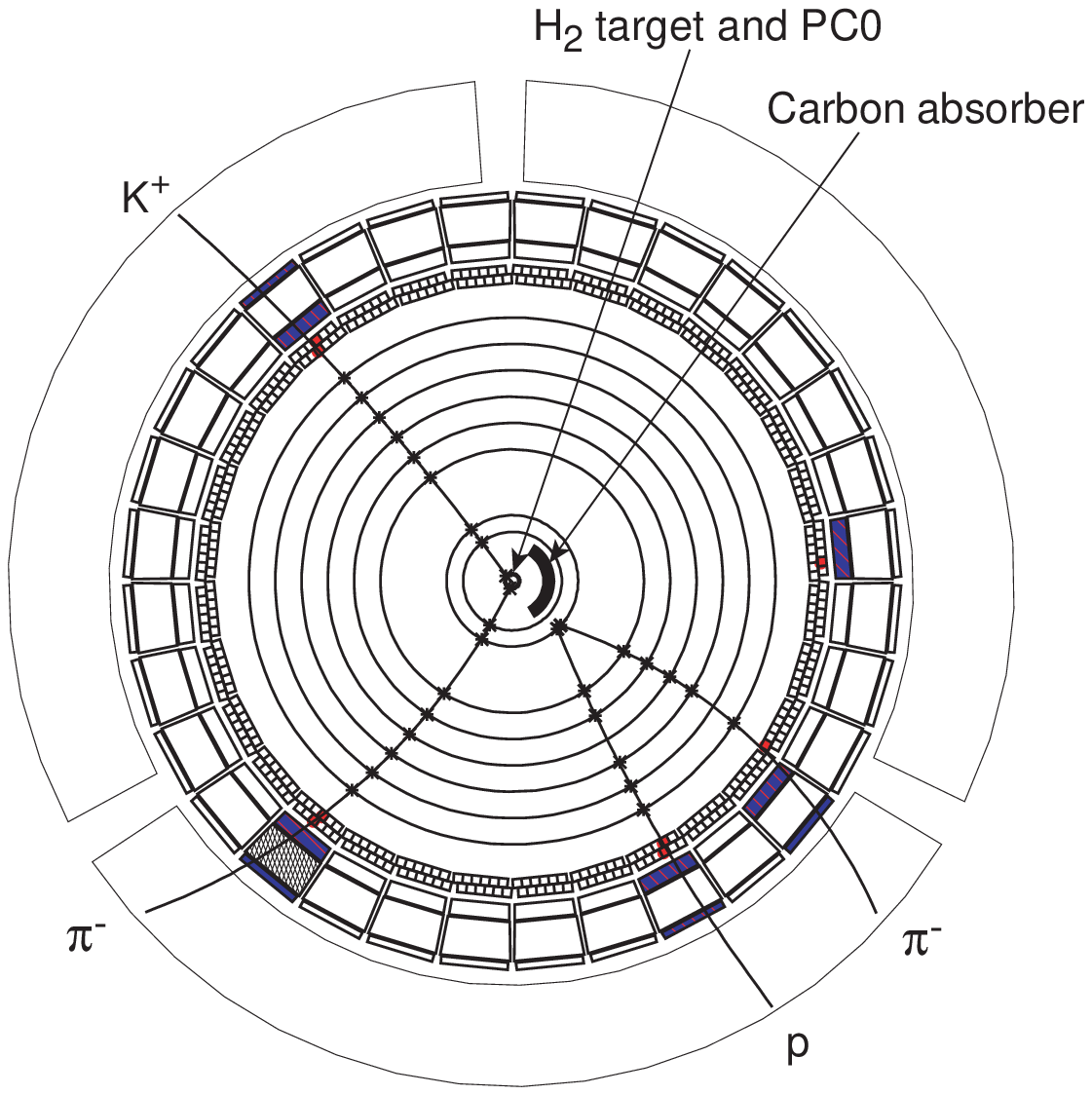}&
    \includegraphics[bb=91 390 459 733,width=7cm,clip]{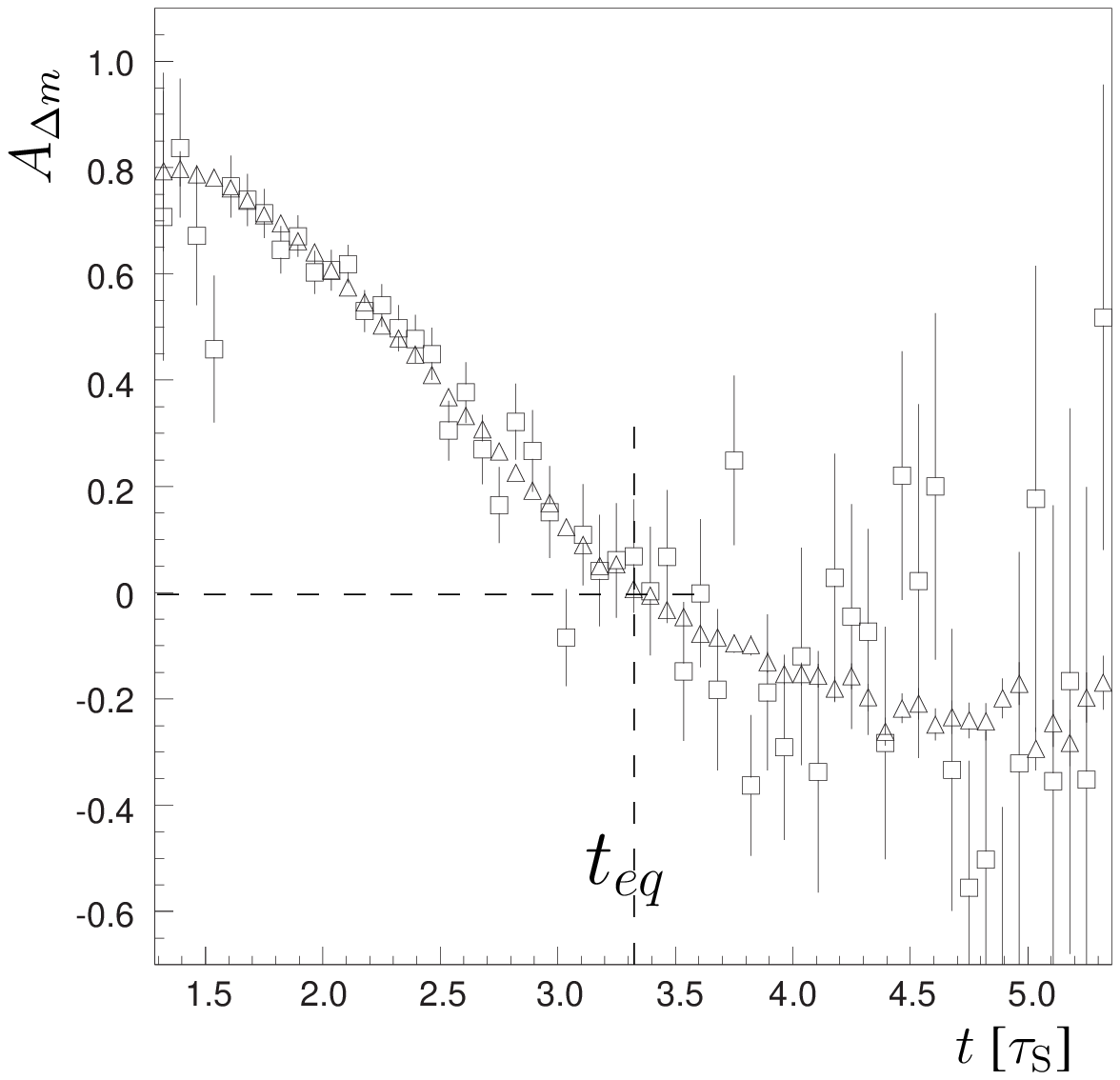}
   \end{tabular}
  \caption{}
  \label{fig:PR41}
 \end{center}
\end{figure}

   Fig.~\ref{fig:PR41} --   
   Zero crossing of \Admexp\ as a manifestation of complementarity. \\ 
   (a) Production of a \ \knb\ and detection of its reaction in the carbon 
   absorber. \\
   The \ppb\ annihilation, \ppb\ \ra\ \knb\ + \kp\ + \pim\ in the H$_2$ target,
   produces the (invisible) \knb , which propagates towards the Carbon 
   absorber, where it creates a (invisible) \Lz\ particle, whose decay products p and \pim\
   from \Lz\ \ra\ p + \pim\ are detected. The information in the 
   picture derived from the charged (visible) particles serves to determine the strangeness of the produced kaon, 
   its individual life time, and its strangeness at the moment
   it produces a reaction in the absorber.
   \\
   The \knb\ in this picture has not changed its (negative) strangeness between target and absorber.
   \\
   (b) Measured asymmetry \Admexp\ between the frequencies of kaons which have changed, and those which have not
   changed their strangeness. Squares: measured, triangles: calculated.
   \\
   The vanishing of \Admexp\ $\propto\ \langle \sigma^3\rangle $ at time $t_{eq}$ may be interpreted in terms of
   the complementarity relation (\ref{unc1}): when $\langle \sigma^3\rangle = 0 $, then a measurement must (and
   does) \it not\rm\ allow one to find out whether the kaon is a \kn\ or a \knb\ (and vice versa).
   \\
   Data from Ref. \cite{dm_a}.


\newpage


\bigskip
Figures 1, 2, 5, 6, 7 and 12 have been reprinted with the 
permission of Elsevier.

Figures 3, 4 and 11 have been reprinted with the permission of Springer.
\end{document}